\documentclass{article}

\usepackage[final,nonatbib]{arxiv}
\usepackage[10pt]{moresize}

\usepackage[numbers]{natbib}

\usepackage{amsmath,amsfonts,bm}

\def\eqref#1{equation~\ref{#1}}

\def\1{\bm{1}}

\def\ra{{\textnormal{a}}}

\DeclareMathAlphabet{\mathsfit}{\encodingdefault}{\sfdefault}{m}{sl}
\SetMathAlphabet{\mathsfit}{bold}{\encodingdefault}{\sfdefault}{bx}{n}

\newcommand{\R}{\mathbb{R}}

\DeclareMathOperator*{\argmin}{arg\,min}

\usepackage{amsmath}
\usepackage{amsthm}
\usepackage{amsfonts}       
\usepackage{nicefrac}       

\usepackage{hyperref}

\usepackage[utf8]{inputenc} 
\usepackage[T1]{fontenc}    
\usepackage{booktabs}       
\usepackage{dcolumn}
\usepackage{siunitx}
\usepackage{amsfonts}       
\usepackage{nicefrac}       
\usepackage{microtype}      

\usepackage{overpic}

\usepackage[]{amsmath}
\usepackage{upgreek}
\usepackage{overpic}
\usepackage[]{multirow}
\usepackage[]{nicefrac}
\usepackage[]{bm}
\usepackage[]{mathrsfs}
\usepackage[]{afterpage}

\usepackage{amssymb}
\usepackage{graphicx}
\usepackage{epstopdf}
\usepackage{subcaption}
\usepackage{multirow}
\usepackage{float}

\usepackage{pgfplots}
\usepackage{mathtools}

\usepackage{tikz}
\usetikzlibrary{matrix,positioning,shapes,shadows,arrows,calc,decorations.pathreplacing}
\usetikzlibrary{arrows.meta}

\definecolor{darkred}{RGB}{228,26,28}
\definecolor{darkblue}{RGB}{44,127,184}
\definecolor{magentaCB}{RGB}{221,28,119}

\definecolor{morange}{RGB}{255, 187, 0}
\definecolor{mblue}{RGB}{ 0, 161, 241}

\newcommand{\dumarg}[1]		{\widetilde{#1}}

\newcommand{\ie}		{\textit{i.e.}}
\newcommand{\eg}		{\textit{e.g.}}

\DeclareMathOperator*{\esp}	{\mathbb{E}}

\mathchardef\mhyphen="2D		

\newcommand{\ldeux}		{L^2}

\newcommand{\lzero}		{L^0}
\newcommand{\lone}		{L^1}

\newcommand{\transpose}		{\mathsf{T}}

\newcommand{\vectorize}[1]	{\boldsymbol{#1}}
\newcommand{\matricize}[1]	{\boldsymbol{\MakeUppercase{#1}}}

\newcommand{\be}	{\begin{equation}}
\newcommand{\ee}	{ \end{equation}}
\newcommand{\bi}	{\begin{itemize}}
\newcommand{\ei}	{ \end{itemize}}
\newcommand{\bea}	{\begin{eqnarray}}
\newcommand{\eea}	{ \end{eqnarray}}
\newcommand{\benum}	{\begin{enumerate}}
\newcommand{\eenum}	{ \end{enumerate}}
\newcommand{\bc}	{\begin{center}}
\newcommand{\ec}	{ \end{center}}
\newcommand{\bealign}	{\begin{align}}
\newcommand{\eealign}	{ \end{align}}

\newcommand{\normLM}[2]		{\left\|{#1}\right\|_{#2}}
\newcommand{\normLMsq}[2]	{\left\|{#1}\right\|_{#2}^2}

\newcommand{\parLM}[1]  {\left\{{#1}\right\}}

\newcommand{\ds}    {\displaystyle}
\newcommand{\bmat}  {\begin{bmatrix}}
\newcommand{\emat}  {\end{bmatrix}}

\newcommand{\obs}       {s}
\newcommand{\bobs}      {\vectorize{\obs}}
\newcommand{\matobs}    {\matricize{\obs}}
\newcommand{\iobs}      {j}
\newcommand{\nobs}      {p}

\newcommand{\nsnap}     {n}
\newcommand{\isnap}     {i}

\newcommand{\qoi}       {x}
\newcommand{\qoiest}    {\est{\qoi}}
\newcommand{\bqoi}      {\vectorize{\qoi}}

\newcommand{\bqoiest}   {\vectorize{\qoiest}}
\newcommand{\nqoi}      {m}

\newcommand{\matqoi}    {\matricize{\qoi}}

\newcommand{\coefs}     {\nu}
\newcommand{\bcoefs}    {\vectorize{\coefs}}
\newcommand{\matcoefs}  {\boldsymbol{N}}

\newcommand{\bcoefslow} {\bcoefs^\star}

\newcommand{\feat}     {\psi}
\newcommand{\bfeat}    {\vectorize{\feat}}

\newcommand{\stocoef}  {\boldsymbol{P}}

\newcommand{\setsensor}{\mathcal{J}}

\newcommand{\mode}      {\phi}
\newcommand{\matmode}   {\boldsymbol{\Phi}}
\newcommand{\bmode}     {\vectorize{\mode}}
\newcommand{\imode}     {j}
\newcommand{\nmodes}    {k}

\newcommand{\measope}   {\boldsymbol{H}}
\newcommand{\obsope}   {\boldsymbol{G}}

\newcommand{\est}[1]	{\widehat{#1}}

\newcommand{\bbias}     {\boldsymbol{b}}
\newcommand{\bz}        {\boldsymbol{z}}
\newcommand{\Wmat}      {\matricize{w}}

\newcommand{\NNclass}   {\mathscr{F}}
\newcommand{\by}        {\boldsymbol{y}}

\usepackage{lipsum}
\usepackage{amsfonts}
\usepackage{graphicx}
\usepackage{epstopdf}
\usepackage{algorithmic}
\ifpdf
\DeclareGraphicsExtensions{.eps,.pdf,.png,.jpg}
\else
\DeclareGraphicsExtensions{.eps}
\fi

\usepackage[verbose=true,letterpaper]{geometry}
\AtBeginDocument{
	\newgeometry{
		textheight=8.8in,
		textwidth=6.3in,
		top=1in,
		headheight=14pt,
		headsep=25pt,
		footskip=30pt
	}
}

\title{Shallow Neural Networks for Fluid Flow Reconstruction with Limited Sensors}

\date{} 					

\author{%
	N. Benjamin Erichson \\
	ICSI and UC Berkeley\\
	\texttt{erichson@berkeley.edu}
	\And
	Lionel Mathelin \\
	Universit\'e Paris-Saclay, LIMSI-CNRS \\
	\texttt{mathelin@limsi.fr}
	\And
	Zhewei Yao \\
	ICSI and UC Berkeley \\
	\texttt{zheweiy@berkeley.edu}
	\And
	Steven L. Brunton \\
	University of Washington \\
	\texttt{sbrunton@uw.edu}
	\And	
	Michael W. Mahoney \\
	ICSI and UC Berkeley\\
	\texttt{mmahoney@stat.berkeley.edu}
	\And
	J. Nathan Kutz \\
	University of Washington \\
	\texttt{kutz@uw.edu}
	}

\begin{document}

\maketitle
\begin{abstract}
In many applications, it is important to reconstruct a fluid flow field, or some other high-dimensional state, from limited measurements and limited data. In this work, we propose a shallow neural network-based learning methodology for such fluid flow reconstruction. Our approach learns an end-to-end mapping between the sensor measurements and the high-dimensional fluid flow field, without any heavy preprocessing on the raw data. No prior knowledge is assumed to be available, and the estimation method is purely data-driven. We demonstrate the performance on three examples in fluid mechanics and oceanography, showing that this modern data-driven approach outperforms traditional modal approximation techniques which are commonly used for flow reconstruction. Not only does the proposed method show superior performance characteristics, it can also produce a comparable level of performance with traditional methods in the area, using significantly fewer sensors. Thus, the mathematical architecture is ideal for emerging global monitoring technologies where measurement data are often limited.
\end{abstract}

\section[Introduction]{Introduction}
\label{sec: introduction}

\begin{figure}[!b]
	\centering
	\DeclareGraphicsExtensions{.png}
	\includegraphics[width=0.95\textwidth]{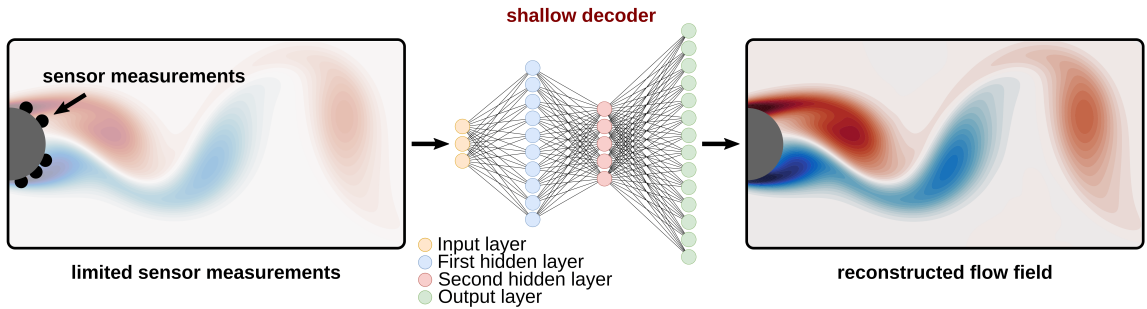}
	\vspace{+0.2cm}
	\caption{Illustration of our \textsc{shallow decoder} which maps a few sensor measurements $\bobs \in \R^{5}$ to the estimated field $\bqoiest \in \R^{78,406}$. In other words, this neural network based learning methodology provides an end-to-end mapping between the sensor measurements and the fluid flow field.}
	\label{fig:overview}
\end{figure}

The ability to reconstruct coherent flow features from limited observation can be critically enabling for applications across the physical and engineering sciences~\cite{Brunton2015amr,rowley2017model,Callaham2018arxiv,manohar2018data,yu2018flowfield}.
For example, efficient and accurate fluid flow estimation is critical for active flow
control, and it may help to craft more fuel-efficient automobiles as well as high-efficiency turbines. The ability to reconstruct important fluid flow features from limited observation is also central in applications as diverse as cardiac bloodflow modeling and climate science~\cite{doi:10.1029/2018MS001472}. All of these applications rely on estimating the structure of fluid flows based on limited sensor measurements.

More concretely, the objective is to estimate the flow field $\bqoi \in \R^{\nqoi}$ from sensor measurements $\bobs \in \R^{p}$, that is, to learn the relationship $\bobs \mapsto \bqoi$.
The restriction of limited sensors gives $\nobs \ll \nqoi$.
The sensor measurements $\bobs$ are collected via a sampling process from the high-dimensional field $\bqoi$. We can describe this process as
\begin{equation}
	\bobs = \measope(\bqoi),
\end{equation}
where $\measope: \R^{\nqoi} \ra \R^{\nobs}$ denotes a measurement operator. Now, the task of flow reconstruction requires the construction of an inverse model that produces the field $\bqoi$ in response to the observations $\bobs$, which we may describe as
\begin{equation}
	\bqoi = \obsope(\bobs),
\end{equation}
where $\obsope: \R^{\nobs} \ra \R^{\nqoi}$ denotes a non-linear forward operator.
However, the measurement operator $ \measope $ may be unknown or highly-nonlinear in practice.
Hence, the problem is often ill-posed, and we cannot directly invert the measurement operator $\measope$ to obtain the forward operator $\obsope$.

Fortunately, given a set of training examples $\{\bqoi_i, \bobs_i\}_i$, we may learn a function $\mathcal{F}$ to approximate the forward operator $\obsope$.
Specifically, we aim to learn a function $\mathcal{F}: \bobs \mapsto \bqoiest$ which maps a limited number of measurements to the estimated state $\bqoiest$:
\begin{equation}
\bqoiest \, =  \, \mathcal{F} \left(\bobs\right),
\end{equation}
%
%
so that the misfit is small, \emph{e.g.}, in an Euclidean sense over all sensor measurements
\begin{equation*}
	\|\mathcal{F}(\bobs) - \obsope(\bobs) \|_2^2 < \epsilon,
\end{equation*}
where $\epsilon$ is a small positive number.
Neural network based inversion is common practice in machine learning~\cite{mccann2017review}, dating back to the late 80's~\cite{zhou1988image}. This powerful learning paradigm is also increasingly used for flow reconstruction~\cite{ling_kurzawski_templeton_2016,carlberg2018recovering,fukami2018super}, prediction~\cite{vlachas2018data,erichson2019physics,azencot2020forecasting,guastoni2019prediction,srinivasan2019predictions}, and simulations~\cite{kim2018deep}.
In particular, deep inverse transform learning is an emerging concept~\cite{mousavi2017learning,jin2017deep,adler2017solving,ye2018deep}, which has been shown to outperform traditional methods in applications such as denoising, deconvolution, and super-resolution.

Here, we explore shallow neural networks (SNNs) to learn the input-to-output mapping between the sensor measurements and the flow field. 
Figure~\ref{fig:overview} shows a design sketch for the proposed framework for fluid flow reconstruction. We can express the network architecture (henceforth called \textsc{shallow decoder} (SD)), more concisely as follows:
\begin{align*}
	\bobs \mapsto \text{first hidden layer} \mapsto \text{second hidden layer} \mapsto \text{output layer} \mapsto \bqoiest.
\end{align*} 
%
%
SNNs are considered to be networks with very few hidden layers. We favor shallow over deep architectures, because the simplicity of SNNs allows faster training, less tuning, and easier interpretation (and also since it works, and thus there is no need to consider deeper architectures).

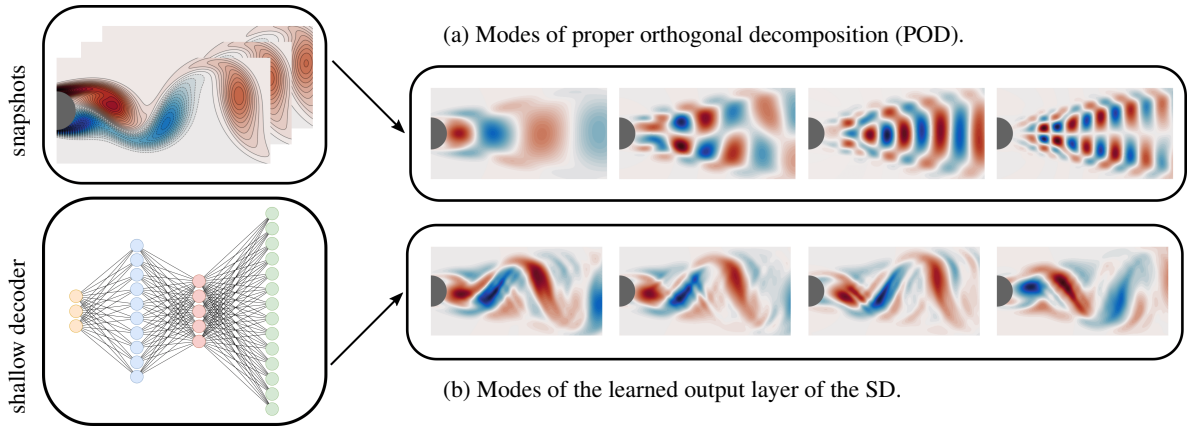
\begin{figure}[!b]
	\centering
	\DeclareGraphicsExtensions{.png}
	\begin{overpic}[width=0.95\textwidth]{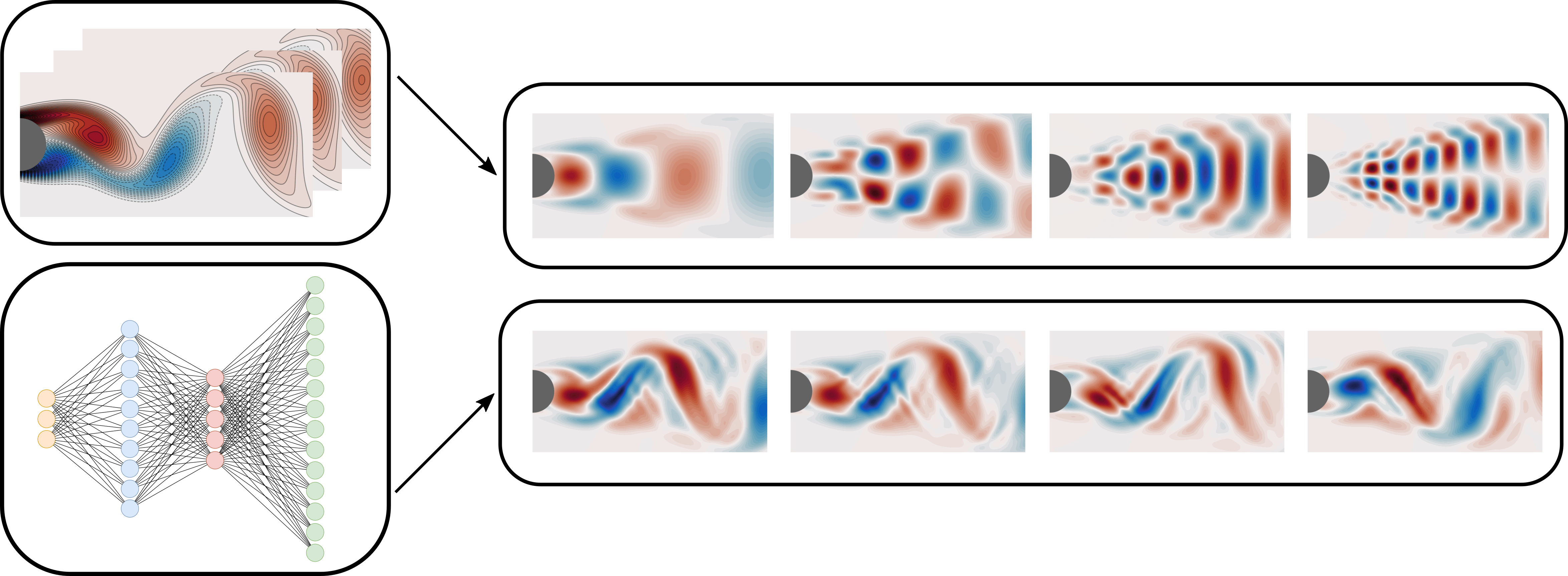} 
		\put(-3,23.5){\rotatebox{90}{\small snapshots}}
		\put(-3,1){\rotatebox{90}{\small {shallow decoder}}}
		\put(35,33.5){\small (a) Modes of proper orthogonal decomposition (POD).}
		\put(35,2.5){\small (b) Modes of the learned output layer of the SD.}
	\end{overpic}
	\vspace*{+0.3cm}	
	\caption{Dominant modes learned by the \textsc{shallow decoder} in contrast to the POD modes. These dominant features show that the SD constructs a reasonable characterization of the flow behind a cylinder. Indeed, by not constraining the modes to be linear and orthogonal, as is enforced with POD, a potentially more interpretable feature space can be extracted from data. Such modes can be exploited for reconstruction of the state space from limited measurements and limited data.}
	\label{fig:acitvation_fun_small}
\end{figure}

There are several advantages of this mathematical approach over traditional scientific computing methods for fluid flow reconstruction~\cite{chaturantabut2010nonlinear,drmac2016new,bui2004aerodynamic,willcox2006unsteady,manohar2018data}. 
First, the SD { considered here features a linear last layer and} provides a supervised joint learning framework for the low-dimensional approximation space of the flow field and the map from the measurements to this low-dimensional space. 
This allows the approximation basis to be tailored not only to the state space but also to the associated measurements, preventing observability issues. 
In contrast, these two steps are disconnected in standard methods (discussed in more detail in Section~\ref{sec:Sec_Formulation}). 
Second, the method allows for flexibility in the measurements, which do not necessarily have to be linearly related to the state, as in many standard methods. 
Finally, the shallow decoder network produces interpretable features of the dynamics, potentially improving on classical proper orthogonal decomposition (POD), also known as principal component analysis (PCA), low-rank features.  
For instance, Figure~\ref{fig:acitvation_fun_small} shows that the basis learned via an SNN exhibits elements resembling physically consistent quantities, in contrast with alternative POD-based modal approximation methods that enforce orthogonality. { The interpretation of the last (linear) layer is as follows: a given mode is constituted by the value of each spatially localized weights connecting the associated given node in the last hidden layer to nodes of the output layer.}
 
Limitations of our approach are standard to data-driven methods, in that the training data should be as representative as possible of the system, in the sense that it should comprise samples drawn from the same statistical distribution as the testing data.

The paper is organized as follows. 
Sec.~\ref{sec:Sec_Formulation} discusses traditional modal approximations techniques. 
%
%
Then, in Sec.~\ref{sec:shallow}, the specific implementation and architecture of our \textsc{shallow decoder} is described.
Results are presented in Sec.~\ref{sec:exp_result} for various applications of interest. 
We aim to reconstruct (a) the vorticity field of a flow behind a cylinder from a handful sensors on the cylinder surface, (b) the mean sea surface temperature from weekly sea surface temperatures for the last 26 years, and (c) the velocity field of a turbulent isotropic flow. 
We show that a very small number of sensor measurements is indeed sufficient for flow reconstruction in these applications. Further, we show that the \textsc{shallow decoder} can handle non-linear measurements and is robust to measurement noise.
The results show significantly improved performance compared to traditional modal approximations techniques.
The paper concludes in Sec.~\ref{sec:Sec_Discussion} with a discussion and outlook of the use of SNNs for more general flow field reconstructions.

\section{Background on high-dimensional state estimation} \label{sec:Sec_Formulation}

The task of { reconstructing} from a limited number of measurements to the high-dimensional state-space is made possible by the fact that the dynamics for many complex systems, or datasets, exhibit some sort of low-dimensional structure. 
This fact has been exploited for state estimation using (i) a tailored basis, such as POD, or (ii) a general basis in which the signal is sparse, \emph{e.g.}, typically a Fourier or wavelet basis will suffice.  
In the former, {\em gappy POD} methods~\cite{everson1995karhunen} have been developed for principled {reconstruction} strategies~\cite{chaturantabut2010nonlinear,drmac2016new,bui2004aerodynamic,willcox2006unsteady,manohar2018data}.  
In the latter, {\em compressive sensing} methods~\cite{candes2006robust,donoho2006compressed,baraniuk2007compressive} serve as a principled technique for reconstruction. 
Both techniques exploit the fact that there exists a basis in which the high-dimensional state vector has a sparse, or compressible, representation. In~\cite{gobalpaper}, a basis is learned such that it leads to a sparse approximation of the high-dimensional state while enforcing observability from the sensors.

Next, we describe standard techniques for the estimation of a state $\bqoi$ from observations $\bobs$, and we discuss observability issues. Established techniques for state reconstruction are based on the idea that a field $\bqoi$ can be expressed in terms of a rank-$k$ approximation
\be \label{eq:DLeq}
\bqoi \approx \bqoiest = \sum_{\imode=1}^{k}{\bmode_{\imode} \, \coefs_{\imode}} = \matmode \, \bcoefs \,,
\ee
where $\ds \parLM{\bmode_{\imode}}_{\imode}$ are the \emph{modes} of the approximation and $\parLM{\coefs_{\imode}}_{\imode}$ are the associated coefficients. 
The approximation space is derived from a given training set using unsupervised learning techniques. A typical approach to determine the approximation modes is POD~\cite{barrault2004empirical,chaturantabut2010nonlinear,drmac2016new,manohar2018data}. Randomized methods for linear algebra enable the fast computation of such approximation modes~\cite{mahoney2011randomized,DM16_CACM,halko2011finding,erichson2016randomized,erichson2018sparse,Erichson2017rdmd}.  
Given the approximation modes $\matmode$, estimating the state $\bqoi$ reduces to determining the coefficients $\bcoefs$ from the sensor measurements $\bobs$ using supervised techniques.
These typically aim to find the minimum-energy or minimum-norm solution that is consistent in a least-squares sense with the measured data.

\subsection{Standard approach: Estimation via POD based methods}

Two POD-based methods are discussed, which we will refer to as \textsc{pod} and \textsc{pod plus} in the following. Both approaches reconstruct the state with POD modes, by estimating the coefficients from sensor information. { The POD modes $\matmode$ are obtained via the singular value decomposition of the mean centered training set $\matqoi = \left(\bqoi_1 \ldots \bqoi_\nsnap\right)$, with typically $\nsnap \le \nqoi$:
\begin{equation}
\matqoi = \boldsymbol{U} \, \boldsymbol{\Sigma} \, \boldsymbol{V}^\transpose \,,
\end{equation}
where the columns of $\boldsymbol{U} \in \R^{\nqoi \times n}$ are the left singular vectors and the columns of $\boldsymbol{V} \in \R^{n \times n}$ are the right singular vectors. The corresponding singular values are the diagonal elements of $ \boldsymbol{\Sigma} \in \R^{n \times n}$. Now, we define the approximation modes as $\matmode := \boldsymbol{U}_k$, by selecting $k$ left singular vectors, with $k\leq p$. Typically, we select the dominant $k$ singular vectors as approximation modes, however, there are exceptions to this rule as discussed below.
}

\subsubsection{Standard POD-based method} 
 Let a linear measurement operator $\measope: \R^{\nqoi} \ra \R^{\nobs}$ describe the relationship between the field and the associated observations, $\bobs = \measope \, \bqoi$. The approximation of the field $\bqoi$ with the approximation modes $\parLM{\bmode_{\imode}}_{\imode}$ is obtained by solving the following equation for {$\bcoefs \in \R^{ n}$}:
\be
\bobs = \measope \, \bqoi \approx \measope \matmode \, \bcoefs \,.
\label{sfromy}
\ee
A standard approach is to simply solve the following least-squares problem 
\be
\bcoefs \in \argmin_{\dumarg{\bcoefs}} \; \normLM{\bobs - \measope \, \matmode \, \dumarg{\bcoefs}}{2}^2 \,. \label{POD_eqreg}
\ee
%
%
The solution with the minimum $\ldeux$-norm is given by:
\be
\bcoefs = \left(\measope \, \matmode\right)^+ \, \bobs \,,
\label{coef_Tikho}
\ee
with the superscript $^+$ denoting the Moore-Penrose pseudo-inverse. In this situation, the high-dimensional state is then estimated as
\be
\bqoi \approx \bqoiest = \matmode \, \bcoefs \,.
\ee
This approach is hereafter referred to as \textsc{POD} { and has been used in previous efforts, \emph{e.g.}, \cite{Murray_Ukeiley_2007, Podvin_2005}.

With a nonlinear measurement operator $\measope$, the problem formulates similarly as a nonlinear least squares problem:
\be
\bcoefs \in \argmin_{\dumarg{\bcoefs}} \; \normLM{\bobs - \measope \left(\matmode \, \dumarg{\bcoefs}\right)}{2}^2 \,. \label{POD_eqreg_nonlin}
\ee
In this case, no closed form solution is available in general and a nonlinear optimization problem must be solved, whose computational burden limits the online (real-time) field reconstruction capability. Further, the solution of the, often ill-posed, problem is not necessarily unique and does not allow for a reliable estimate. In contrast, the shallow decoder is trained end-to-end and essentially learns to associate measurements to the right solution (see Section~\ref{sec:shallow} for details) .
}

\subsubsection{Improved POD-based method} 
{
The standard POD-based method has several shortcomings. First, the least-squares problem formulated in Eq.~\ref{POD_eqreg} can be underspecified. Thus, it is favorable to introduce some bias in order to reduce the variance by means of regularization. Ridge regularization is the most popular regularization technique for reducing the variance of the estimator:
\be
\bcoefs \in \argmin_{\dumarg{\bcoefs}} \; \normLM{\bobs - \measope \, \matmode \, \dumarg{\bcoefs}}{2}^2 + \alpha \, \normLM{\dumarg{\bcoefs}}{2}^2 \,,
\ee
where $\alpha > 0$ is the penalization parameter. Typically, this parameter is determined by $k$-fold cross-validation. An alternative approach to reduce the variance is to select a subset of the POD modes, {\itshape i.e.}, only a few of the estimated coefficients are non-zero. The so-called Least Absolute Shrinkage and Selection Operator (LASSO) for least-squares~\cite{tibshirani1996regression,hastie2009elements} can be formulated as:
\be
\bcoefs \in \argmin_{\dumarg{\bcoefs}} \; \normLM{\bobs - \measope \, \matmode \, \dumarg{\bcoefs}}{2}^2 + \beta \, \normLM{\dumarg{\bcoefs}}{1} \,,
\ee
where $\beta > 0$ controls the amount of sparsity. 
One can also combine both LASSO and ridge regularization, resulting in the so-called ElasticNet~\cite{zou2005regularization,hastie2009elements} regularizer:
\be
\bcoefs \in \argmin_{\dumarg{\bcoefs}} \; \normLM{\bobs - \measope \, \matmode \, \dumarg{\bcoefs}}{2}^2 + \alpha \, \normLM{\dumarg{\bcoefs}}{2}^2 + \beta \, \normLM{\dumarg{\bcoefs}}{1} \,.
\ee
This regularization scheme often shows an improved predictive performance in practice, 
however, it requires that the user fiddles around with two tuning parameters $\alpha$ and $\beta$.

Yet another approach is to use a shrinkage estimator that only retains the high variance POD modes, {\itshape i.e.}, an estimator that selects a subset of all the POD modes that is used for solving the least squares problem. More concretely, we formulate the following constrained problem:
\be
\bcoefs \in \argmin_{\dumarg{\bcoefs}} \; \normLM{\bobs - \measope \, \matmode \, \dumarg{\bcoefs}}{2}^2 \quad \text{s.t.} \quad \matmode_{(n-k)}\dumarg{\bcoefs} =\boldsymbol{0} \,,
\ee
where $\matmode_{(n-k)} = \{\bm{\phi}_{k+1},\dots,\bm{\phi}_{n}\}$. Here, $k \le \nsnap$ refers to the number of selected POD modes, reordered with indices $\left\{1, 2, \ldots, k\right\}$. This hard threshold regularizer constraints the solution to the column space of the selected POD modes and is also known as Principal Component Regression (PCR)~\cite{hastie2009elements}.
In contrast to the smooth shrinkage effect of ridge regularization, the hard threshold regularizer has a discrete shrinkage effect that nullifies the contributions of some of the low variance modes completely. 
However, based on our experiments, both ridge regression and the hard threshold shrinkage estimator perform on par for the task of flow field reconstruction. 
This said, the ElasticNet regularizer might lead to a better predictive accuracy, since it can select the POD modes that are most useful for prediction, rather than only selecting the high-variance POD modes. It is known that the POD modes with low variances may also be important for predictive tasks~\cite{frank1993statistical,jolliffe1982note} and could help to further improve the performance of the POD-based methods.
}

Another shortcoming of the POD-based approach is that it requires explicit knowledge of the observation operator $\measope$ and is subjected to ill-conditioning of the least-squares problem. These limitations render this ``vanilla flavored'' approach often impractical in many situations, and they motivate an alternative formulation.
The idea is to learn the map between coefficients and observations without explicitly referring to $\measope$. It can be implicitly described by a, possibly nonlinear, operator $\stocoef: \R^{\nmodes} \ra \R^{\nobs}$ typically determined offline by minimizing the Bayes risk, defined as the misfit in the $\ldeux$-sense:
\be
\stocoef \in \argmin_{\dumarg{\stocoef}} \; {\esp{}_{\mu_{\bobs, \bcoefs}} \left[\normLMsq{\bobs - \dumarg{\stocoef} \, \bcoefs}{2}\right]} \,, \label{Eq_Bayesrisk}
\ee
where $\mu_{\bobs, \bcoefs}$ is the joint probability measure of { the observations $\bobs$ and the coefficients $\bcoefs$ obtained by projecting the field onto the (orthonormal) POD modes, $\bcoefs = \matmode^\transpose \, \bqoi$. This step only relies on information from the training set and is thus performed offline.}

We assume the training set is representative of the underlying system, in the sense that it should contain independent samples  drawn from the stationary distribution of the physical system at hand. The Bayes risk is then approximated by an empirical estimate, and the operator $\stocoef$ is determined as
\be
\stocoef \in \argmin_{\dumarg{\stocoef}} \; {\sum_{i=1}^{\nsnap}{\normLMsq{\bobs_i - \dumarg{\stocoef} \, \bcoefs_i}{2}}} \,. \label{Eq_Ptraining}
\ee
When the measurement operator $\measope$ is linear, $\stocoef$ is then an empirical estimate of $\measope \, \matmode$, the contribution of the basis modes $\parLM{\bmode_{\imode}}_{\imode}$ to the measurements $\bobs$.
{ This formulation was already considered in our previous work, \textit{e.g.}, \cite{gobalpaper}, and brings flexibility in the properties of the map $\stocoef$ compared to the closed-form solution in Eq.~\ref{coef_Tikho}}. For instance, regularization by sparsity can be enforced in $\stocoef$, via $\lzero$- or $\lone$-penalization.
Expressing Eq.~\ref{Eq_Ptraining} in matrix form yields:
\be
\stocoef \in \argmin_{\dumarg{\stocoef} \in \R^{\nobs \times \nmodes}} \; {\normLM{\matobs - \dumarg{\stocoef} \, \matcoefs}{F}^2} \,, \label{Eq_Ptraining2}
\ee
where $\matobs \in \R^{\nobs \times \nsnap}$ and $\matcoefs \in \R^{\nmodes \times \nsnap}$ respectively refers to the training data measurements $\parLM{\bobs_i}_i$ and coefficients $\parLM{\bcoefs_i}_i$. It immediatly follows
\bea
\stocoef \,  = \, \matobs \, \matcoefs^+ \, = \, \matobs \, \left(\matmode^+ \, \matqoi\right)^+ \, = \, \matobs \boldsymbol{V} \boldsymbol{\Sigma}^+ \,,
\eea
and the online approximation obtained by \textsc{pod plus} is finally given by the solution to the following least-squares problem
\be
\bcoefs \in \argmin_{\dumarg{\bcoefs}} \; \normLM{\bobs - \stocoef \, \dumarg{\bcoefs}}{2}^2 \,.
\ee
However, $\bcoefs \in \R^{\nmodes}$ is typically higher-dimensional than $\bobs \in \R^{\nobs}$, and thus the problem is ill-posed. We then make use of the popular Tikhonov regularization, selecting the solution with the minimum $\ldeux$-norm.
This results in a ridge regression problem formulated as:
\be
\bcoefs \in \argmin_{\dumarg{\bcoefs}} \; \normLM{\bobs - \stocoef \, \dumarg{\bcoefs}}{2}^2 + \lambda \, \normLM{\dumarg{\bcoefs}}{2}^2,
\ee
with $\lambda > 0$.
As will be seen in the examples below, penalization of the magnitude of the coefficients can significantly improve the performance of the POD approach.

\subsection{Observability issue}\label{sec:observability}

The above techniques are standard in the scientific computing literature for flow reconstruction, but they bear a severe limitation. Indeed, since it is derived in an unsupervised fashion from the set of instances $\parLM{\bqoi_{\isnap}}_{\isnap}$, the approximation basis $\ds \parLM{\bmode_{\imode}}_{\imode}$ is \emph{agnostic} to the measurements $\bobs$. In other words, the approximation basis is determined with no supervision by the measurements.
To illustrate the impact of this situation, let $\bcoefslow = \matmode^+ \, \bqoi$ be the least-squares estimate of the approximation coefficients for a given field $\bqoi$. The difference between the least-square estimate coefficients $\bcoefslow$ and the coefficients $\bcoefs$ obtained from the linear sensor measurements $\bobs$ writes
\be
\bcoefslow - \bcoefs = \left(\matmode^+ - \left(\measope \, \matmode\right)^+ \, \measope\right) \,  \bqoi \,,
\ee
and the error in the reconstructed field is obtained immediately:
\be
\normLM{\bqoi - \bqoiest}{} = \normLM{\left(\boldsymbol{I} - \matmode \, \left(\measope \, \matmode\right)^+ \, \measope\right) \,  \bqoi}{} \,,
\ee
where $\boldsymbol{I}$ is the identity matrix of suitable dimension.

The error in the reconstructed field is seen to depend on both the approximation basis $\matmode$ and the measurement operator $\measope$. The measurement operator is entirely defined by the sensor locations, and it does not depend on the basis considered to approximate the field. Hence, to reduce (the expectation of) the reconstruction error, the approximation basis must be informed \emph{both} by the dataset $\parLM{\bqoi_i}_i$ \emph{and} the sensors available, through $\measope$.
For example, poorly located sensors will lead to a large set of $\bqoi_i$ to lie in the nullspace of $\measope$, preventing their estimation, while the coefficients of certain approximation modes may be affected by the observation $\measope \, \bqoi_i$ of certain realizations $\bqoi_i$ being severely amplified by $\left(\measope \, \matmode\right)^+$ if the approximation basis is not carefully chosen.

This remark can be interpreted in terms of the control theory concept of \emph{observability} of the basis modes by the sensors. Most papers in the literature focus their attention on deriving an approximation basis leading to a good representation~\cite{bui2004aerodynamic,willcox2006unsteady,manohar2018data}, \ie{}, such that the training set is well approximated in the $\nmodes$-dimensional basis $\ds \parLM{\bmode_{\imode}}_{\imode}$, $\bqoi \approx \matmode \, \bcoefs$. But \emph{how well} the associated coefficients $\bcoefs = \bcoefs\left(\bobs\right)$ are informed by the measurements is usually overlooked when deriving the basis. In practice, the decoupling between learning an approximation basis and learning a map to the associated coefficients often leads to a performance bottleneck in the estimation procedure. Enforcing observability of the approximation basis by the sensors is key to a good recovery performance and can dramatically improve upon unsupervised methods, as shown in \cite{gobalpaper}.

\section{Shallow neural networks for flow reconstruction}\label{sec:shallow}

%
%

{
Shallow learning techniques are widely used for flow reconstruction. For instance, the approximation based approach for flow reconstruction, outlined in Section~\ref{sec:Sec_Formulation}, can be considered to have two levels of complexity. The first level is concerned with computing an approximation basis, while the second level performs a linear weighted combination of the basis elements to estimate the high-dimensional flow field. 
Such shallow learning techniques are easy to train and tune. In addition, the levels are often physically meaningful, and they may provide some interesting insights into the underlying mechanics of the system under consideration.

In the following, we propose a simple SSN as an alternative to traditional methods, which are typically \emph{very} shallow, for flow reconstruction problems. 
Our proposed shallow decoder adds only one or two additional layers of complexity to the problem.
}

\subsection{A shallow decoder for flow reconstruction}  \label{sec:Sec_Decoder}

We can define a fully-connected neural network (NN) with $K$ layers as a nested set of functions
\begin{equation}
	\mathcal{F}(\bobs; \boldsymbol{W}) := R(\boldsymbol{W}^K R(\boldsymbol{W}^{K-1} \cdots R(\boldsymbol{W}^1 \bobs))) \,,
\end{equation}
where $R(\cdot):\mathbb{R} \rightarrow \mathbb{R}$ denotes a coordinate-wise scalar (non-linear) activation function and $\boldsymbol{W}$ denotes a set of $\left\{\boldsymbol{W}^k\right\}_k$ weight matrices, $k=1,...,K$, with {appropriate} dimensions. 
NN-based learning provides a flexible framework for estimating the relationship between quantities from a collection of samples. Here, we consider SNNs, which are considered to be networks with very few, often only one, or even no, hidden layers, \ie{}, $K$ is very small.

In the following, an estimate of a vector $\by$ is denoted as $\widehat{\by}$, while $\widetilde{\by}$ denotes dummy vectors upon which one optimizes.
Relying on a training set $\{\bqoi_i, \bobs_i \}_{i=1}^{n}$, with $n$ examples $\bqoi_i$ and corresponding sensor measurements $\bobs_i$, we aim to learn a function $\mathcal{F}: \bobs \mapsto \bqoiest$ belonging to a class of neural networks $\NNclass$ which minimizes the misfit in an Euclidean sense, over all sensor measurements
\begin{equation}
\mathcal{F} \, \in \, \argmin_{\dumarg{\mathcal{F}} \in \NNclass} \, \sum_{i=1}^{n} \, \normLM{\bqoi_i  \, - \, \dumarg{\mathcal{F}}\left(\bobs_i\right)}{2}^2 \,.
\end{equation}

We assume that only a small number of training examples is available. 
Further, no prior information is assumed to be available, and the estimation method is purely data-driven.
Importantly, we assume no knowledge about the underlying measurement operator which is used to collect the sensor measurements.
Further, unlike traditional methods for flow reconstruction, this NN-based learning methodology allows the joint learning of \emph{both} the modes and the coefficients.  

\subsection{Architecture}

We now discuss some general principles guiding the design of a good network architecture for flow reconstruction.  
These considerations lead to the following nested nonlinear function
\begin{equation}
\mathcal{F}(\bobs) \, = \, \boldsymbol{\Omega}(\bcoefs(\boldsymbol{\psi}(\bobs))) \,.
\end{equation} 
%
%
%
The architecture design is guided by the paradigm of simplicity. Indeed, the architecture should enable fast training, little tuning, and offer an intuitive interpretation. 

Recall that the interpretability of the flow field estimate is favored by representing it in a basis of moderate size, whose modes can be identified with spatial structures of the field. 
This means, the estimate can be represented as a linear combination of $k$ modes $\parLM{\bmode_{\imode}}_{\imode}$, weighted by coefficients $\parLM{\coefs_\imode}_\imode$, see Eq.~\ref{eq:DLeq}.
These modes are a function of the inputs. This naturally leads to consider a network in which the output $\bqoiest$ is given by a linear, fully connected, last layer of $k$ inputs, interpreted as~$\bcoefs$. 
These coefficients are informed by the sensor measurements $\bobs$ in a nonlinear way. 

The nonlinear map $\bobs \mapsto \bcoefs$ can be described by a hidden layer, whose outputs $\bfeat$ are hereafter termed \emph{measurement features}, in analogy with kernel-based methods, where raw measurements $\bobs$ are nonlinearly lifted as extended measurements to a higher-dimensional space.
In this architecture, the measurement features $\bfeat$ essentially describe nonlinear combinations of the input measurement~$\bobs$. The nonlinear combinations are then mapped to the coefficients $\bcoefs$ associated with the modes $\bmode$.
While the size of the output layer is that of the discrete field $\bqoi$, the size of the last hidden layer ($\bcoefs$) is \emph{chosen} and defines the size $k$ of the dictionary $\matmode$. This size can be estimated from the data $\parLM{\bqoi_i}_i$ by dimensionality estimation techniques~\cite{Granata_16, Facco_17}.
Restricting the description of the training data to a low-dimensional space is of potential interest to practitioners who may interpret the elements of the resulting basis in a physically meaningful way. 
The additional structure allows one to express the field of interest in terms of modes that practitioners may interpret, \ie{}, relate to some physics phenomena such as traveling waves, instability patterns (\eg{}, Kelvin-Helmholtz), etc. 

In contrast, the size of the first hidden layer describing $\bfeat$ is essentially driven by the size of the input layer ($\bobs$) and the number of nonlinear combinations used to nonlinearly inform the coefficients $\bcoefs$.
The general shape of the network then bears flexibility in the hidden layers. A popular architecture for decoders consists of non decreasing layer sizes, so as to increase continuously the size of the representation from the low-dimensional observations to the high-dimensional field.
We can model $\mathcal{F}$ as a shallow neural network with two hidden layers $\boldsymbol{\psi}$ and $\bcoefs$, followed by a linear output layer $\boldsymbol{\Omega}$.

Two types of hidden layers, namely fully-connected (FC) and convolution layers can be considered. 
The power of convolution layers is key to the success of recent deep learning architectures in computer vision. However, in our problem, we favor fully-connected layers. The reason are as follows: (i) our sensor measurements have no spatial ordering; (ii) depending on the number of filters, convolution layers require a large number of examples for training, while we assume that only a small number of examples are available for training; (iii) potential dynamical systems that we consider evolve on a curved domain which is typically represented using an unstructured grid. Thus, the first and second hidden layers take the form 
\[
\bz^{\boldsymbol{\psi}} = \boldsymbol{\psi}(\bobs) := R(\Wmat^{\boldsymbol{\psi}} \bobs + \bbias^{\boldsymbol{\psi}}) \,,
\]
and
\[
\bz^{\bcoefs} = \bcoefs(\bz^{\boldsymbol{\psi}}) := R(\Wmat^{\bcoefs} \bz^{\boldsymbol{\psi}} + \bbias^{\bcoefs}) \,,
\]
where $\Wmat$ denotes a dense weight matrix and $\bbias$ is a bias term. The function $R(\cdot)$ denotes an activation function used to introduce nonlinearity into the model as discussed below. The final linear output layer simply takes the form of
\[
\bqoiest = \boldsymbol{\Omega}(\bz^{\bcoefs}) := \boldsymbol{\Phi} \bz^{\bcoefs} + \bbias^{\boldsymbol{\Phi}} \,,
\]
where we interpret the columns of the weight matrix $\boldsymbol{\Phi}$ as modes. 
In summary, the architecture of our shallow decoder can be outlined as
\begin{align*}
\bobs \mapsto  \boldsymbol{\psi}(\bobs) \mapsto \bcoefs(\bz^{\boldsymbol{\psi}}) \mapsto \boldsymbol{\Omega}(\bz^{\bcoefs}) \equiv \bqoiest \,.
\end{align*}  
Depending on the dataset, we need to adjust the size of each layer. Here, we use narrow rather than wide layers. 
%
Prescribing the size of the output layer restricts the dimension of the space in which the estimation lies, and it effectively regularizes the problem, \eg{}, filtering-out most of the noise which is not living in a low-dimensional space. 

The rectified linear unit (ReLU) activation function is among the most popular choices in computer vision applications, owing to its favorable properties~\cite{glorot2011deep}. The ReLU activation
is defined as the positive part of a signal $\bz$:
\begin{equation}
R(\bz) := \max(\bz,\boldsymbol{0}) \,.
\end{equation}
The transformed input signal is also called activation. 
While the ReLU activation function performs best on average in our experiments, there are other choices. For instance, we have also considered the Swish~\cite{agostinelli2014learning} activation function.

\subsection{Regularization}

{Overfitting is a common problem in machine learning and occurs if a function fits a limited set of data points too closely.} In particular, this is a problem for deep neural networks which often have more neurons (trainable parameters) than can be justified by the limited amount of training examples which are available.
There is increasing interest in characterizing and understanding generalization and overfitting in NNs~\cite{poggio2017theory,bartlett2017spectrally}.
Hence, additional constraints are required to learn a function which generalizes to new observations that have not been used for training.
Standard strategies to avoid overfitting include early stopping rules, and weight penalties ($\ldeux$ regularization) to regularize the complexity of the function (network). In addition to these two strategies, we use also batch normalization (BN)~\cite{ioffe2015batch} and dropout layers (DL)~\cite{srivastava2014dropout} to improve the convergence and robustness of the shallow decoder. This yields the following architecture:
\begin{align*}
\bobs \mapsto  \boldsymbol{\psi}(\bobs)  \mapsto BN  \mapsto DL \mapsto \bcoefs(\bz^{\boldsymbol{\psi}}) \mapsto BN \mapsto \boldsymbol{\Omega}(\bz^{\bcoefs}) \equiv \bqoiest \,.
\end{align*}
Regularization, in its various forms, requires one to ``fiddle'' with a large number of knobs (\emph{i.e.}, hyper-parameters).
However, we have found that SNNs are less sensitive to the particular choice of parameters; hence, SNNs are easier to tune.

\paragraph{Batch normalization.} BN is a technique to normalize (mean zero and unit standard deviation) the activation. From a statistical perspective, BN eases the effect of internal covariate shifts~\cite{ioffe2015batch}. In other words, BN accounts for the change of distribution of the output signals (activation) across different mini batches during training. Each BN layer has two parameters which are learned during the training stage. This simple, yet effective, prepossessing step allows one to use higher learning rates for training the network. In addition it also reduces overfitting owing to its regularization~effect. 

\paragraph{Dropout layer.} DL helps to improve the robustness of a NN. The idea is to switch off (drop) a small fraction of randomly chosen hidden units (neurons) during the training stage. This strategy can be seen as some form of regularization which also helps to reduce interdependent learning between the units of a fully connected layer.
In our experiments the drop ratio is set to $p=10\%$.

\subsection{A note on overparameterized networks}

{The expressive power of NNs can be seen as a function of the depth (\ie, number of hidden layers) and the width (\ie, number of neurons per hidden layer) of the architecture~\cite{lu2017expressive}. 
	Shallow networks typically tend to compensate for the reduced depth by increasing the width of the hidden layers.  
	In turn, this can lead to shallow architectures that have more parameters than a comparable deep and narrow architecture for the same problem.
	However, such (potentially) overparameterized networks do not necessarily perform worse. On the contrary, recent theory suggests that it can be easier to train very overparameterized models with stochastic gradient descent (SGD)~\cite{keskar2016large,allen2019learning}.

	This may be surprising, since conventional ML wisdom states that overparamerized models tend to overfit and show poor generalization performance. However, recent results show that overparamerized models trained to minimum norm solutions can indeed preserve the ability to generalize well~\cite{zhang2016understanding,hardt2015train,belkin2018reconciling,radhakrishnan2019memorization,derezinski2019exact}.
}

\subsection{Optimization}

Given a training set with $n$ targets $\parLM{\bqoi_i}_i$ and corresponding sensor measurements $\parLM{\bobs_i}_i$, we minimize the misfit between the reconstructed quantity $\bqoiest = \mathcal{F}(\bobs)$ and the observed quantity $\bqoi$, in terms of the squared $\ldeux$-norm
\[
\mathcal{F} \in \argmin_{\mathcal{\widetilde{F}}} \sum_{i=1}^{n} \normLM{\bqoi_i - \mathcal{\widetilde{F}}\left(\bobs_i\right)}{2}^2 + \lambda \| \boldsymbol{W}^i \|_2^2  \,.
\]
The second term on the right hand side introduces $\ldeux$ regularization to the weight matrices, which is controlled via the parameter $\lambda > 0$.
It is well-known that $\ldeux$-norm is sensitive to outliers; and 
the $\lone$-norm can be used as a more robust loss function.
We use the ADAM optimization algorithm~\cite{kingma2014adam} to train the shallow decoder, with learning rate $10^{-2}$ and  weight decay $10^{-4}$ (also known as $\ldeux$ regularization). The learning rate, also known as step size, controls how much we adjust the weights in each epoch. The weight decay parameter is important since it allows one to regularize the complexity of the network. In practice, we can improve the performance by changing the learning rate during training. We decay the learning rate by a factor of $0.9$ after $100$ epochs. Indeed, the reconstruction performance in our experiments is considerably improved by this dynamic scheme, compared to a fixed parameter setting.
In our experiments, ADAM shows a better performance than SGD with momentum~\cite{sutskever2013importance} and averaged SGD~\cite{Polyak:1992:ASA:131092.131098}.
The hyper-parameters can be fine tuned in practice, but our choice of parameters works reasonably well for several different examples. Note that we use the method described by~\cite{he2015delving} in order to initialize the weights. This initialization scheme is favorable, in particular because the output layer is high-dimensional.

\section{Empirical evaluation}\label{sec:exp_result}

We evaluate our methods on three classes of data.
First, we consider a periodic flow behind a circular cylinder, as a canonical example of fluid flow.
Then, we consider the weekly mean sea surface temperature (SST), as a second and more challenging example. 
Finally, the third and most challenging example we consider is a forced isotropic turbulence flow.

\begin{figure}[!t]
	\centering
	\begin{subfigure}[t]{0.4\textwidth}
		\centering
		\DeclareGraphicsExtensions{.pdf}
		\includegraphics[width=1\textwidth]{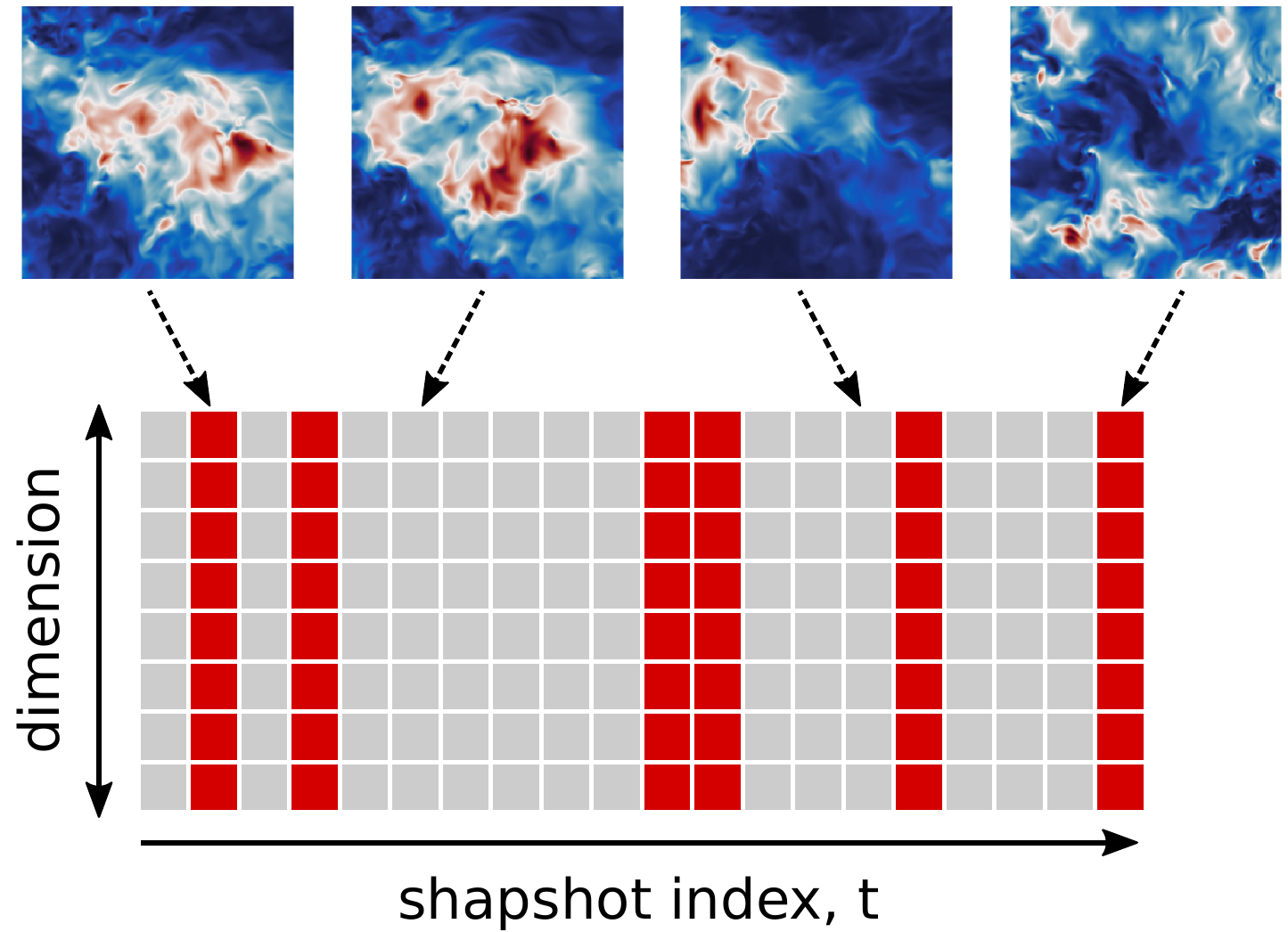}
		\caption{Within sample prediction.}
	\end{subfigure}
	~
	\begin{subfigure}[t]{0.4\textwidth}
		\centering
		\DeclareGraphicsExtensions{.pdf}
		\includegraphics[width=1\textwidth]{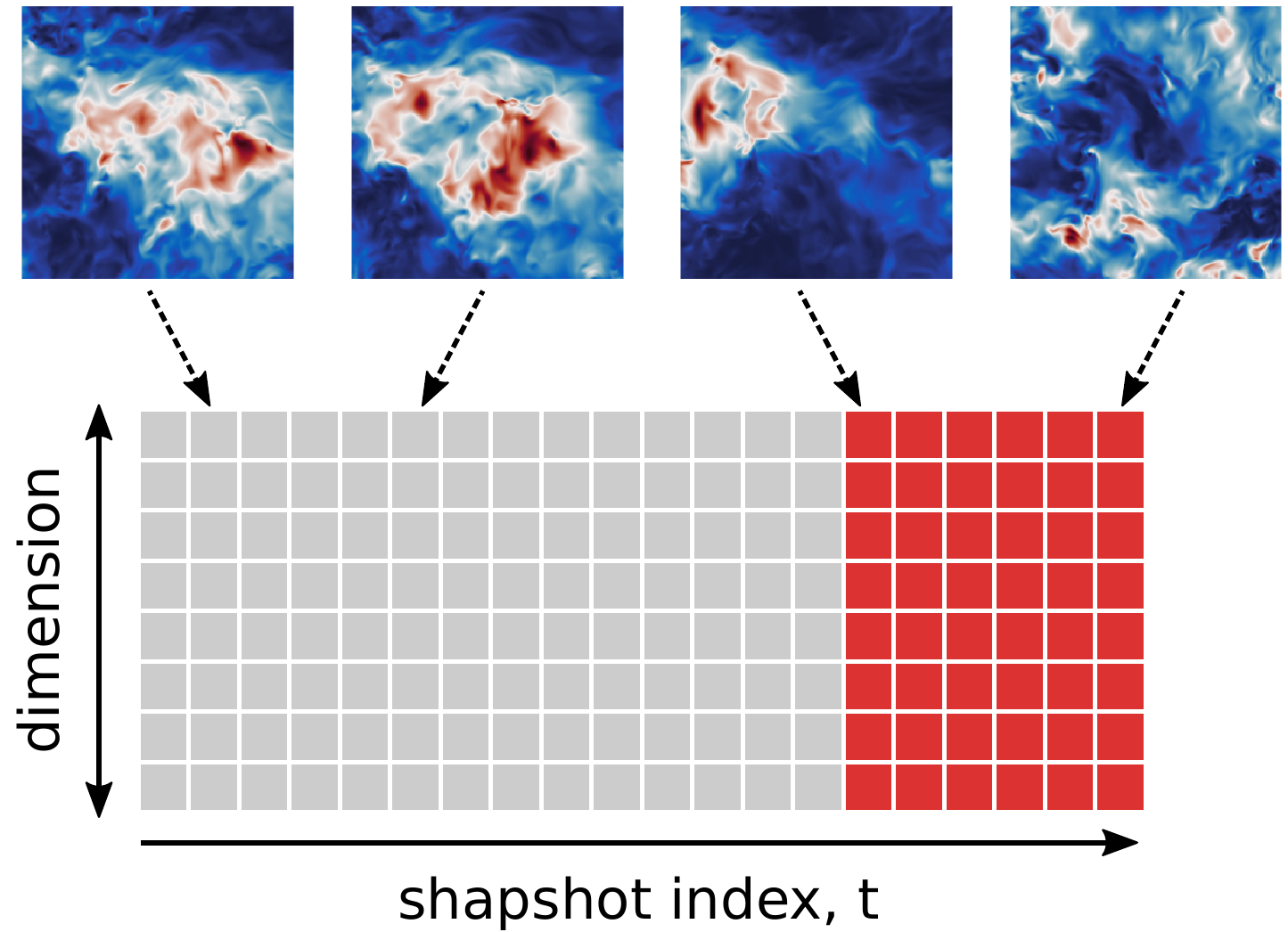}
		\caption{Out of sample prediction.}
	\end{subfigure}	
	~	
	\caption{Two different training and test set configurations, showing (a) a within sample prediction task and (b) an out of sample prediction task. Here, the gray columns indicate snapshots used for training, while the red columns indicate snapshots used for testing. }
	\label{fig:training_test_config}
\end{figure}

As discussed in Section~\ref{sec: introduction}, the \textsc{shallow decoder} requires that the training data represent the system, in the sense that they should comprise samples drawn from the same statistical distribution as the testing data.  Indeed, this limitation is standard to data-driven methods, both for flow reconstruction and also more generally. 
Hence, we are mainly concerned with exploring reconstruction performance and generalizability for \emph{within sample prediction} rather than for \emph{out of sample prediction} tasks.
In our third example, however, we demonstrate the limitations of the \textsc{shallow decoder}, illustrating difficulties that arise when one tries to extrapolate, rather than interpolate, the flow field.
Figure~\ref{fig:training_test_config} illustrates the difference between the two types of tasks. 

In the first two example classes of data, the sensor information is a subset of the high-dimensional flow field, \ie, the measurement operator $\measope \in \R^{\nobs \times \nqoi}$ only has one non-zero entry in rows corresponding to the index of a sensor location. Letting $\setsensor \in \left[1, \nqoi\right]^\nobs \subset \mathbb{N}^\nobs$ be the set of indices indexing the spatial location of the sensors, the measurement operator is such that
\be
\bobs = \measope \, \bqoi = \bqoi_{\setsensor} \,,     \label{Eq_obsassubsetqoi}
\ee
that is, the observations are simply point-wise measurements of the field of interest. In the above equation, $\bqoi_{\setsensor}$ is the restriction of $\bqoi$ to its entries indexed by $\setsensor$.
In this paper, no attempt is made to optimize the location of the sensors. In practical situations, they are often given or constrained by other considerations (wiring, intrusivity, manufacturing, etc.). We use random locations in our examples.
The third example class of data demonstrates the SD using sub-gridscale measurements. 

The error is quantified in terms of the normalized root-mean-square residual error
\begin{equation}
	\text{NME} = \frac{\left\|\bqoi - \bqoiest\right\|_2}{\left\|\bqoi\right\|_2} \,,
\end{equation}
denoted in the following as ``NME.'' However, this measure can be misleading if the empirical mean is dominating. Hence, we consider also a more sensitive measure which quantifies the reconstruction accuracy of the deviations around the empirical mean. We define this measure as
\begin{equation}
\text{NFE} = \frac{\left\|\bqoi' - \bqoiest' \right\|_2}{\left\|\bqoi'\right\|_2} \,,
\end{equation}
where $\bqoi'$ and $\bqoiest'$ are the fluctuating parts around the empirical mean. In our experiments, we average the errors over $30$ runs for different sensor distributions.

\subsection{Fluid flow behind the cylinder}

The first example we consider is the fluid flow behind a circular cylinder, at Reynolds number $100$, based on cylinder diameter, a canonical example in fluid dynamics~\cite{Noack2003jfm}. 
The flow is characterized by a periodically shedding wake structure and exhibits smooth, large scale, patterns. 
A direct numerical simulation of the two-dimensional Navier-Stokes equations is achieved via the immersed boundary projection method~\cite{taira:07ibfs,taira:fastIBPM}.  
In particular, we use the fast multidomain method~\cite{taira:fastIBPM}, which simulates the flow on five nested grids of increasing size, with each grid consisting of $199\times 449$ grid points, covering a domain of $4\times 9$ cylinder diameters on the finest domain.  
We collect $151$ snapshots in time, sampled uniformly in time and covering several periods of vortex shedding.
For the following experiment, we use cropped snapshots of dimension $199\times 384$ on the finest domain, as we omit the spatial domain upstream to the cylinder. Further, we split the dataset into a training and test set so that the training set comprises the first $100$ snapshots, while the remaining $51$ snapshots are used for validation. Note that different splittings (interpolation and extrapolation) yield nearly the same results since the flow is periodic.

\subsubsection{Varying numbers of random structured point-wise sensor measurements}
We investigate the performance of the \textsc{shallow decoder} using varying numbers of sensors. A realistic setting is considered in that the sensors can only be located on a solid surface. The retained configuration aims at reconstructing the entire vorticity flow field from information at the cylinder surface only.
The results are averaged over different sensor distributions on the cylinder downstream-facing surface and are summarized in Table~\ref{tab:cylinder_summary}. 
Further, to contextualize the precision of the algorithms, we also state the standard deviation in parentheses.

\begin{figure}[!b]
	\centering
	\begin{subfigure}[t]{1\textwidth}
		\centering
		\DeclareGraphicsExtensions{.pdf}
		\begin{overpic}[width=0.4\textwidth]{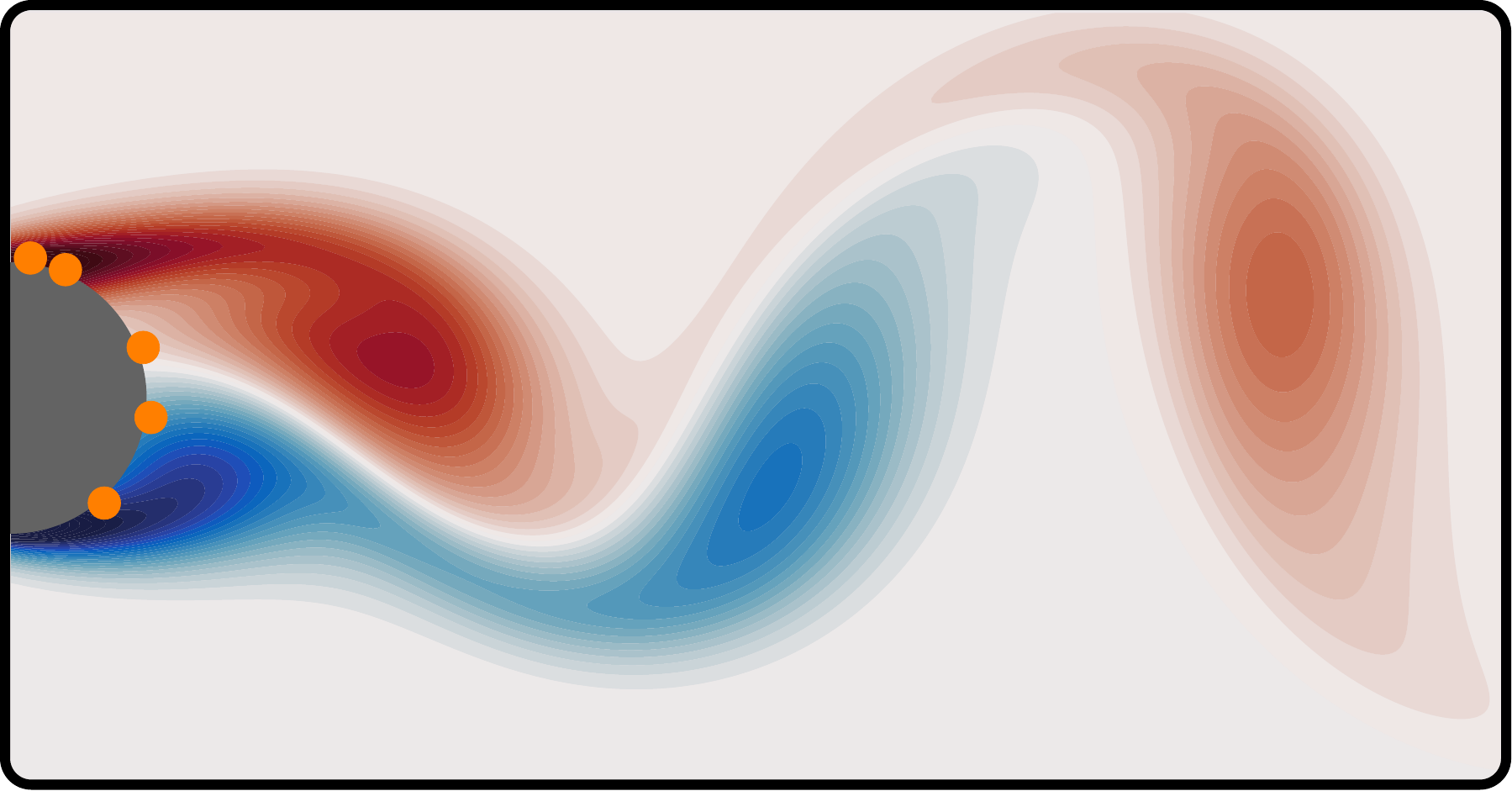} 
			\put(-6,16){\rotatebox{90}{\scriptsize (a) Truth}}
		\end{overpic}		
		\includegraphics[width=0.4\textwidth]{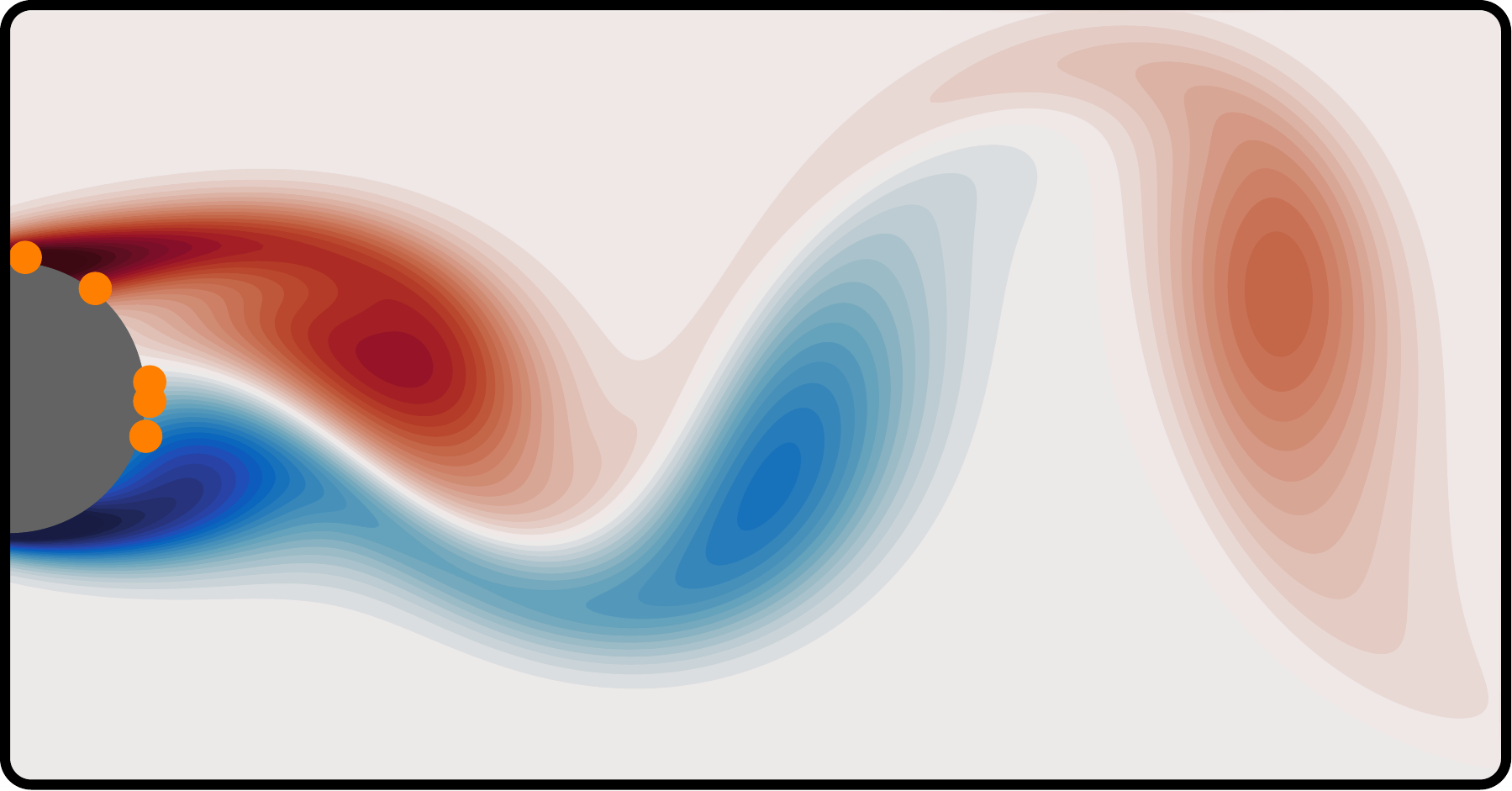}
	\end{subfigure}
	
	\vspace{+0.2cm}
	\centering
	\begin{subfigure}[t]{1\textwidth}
		\centering
		\DeclareGraphicsExtensions{.pdf}
		\begin{overpic}[width=0.4\textwidth]{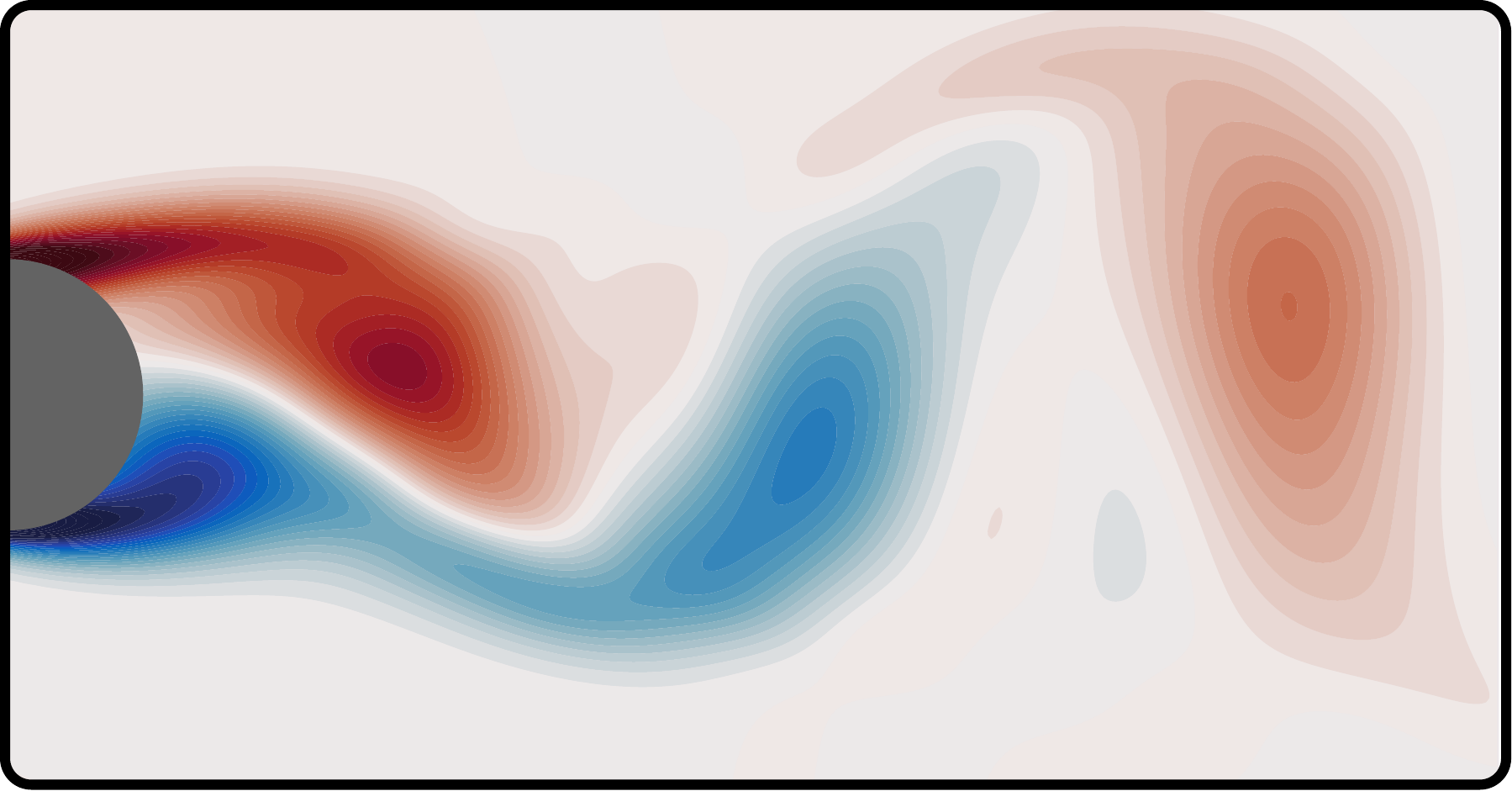} 
			\put(-6,18){\rotatebox{90}{\scriptsize (b) \textsc{pod}}}
		\end{overpic}			
		\includegraphics[width=0.4\textwidth]{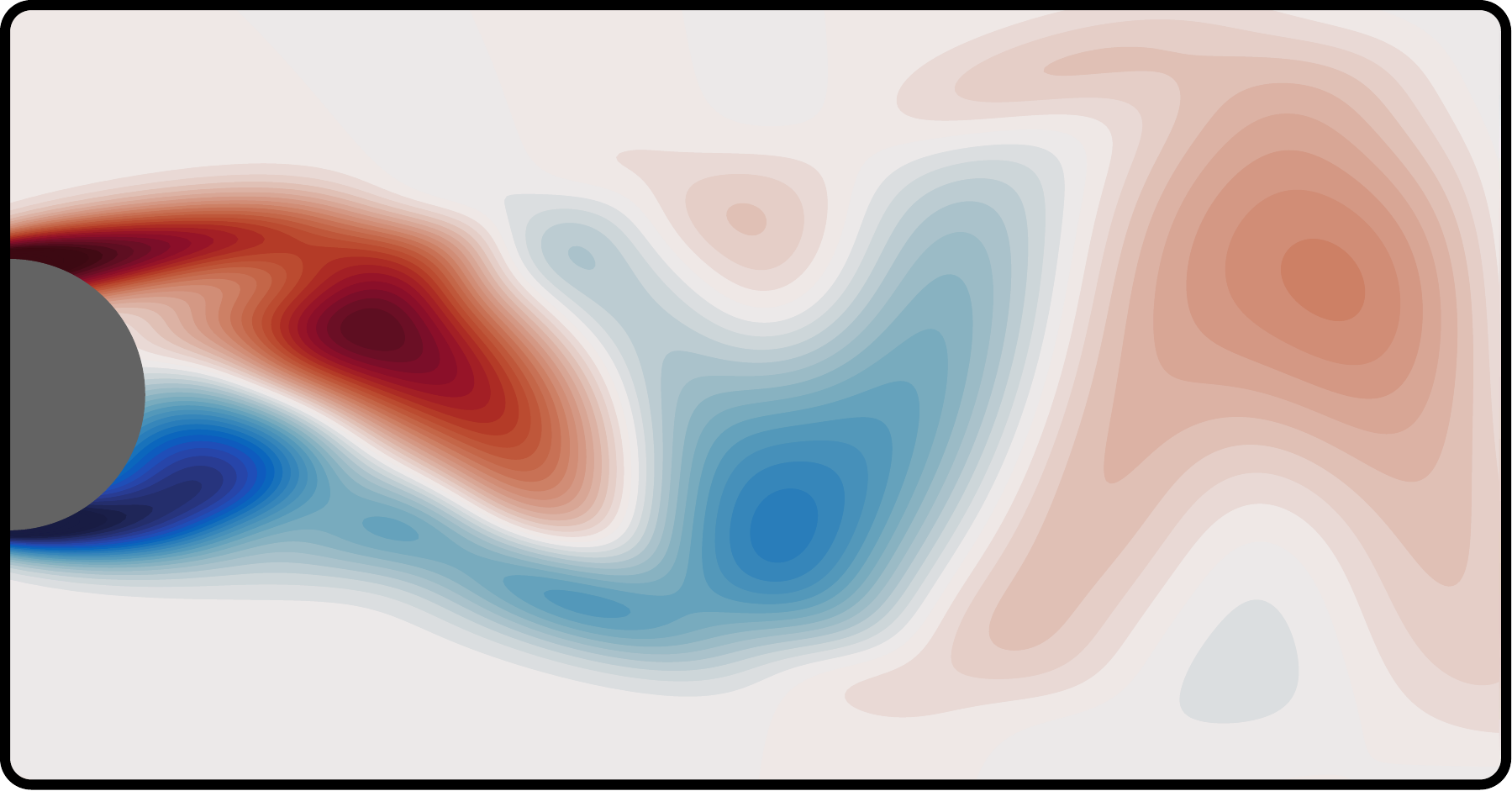}
	\end{subfigure}
	
	\vspace{+0.2cm}
	\centering	
	\begin{subfigure}[t]{1\textwidth}
		\centering
		\DeclareGraphicsExtensions{.pdf}
		\begin{overpic}[width=0.4\textwidth]{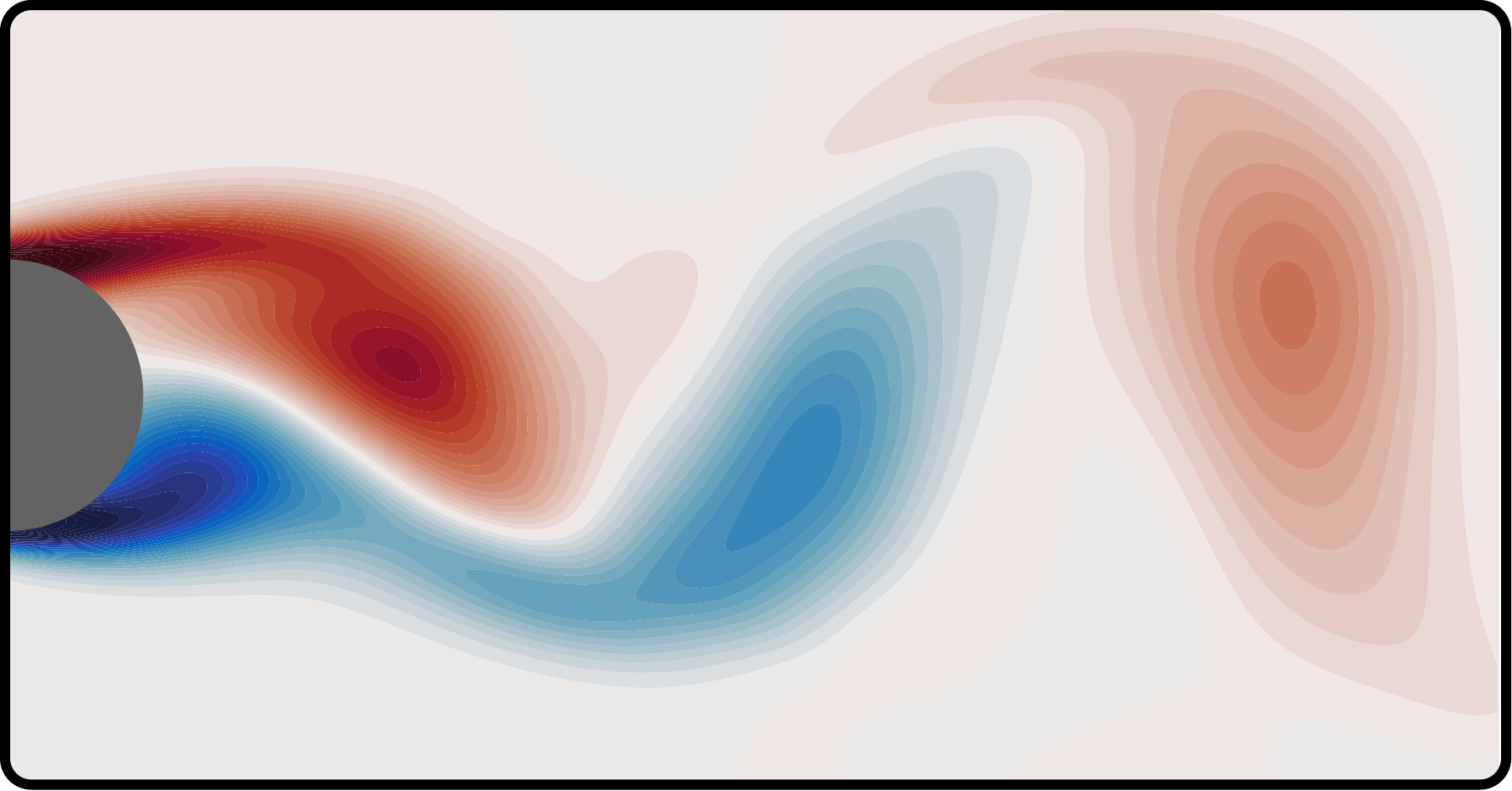} 
			\put(-6,13.5){\rotatebox{90}{\scriptsize (c) \textsc{pod plus}}}
		\end{overpic}			
		\includegraphics[width=0.4\textwidth]{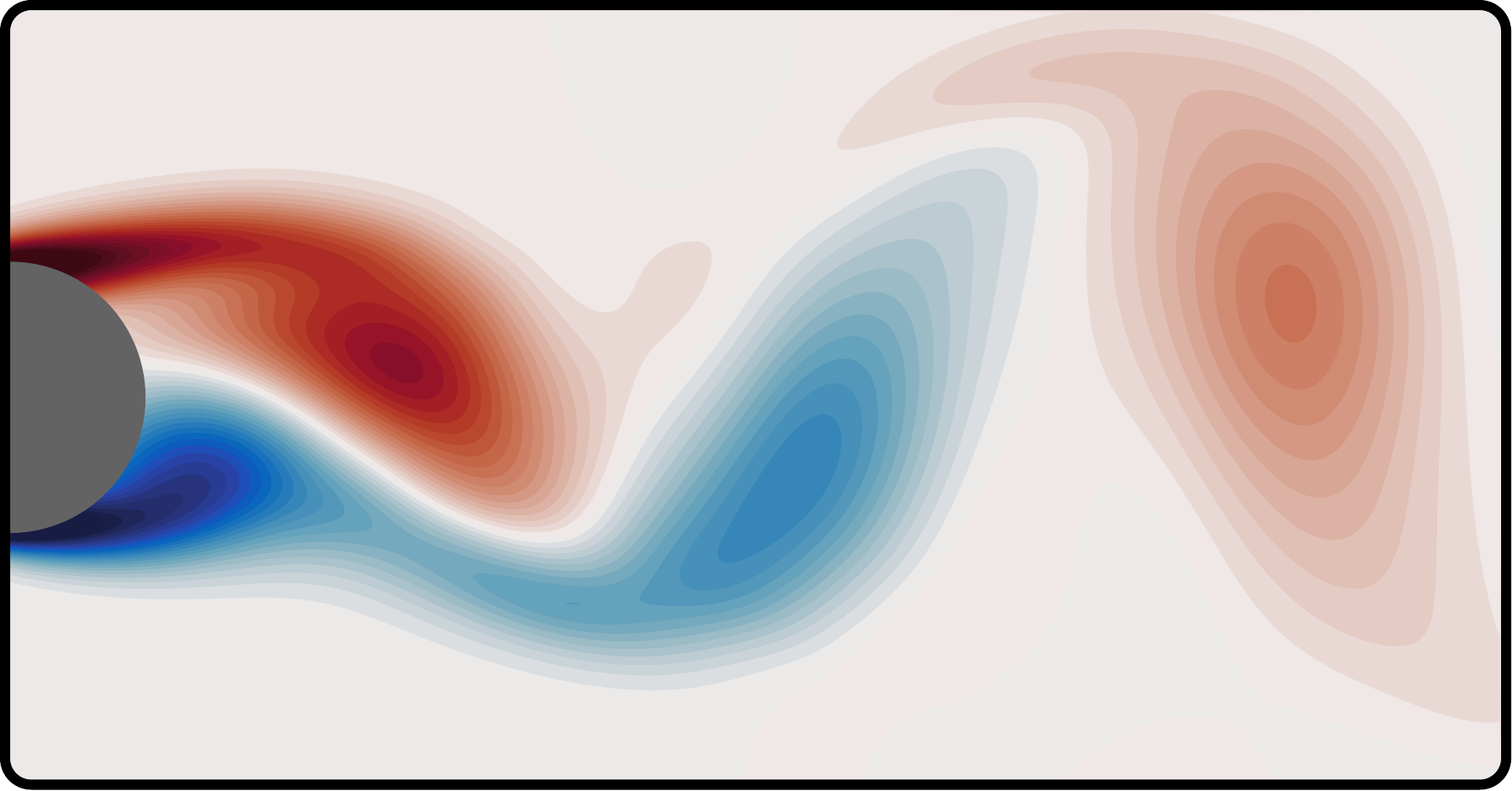}
	\end{subfigure}	
	
	\vspace{+0.2cm}
	\centering
	\begin{subfigure}[t]{1\textwidth}
		\centering
		\DeclareGraphicsExtensions{.pdf}
		\begin{overpic}[width=0.4\textwidth]{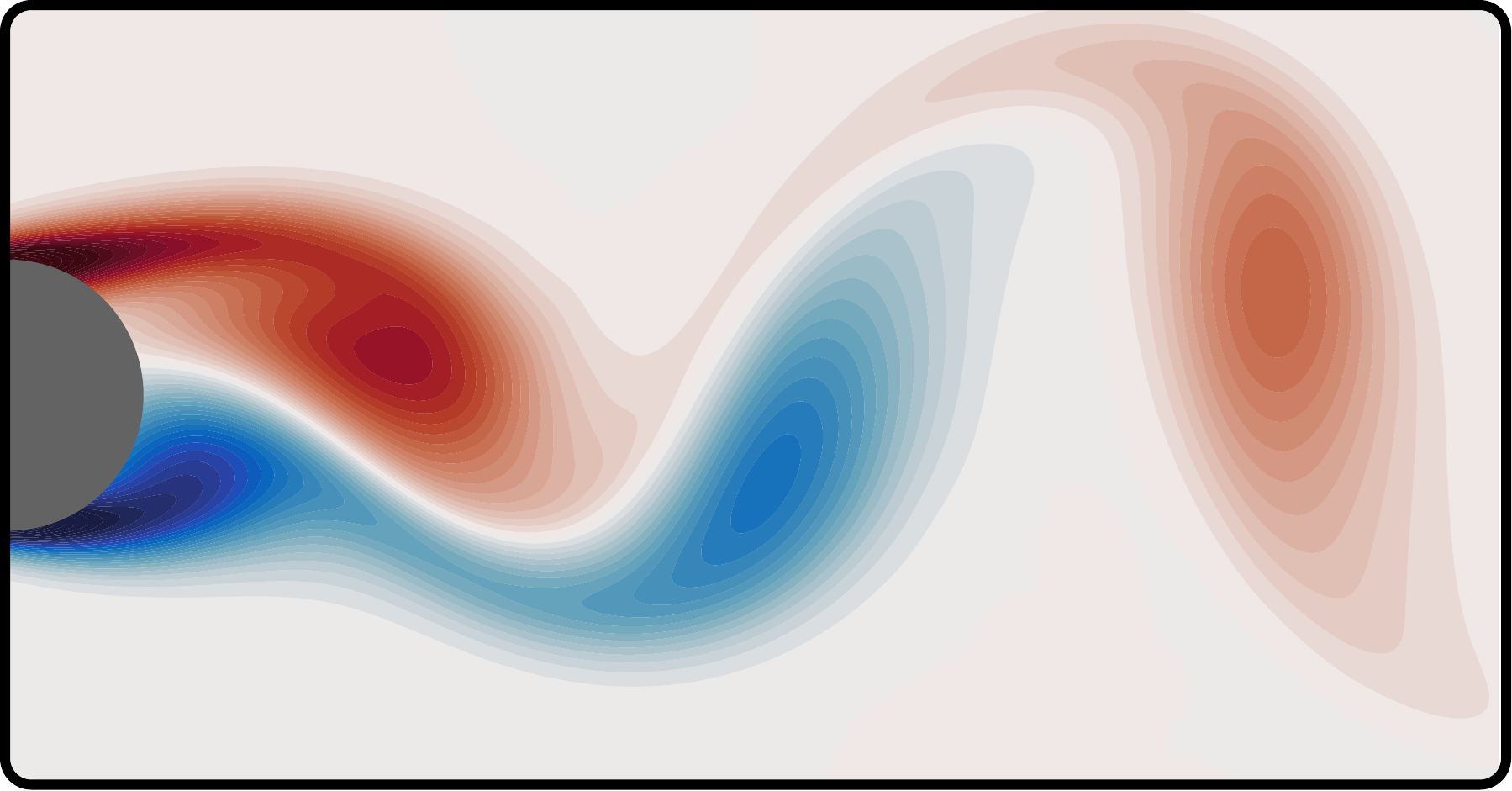} 
			\put(-6,3.0){\rotatebox{90}{\scriptsize (d) \textsc{shallow decoder}}}
		\end{overpic}		
		\includegraphics[width=0.4\textwidth]{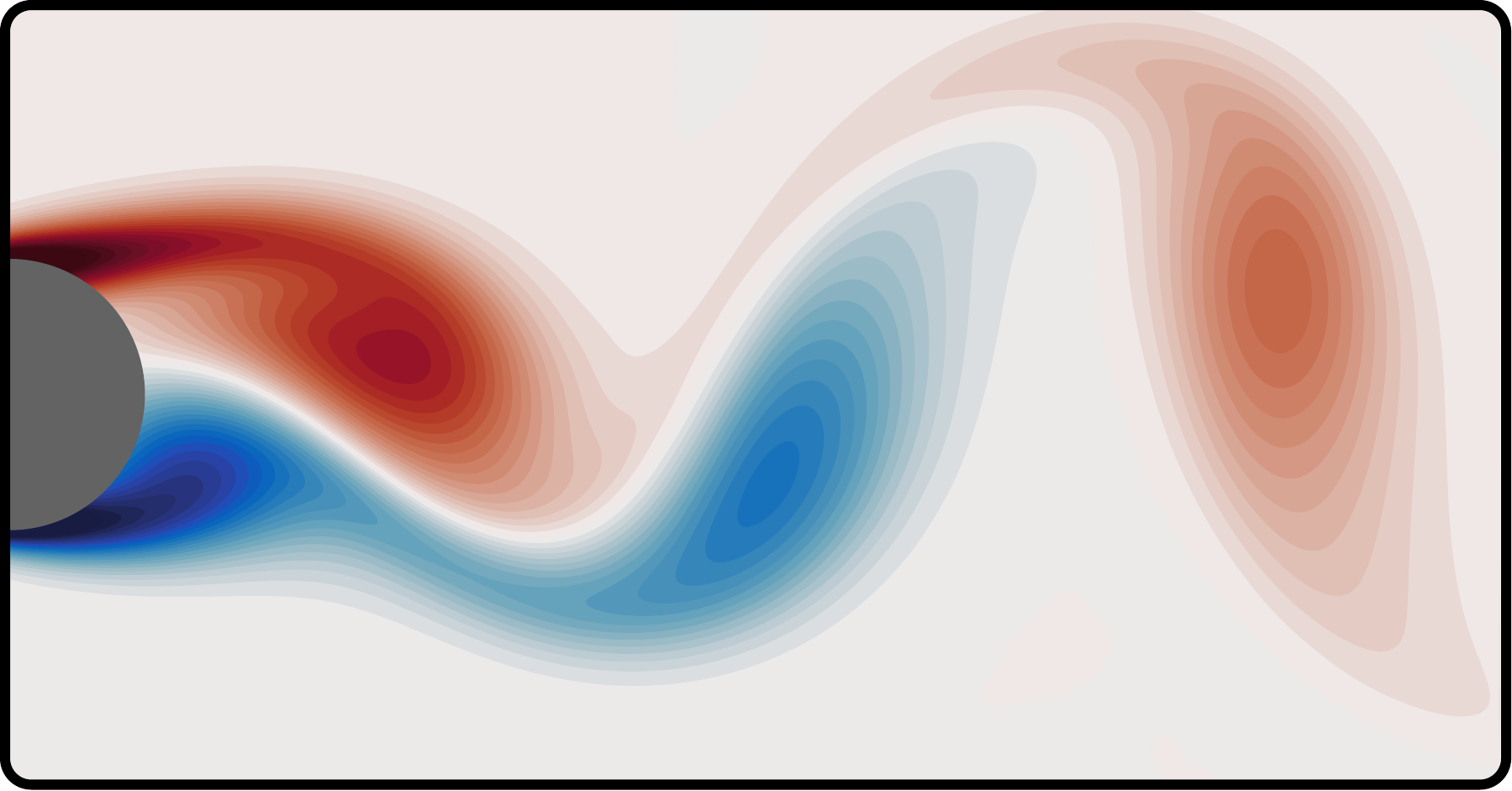}
	\end{subfigure}		
	\caption{Visual results for the canonical flow for two different sensor distributions. In (a) the target snapshots and the specific sensor configurations (here using $5$ sensors) are shown. Depending on the sensor distribution, the POD-based method is not able to accurately reconstruct the high-dimensional flow field, as shown in (b). The regularized \textsc{pod plus} method performs slightly better, as shown in (c). The \textsc{shallow decoder} yields an accurate flow reconstruction, as shown in (d).}
	\label{fig:cylinder_vis}
\end{figure}

\begin{table}[!b]
	\centering
	\scalebox{0.95}{
		\begin{tabular}{lcccccccc} \toprule
			& {Sensors}  & \multicolumn{2}{c}{Training Set}  &  \multicolumn{2}{c}{Test Set} \\ 
			\cmidrule{3-4}  \cmidrule{5-6} 
			& 	& {NME} & {NFE}   & {NME} & {NFE}
			\\
 		
			\midrule
			{\textsc{pod}}  			& 5 & {0.465 (0.39)}  & {0.675 (0.57)}  & {0.488 (0.41)}  & {0.698 (0.59)} \\
			
			{\textsc{pod} ($k^*=4$)}  	& 5 & {0.217 (0.02)} & {0.325 (0.01)} & {0.227 (0.03)} & {0.324 (0.04)} \\    					
			
			{\textsc{pod plus} ($\alpha=1\text{e-}8$)}  	& 5 & {0.198 (0.02)}  & {0.288 (0.03)}  & {0.203 (0.02)}  & {0.291 (0.03)} \\    
			
			{\textsc{shallow decoder}}  	& 5 & {0.003 (0.00)}  & {0.004 (0.00)}  & {0.006 (0.00)}  & {0.008 (0.00)} \\    
			
			\midrule    
			{\textsc{pod}}  		  & 10 & {0.346 (1.54)} & {0.502 (2.23)} & {0.379 (1.70)} & {0.542 (2.43)} \\
			
			{\textsc{pod} ($k^*=8$)}  	& 10 & {0.049 (0.00)} & {0.071 (0.01)} & {0.051 (0.01)} & {0.072 (0.01)} \\    			
			
			{\textsc{pod plus} ($\alpha=1\text{e-}13$)}    & 10 & {0.035 (0.01)} & {0.050 (0.02)} & {0.035 (0.01)} & {0.050 (0.02)} \\    
			
			{\textsc{shallow decoder}}  & 10 & {0.002 (0.00)} & {0.003 (0.00)} & {0.005 (0.00)} & {0.007 (0.00)} \\    
			\midrule    

			{\textsc{pod}}  		  & 15 & {0.441 (1.81)} & {0.639 (2.63)} & {0.574 (2.44)} & {0.821 (3.49)} \\
			
			{\textsc{pod} ($k^*=12$)}  	& 15 & {0.015 (0.00)} & {0.023 (0.01)} & {0.016 (0.01)} & {0.023 (0.01)} \\    			
			
			{\textsc{pod plus} ($\alpha=1\text{e-}12$)}    & 15 & {0.016 (0.01)}& {0.023 (0.01)} & {0.016 (0.01)} & {0.022 (0.01)} \\    
			
			{\textsc{shallow decoder}}  & 15 & {0.002 (0.00)} & {0.003 (0.00)} & {0.005 (0.00)} & {0.007 (0.00)} \\

			\bottomrule 
	\end{tabular}}\vspace{+0.2cm}
	\caption{Performance for the flow past cylinder for a varying number of sensors. Results are averaged over $30$ runs with different sensor distributions, with standard deviations in parentheses. { The parameter $k^*$ indicates the number of modes that were used for flow reconstruction by the \textsc{pod} method, and $\alpha$ refers to the strength of ridge regularization applied to \textsc{pod plus}.}}
	\label{tab:cylinder_summary}
\end{table}

The \textsc{shallow decoder} shows an excellent flow reconstruction performance compared to traditional methods. Indeed, the results show that very few sensors are already sufficient to get an accurate approximation. Further, we can see that the \textsc{shallow decoder} is insensitive to the sensor location, \ie{}, the variability of the performance is low when different sensor distributions on the cylinder surface are used. In stark contrast, this simple setup poses a challenge for the \textsc{pod} method {without regularization}, which is seen to be highly sensitive to the sensor configuration.
This is expected since poorly located sensors lead to a large probability that the vorticity field $\bqoi_i$ lies in the nullspace of $\measope$, preventing its estimation, as discussed in Section~\ref{sec:Sec_Formulation}.
While regularization can improve the robustness slightly, the POD-based methods still require about at least $15$ sensors to provide accurate estimations for the high-dimensional flow field. ({ Here, we list results for the \textsc{pod} method with hard-threshold regularization and \textsc{pod plus} method with ridge regularization. The number of retained components (hard-threshold), that were used for flow reconstruction, is indicated by $k^*$ and the strength of ridge regularization is denoted by the parameter $\alpha$. See Appendix~\ref{sec:tuning} for more details.}) In contrast, the \textsc{shallow decoder} exhibits a good performance with as few as 5 sensors. Note that the traditional methods could benefit from optimal sensor placement~\cite{manohar2018data}; however, this is beyond the scope of this paper.

Figure~\ref{fig:cylinder_vis} provides visual results for two specific sensor configuration using $5$ sensors. The second configuration is challenging for \textsc{pod}, which fails to provide an accurate reconstruction. \textsc{pod plus} provides a more accurate reconstruction of the flow field. The \textsc{shallow decoder} outperforms the traditional methods in both situations.

\subsubsection{Non-linear sensor measurements}
So far, the sensor information consisted of pointwise measurements of the local flow field so that the $\iobs$-th measurement is given by $\bobs^{\left(\iobs\right)} = \measope_\iobs \, \bqoi = \delta_{\tau_\iobs} \left[\bqoi\right] = \bqoi^{\left(\iobs\right)}$, $\iobs = 1, \ldots, \nobs$, with $\delta_{\tau_\iobs}$ a Dirac distribution centered at the location of the $\iobs$-th sensor and $\bobs^{\left(\iobs\right)}$ and $\bqoi^{\left(\iobs\right)}$ the $\iobs$-th component of $\bobs$ and $\bqoi$ respectively. We now consider \emph{nonlinear} measurements to demonstrate the flexibility of the \textsc{shallow decoder}. Here, we consider the simple setting of squared sensor measurements: $\bobs^{\left(\iobs\right)} = \left(\bqoi \odot \bqoi\right)^{\left(\iobs\right)}$, where $\odot$ denotes the Hadamard product. Table~\ref{tab:cylinder_transform_summary} provides a summary of the results, using $10$ sensors. 
The \textsc{shallow decoder} is agnostic to the functional form of the sensor measurements, and it achieves nearly the same performance as in the linear case above, i.e., the error for the test set increases less than $1\%$ compared to the linear case in Table~\ref{tab:cylinder_summary}.

\begin{table}[!b]
	\centering
	\scalebox{0.95}{
		\begin{tabular}{lcccccccc} \toprule
			& {Sensors}  & \multicolumn{2}{c}{Training Set}  &  \multicolumn{2}{c}{Test Set} \\ 
			\cmidrule{3-4}  \cmidrule{5-6} 
			& 	& {NME} & {NFE}   & {NME} & {NFE}
			\\
			\midrule
			{\textsc{pod}}  	& 10 & -  & -  & -  & -  
			\\
			{\textsc{pod plus} ($\alpha=5\text{e-}4$)}  	& 10 & {0.676 (0.00)}  & {0.981 (0.00)}  & {0.682 (0.09)}  & {0.974 (0.00)} \\    
			
			{\textsc{shallow decoder}}  & 10 & {0.002 (0.00)}  & {0.003 (0.00)}  & {0.006 (0.00)}  & {0.009 (0.01)} \\    
			
			\bottomrule 
	\end{tabular}}\vspace{+0.2cm}
	\caption{Performance for estimating the flow behind a cylinder using nonlinear sensor measurements. Results are averaged over $30$ runs with different sensor distributions, with std. dev. in parentheses. The standard POD-based method fails for this task. \textsc{pod plus} is able to reconstruct the flow field, yet the estimation quality is poor. In contrast, the SD method performs well. }
	\label{tab:cylinder_transform_summary}
\end{table}

\begin{table}[!b]
	\centering
	\scalebox{0.95}{
		\begin{tabular}{lcccccccc} \toprule
			& {SNR}  & \multicolumn{2}{c}{Training Set}  &  \multicolumn{2}{c}{Test Set} \\ 
			\cmidrule{3-4}  \cmidrule{5-6} 
			& 	& {NME} & {NFE}   & {NME} & {NFE}
			\\
			\midrule
			{\textsc{pod}}  			& 10 & {9.171 (14.7)} & {12.69 (20.4)} & {8.746 (12.9)} & {11.93 (17.6)} \\
			
			{\textsc{pod} ($k^*=2$)}  	& 10 & {0.461 (0.02)} & {0.638 (0.03)} & {0.468 (0.02)} & {0.639 (0.02)} \\ 			
			
			{\textsc{pod plus} ($\alpha=5\text{e-}5$)}  	& 10 & {0.468 (0.02)} & {0.648 (0.02)} & {0.472 (0.02)} & {0.644 (0.2)} \\        
			
			{\textsc{shallow decoder}}  & 10 & {0.138 (0.02)} & {0.201 (0.02)} & {0.278 (0.04)} & {0.397 (0.05)} \\   
			
			\midrule    
			{\textsc{pod}}  			& 50 & {4.837 (3.08)} & {6.946 (4.42)} & {4.520 (2.75)} & {6.390 (3.89)} \\
			
			{\textsc{pod} ($k^*=2$)}  	& 50 & {0.342 (0.01)} & {0.492 (0.01)} & {0.349 (0.01)} & {0.493 (0.01)} \\ 		
			
			{\textsc{pod plus} ($\alpha=1\text{e-}5$)}  	& 50 & {0.370 (0.03)} & {0.539 (0.04)} & {0.371 (0.02)} & {0.524 (0.03)} \\    
			
			{\textsc{shallow decoder}}  & 50 & {0.134 (0.02)} & {0.198 (0.02)} & {0.173 (0.02)} & {0.247 (0.03)} \\    
			
			\bottomrule 
	\end{tabular}}

	\caption{Performance for estimating the flow behind a cylinder in presence of white noise, using $10$ sensors. Results are averaged over $30$ runs with different sensor distributions, with std. dev. in parentheses. \textsc{pod} fails for this task, while \textsc{pod plus} shows a better performance. The SD shows to be robust to noisy sensor measurements and outperforms the traditional techniques. {The parameter $k^*$ indicates the number of modes that were used for flow reconstruction by the \textsc{pod} method, and the parameter $\alpha$ refers to the strength of ridge regularization applied to the the \textsc{pod plus} method.}}
	\label{tab:cylinder_noisy_summary}
\end{table}

\begin{figure}[!t]
	\centering
	\begin{subfigure}[t]{0.4\textwidth}
		\centering
		\DeclareGraphicsExtensions{.pdf}
		\includegraphics[width=1\textwidth]{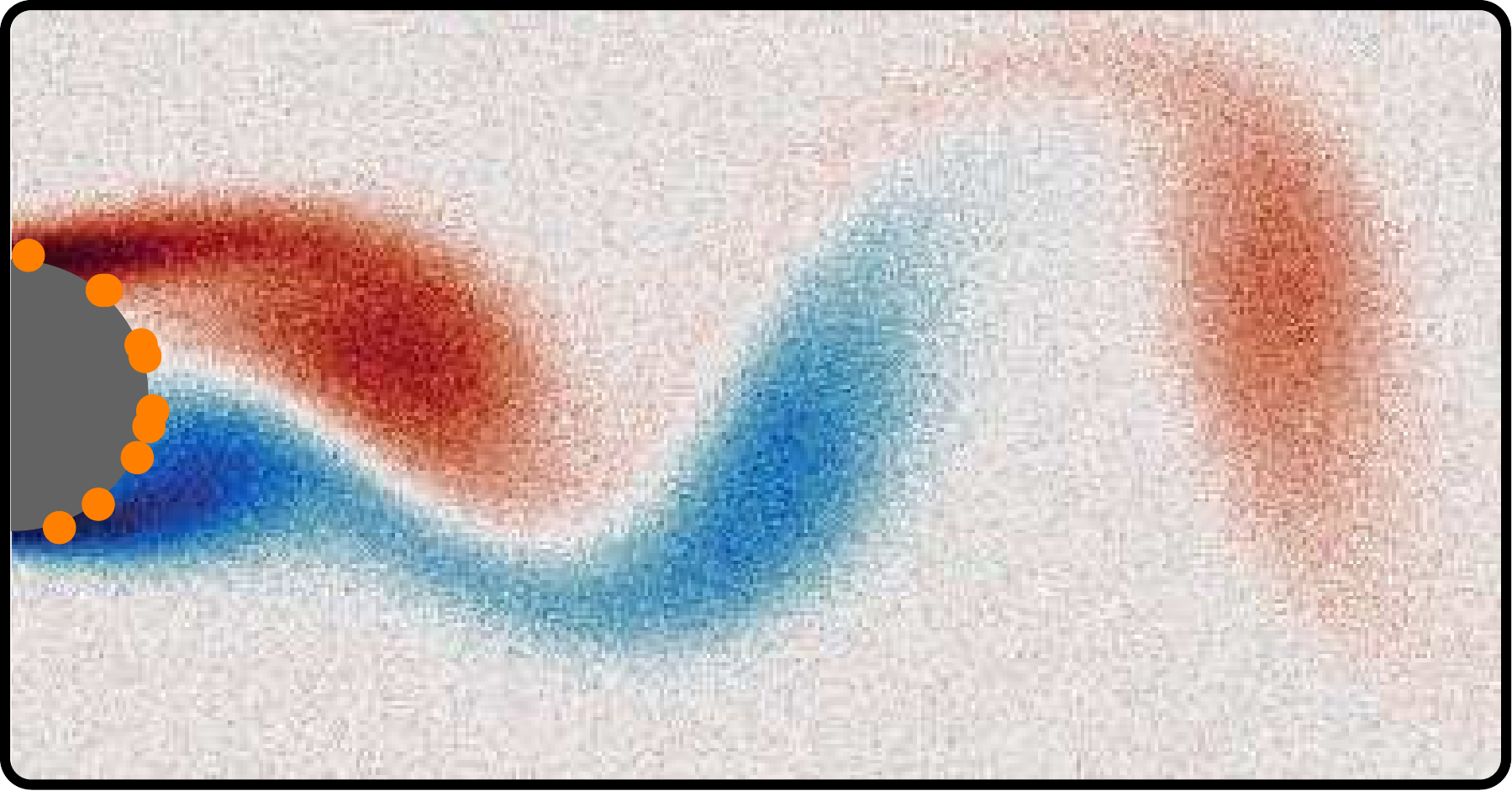}
		\caption{Truth}
	\end{subfigure}
	~	
	\begin{subfigure}[t]{0.4\textwidth}
		\centering
		\DeclareGraphicsExtensions{.pdf}
		\includegraphics[width=1\textwidth]{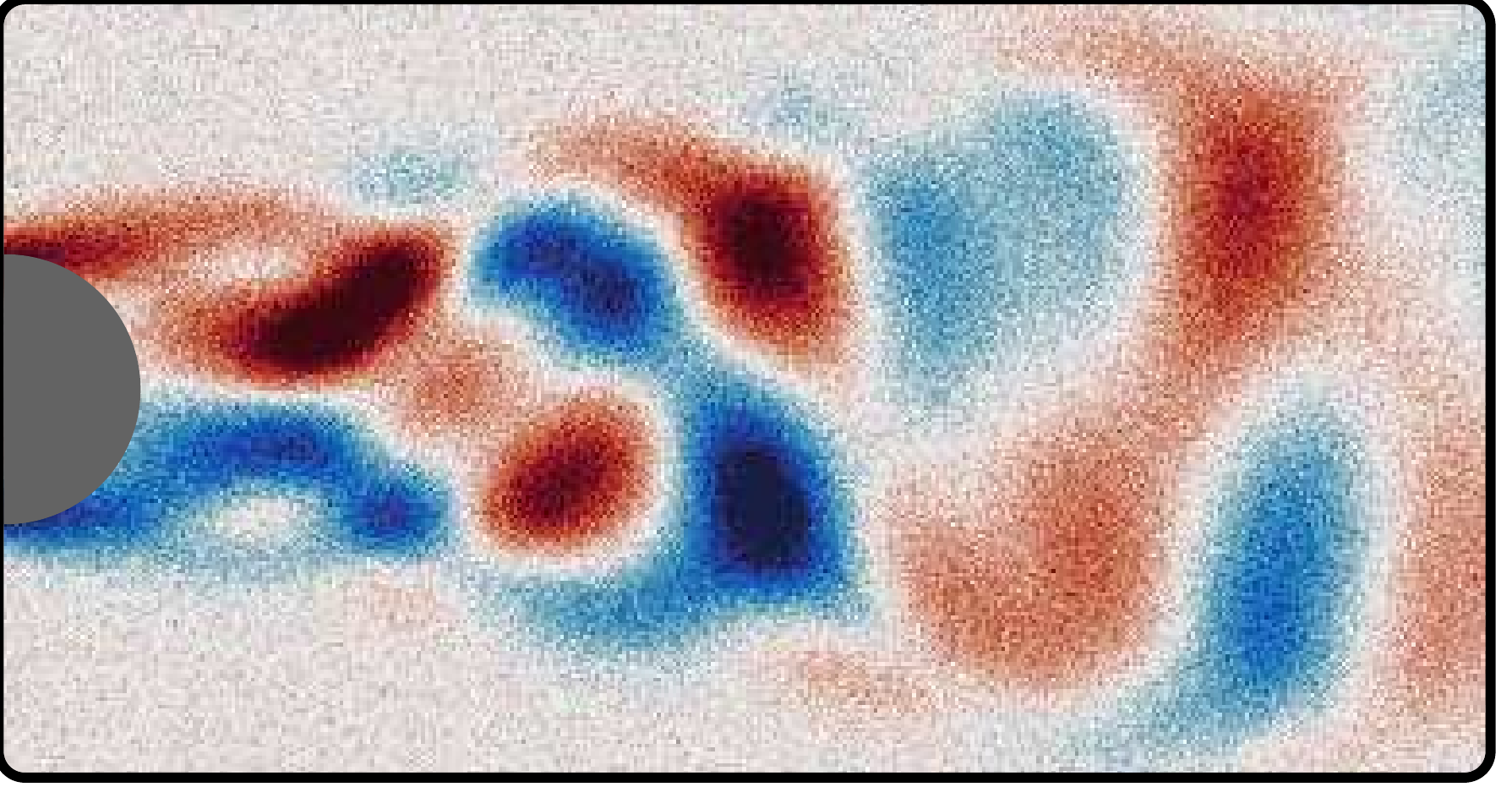}
		\caption{\textsc{pod}}
	\end{subfigure}	
	
	\begin{subfigure}[t]{0.4\textwidth}
		\centering
		\DeclareGraphicsExtensions{.pdf}
		\includegraphics[width=1\textwidth]{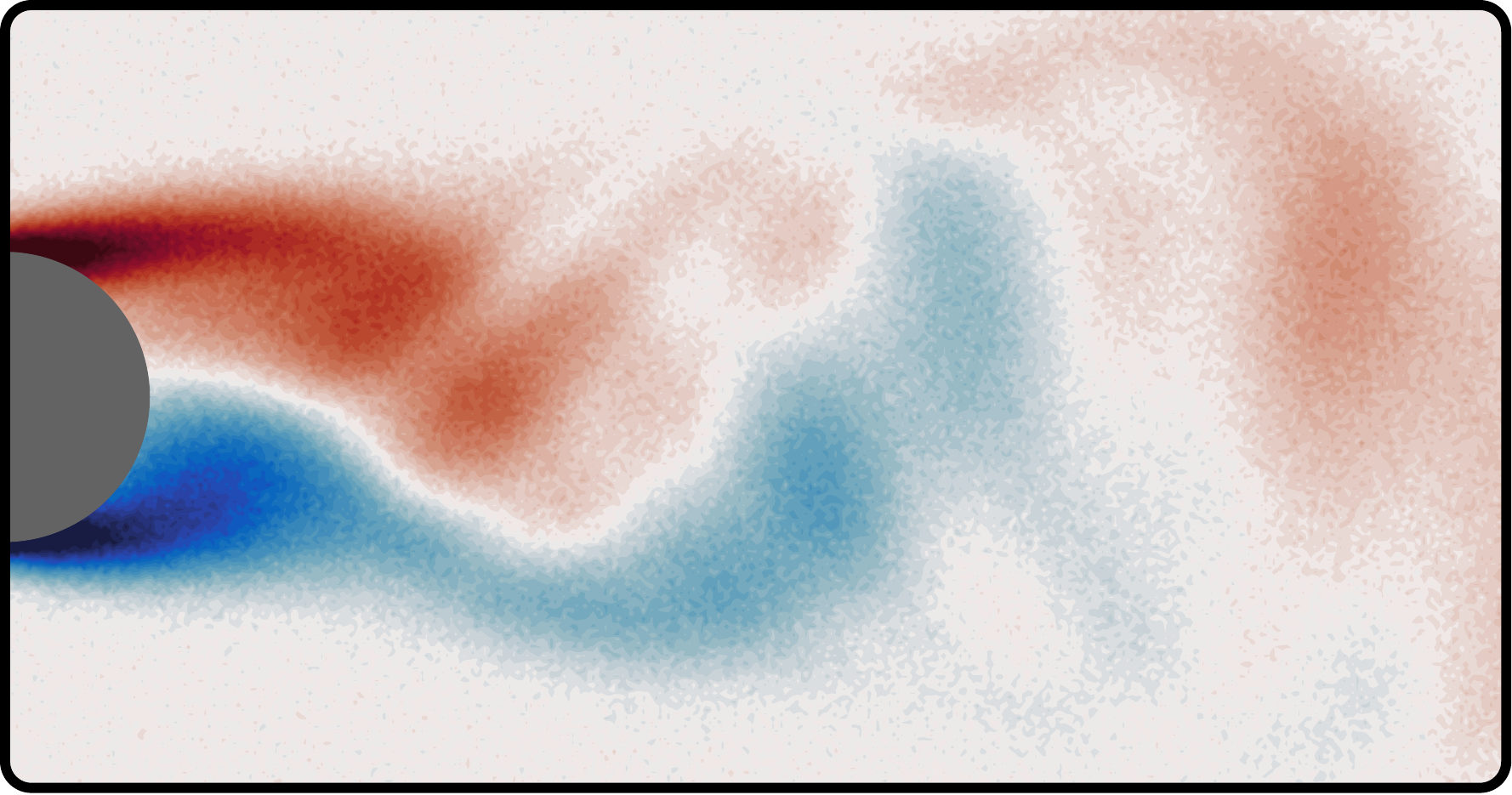}
		\caption{\textsc{pod plus}}
	\end{subfigure}	
	~		
	\begin{subfigure}[t]{0.4\textwidth}
		\centering
		\DeclareGraphicsExtensions{.pdf}
		\includegraphics[width=1\textwidth]{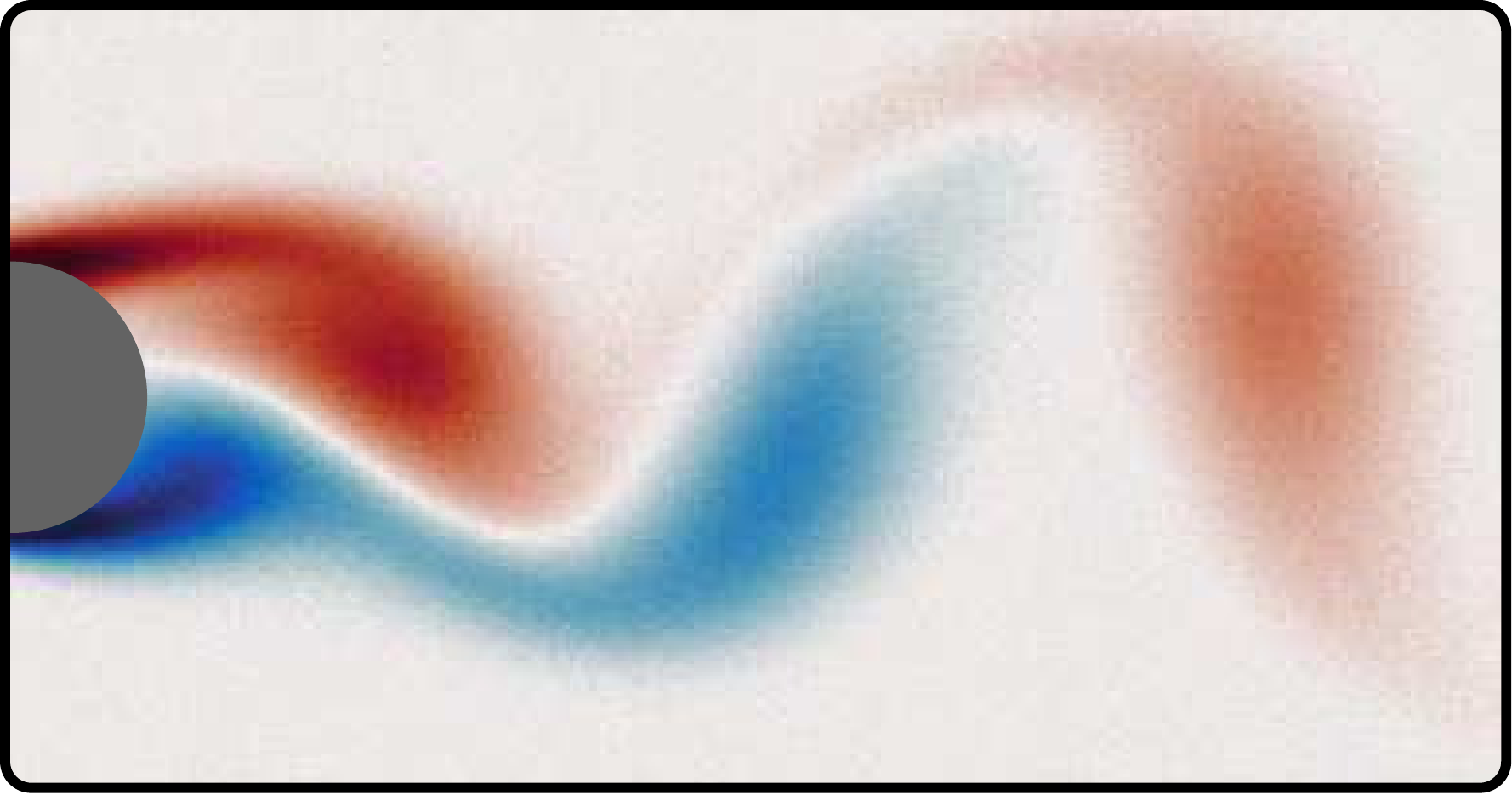}
		\caption{Shallow Decoder}
	\end{subfigure}		
	
	\caption{Visual results for the noisy flow behind the cylinder. Here the signal-to-noise ratio is $10$. In (a) the target snapshot and the corresponding sensor configuration (using $10$ sensors) is shown.  Both, \textsc{pod} and \textsc{pod plus} are not able to reconstruct the flow field, as shown in (b) and (d). The SD is able to reconstruct the coherent structure of the flow field, as shown in (d).}
	\label{fig:cylinder_noisy_vis}
\end{figure}

\subsubsection{Noisy sensor measurements}
To investigate further the robustness and flexibility of the \textsc{shallow decoder}, we consider flow reconstruction in the presence of additive white noise. While this is not of concern when dealing with flow simulations, it is a realistic setting when dealing with flows obtained in experimental studies. Table~\ref{tab:cylinder_noisy_summary} lists the results for both a high and low noise situation with linear measurements. By inspection, the performance of the \textsc{shallow decoder} outperforms classical techniques. In the high noise case, with a signal-to-noise ratio (SNR) of $10$, the average relative reconstruction error for the test set is about $27\%$ for the \textsc{shallow decoder}. For a SNR of $50$, the relative error is as low as $17\%$. Note that we here use an additional dropout layer (placed after the first fully-connected layer) to improve the robustness of the \textsc{shallow decoder}. 
In contrast, standard \textsc{pod} fails in both situations. Again, the \textsc{pod plus} method shows improved results over the standard \textsc{pod}. However, the visual results in Figure~\ref{fig:cylinder_noisy_vis} show that the reconstruction quality of the \textsc{shallow decoder} is favorable.
The \textsc{shallow decoder} shows a clear advantage and a denoising effect. Indeed the reconstructed snapshots allow for a meaningful interpretation of the underlying structure.

\subsubsection{Summary of empirical results for the flow behind the cylinder}

{The empirical results show that the advantage of the \textsc{shallow decoder} compared to the traditional POD based techniques is pronounced, even for a simple problem such as the flow behind the cylinder. It can be seen, that the performance of the traditional techniques is patchy, \ie{}, the reconstruction quality is highly sensitive to the sensor location.
While regularization can mitigate a poor sensor placement design, a relatively larger number ($>15$) of sensors is required in order to achieve an accurate reconstruction performance. 
More challenging situations such as nonlinear measurements and sensor noise pose a challenge for the traditional techniques, while the \textsc{shallow decoder} shows to be able to reconstruct dominant flow features in such situations.
The computational demands required to train the \textsc{shallow decoder} are minimal, \emph{e.g.}, the time for training on a modern GPU remains below two minutes for this example.
}

\begin{table}[!b]
	\centering
	\scalebox{0.95}{
		\begin{tabular}{lcccccccc} \toprule
			& {Sensors}  & \multicolumn{2}{c}{Training Set}  &  \multicolumn{2}{c}{Test Set} \\ 
			\cmidrule{3-4}  \cmidrule{5-6} 
			& 	& {NME} & {NFE}   & {NME} & {NFE}
			\\
			\midrule
			{\textsc{pod}}  			& 32 & {0.637 (0.59)} & {5.915 (5.56)} & {0.649 (0.62)} & {6.04 (5.77)} \\
			
			{\textsc{pod} ($k^*=5$)}  	& 32 & {0.036 (0.00)} & {0.342 (0.01)} & {0.037 (0.00)} & {0.344 (0.01)} \\        			
			
			{\textsc{pod plus} ($\alpha=1\text{e-}5$)}  	& 32 & {0.036 (0.00)} & {0.341 (0.01)} & {0.037 (0.00)} & {0.343 (0.01)} \\        
			
			{\textsc{shallow decoder}}  & 32 & {0.009 (0.00)} & {0.088 (0.00)} & {0.014 (0.00)} & {0.128 (0.00)} \\   
			
			\midrule    
			{\textsc{pod}}  			& 64 & {0.986 (1.34)} & {9.183 (12.5)} & {1.007 (1.36)} & {9.344 (12.7)} \\
			
			{\textsc{pod} ($k^*=14$)}  	& 64 & {0.032 (0.00)} & {0.298 (0.01)} & {0.032 (0.00)} & {0.301 (0.01)} \\   			
			
			{\textsc{pod plus} ($\alpha=5\text{e-}5$)}  	& 64 & {0.032 (0.00)} & {0.301 (0.00)} & {0.032 (0.00)} & {0.301 (0.00)} \\    
			
			{\textsc{shallow decoder}}  & 64 & {0.009 (0.00)} & {0.085 (0.00)} & {0.012 (0.00)} & {0.118 (0.00)} \\    
			
			\bottomrule 
	\end{tabular}}\vspace{+0.2cm}
	\caption{Performance for estimating the SST dataset for varying numbers of sensors. Results are averaged over $30$ runs with different sensor distributions, with standard deviations in parentheses. The SD outperforms the traditional techniques and shows to be highly invariant to the sensor location. { The parameter $k^*$ indicates the number of modes that were used for flow reconstruction by the \textsc{pod} method, and $\alpha$ refers to the strength of ridge regularization applied to \textsc{pod plus}.}}
	\label{tab:sst_summary}
\end{table}
\begin{figure}[!b]
	
	\centering
	\begin{subfigure}[t]{0.49\textwidth}
		\centering
		\DeclareGraphicsExtensions{.png}
		\includegraphics[width=1\textwidth]{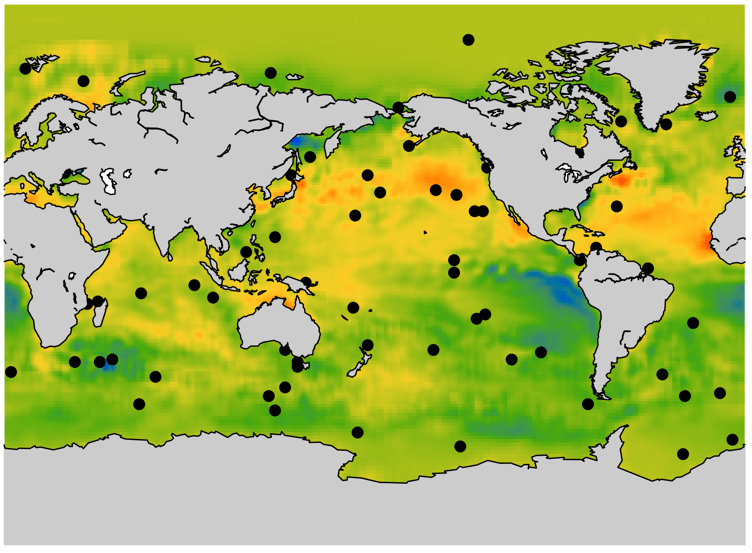}
		\caption{Truth}
	\end{subfigure}
	\hspace{-0.3cm}
	~		
	\begin{subfigure}[t]{0.49\textwidth}
		\centering
		\DeclareGraphicsExtensions{.png}
		\includegraphics[width=1\textwidth]{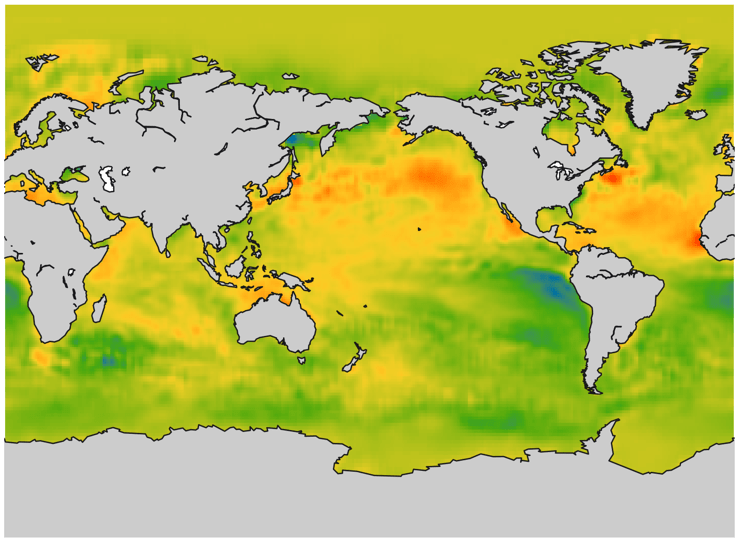}
		\caption{Shallow Decoder}
	\end{subfigure}		
	
	\caption{Visual results for the SST dataset. In (a), the high-dimensional target and the sensor configurations (using $64$ sensors) are shown; and in (b), the results of the Shallow Decoder are shown.  Note that we show here the mean centered snapshot. The \textsc{shallow decoder} shows an excellent reconstruction quality for the fluctuations around the mean with an error as low as $12\%$.}
	\label{fig:sst_vis}
\end{figure}

\subsection{Sea surface temperature using random point-wise measurements}

The second example we consider is the more challenging sea surface temperature (SST) dataset. Complex ocean dynamics lead to rich flow phenomena, featuring interesting seasonal fluctuations. While the mean SST flow field is characterized by a periodic structure, the flow is non-stationary.
The dataset consists of the weekly sea surface temperatures for the last 26 years, publicly available from the National Oceanic \& Atmospheric Administration (NOAA).%
%
The data comprise $1483$ snapshots in time with spatial resolution of $180\times 360$. For the following experiments, we only consider $44,219$ measurements, by excluding measurements corresponding to the land masses. Further, we create a training set by selecting $1100$ snapshots at random, while the remaining snapshots are used for validation.

We consider the performance of the \textsc{shallow decoder} using varying numbers of random sensors scattered across the spatial domain. The results are summarized in Table~\ref{tab:sst_summary}. 
We observe a large discrepancy between the NME and NFE error. This is because the long-term annual mean field accounts for the majority of the spatial structure of the field. Hence, the NME error is uninformative with respect to the performance of reconstruction methods. In terms of the NFE error the POD based reconstruction techniques is shown to fail to reconstruct the high-dimensional flow field using limited sensor measurements. In contrast, the \textsc{shallow decoder} demonstrates an excellent reconstruction performance both using $32$ and $64$ measurements.
Figure~\ref{fig:sst_vis} shows visual results to support these quantitative findings.

\subsection{Turbulent flow using sub-gridscale measurements}\label{sec:result_additional}

The final example we consider is the velocity field of a turbulent isotropic flow. Unlike the previous examples, the isotropic turbulent flow is non-periodic in time and highly non-stationary.
Thus, this dataset poses a challenging task.
Here, we consider data from a forced isotropic turbulence flow generated with a direct numerical simulation using $1,024^3$ points in a triply periodic $\left[0, 2\pi\right]^3$ domain. For the following experiments, we are using $800$ snapshots for training and $200$ snapshots for validation. The data spread across about one large-eddy turnover time.
The data is provided as part of the Johns Hopkins Turbulence Database~\cite{li2008public}.

\begin{figure}[!b]
	
	\centering
	\begin{subfigure}[t]{0.31\textwidth}
		\centering
		\DeclareGraphicsExtensions{.pdf}
		\includegraphics[width=1\textwidth]{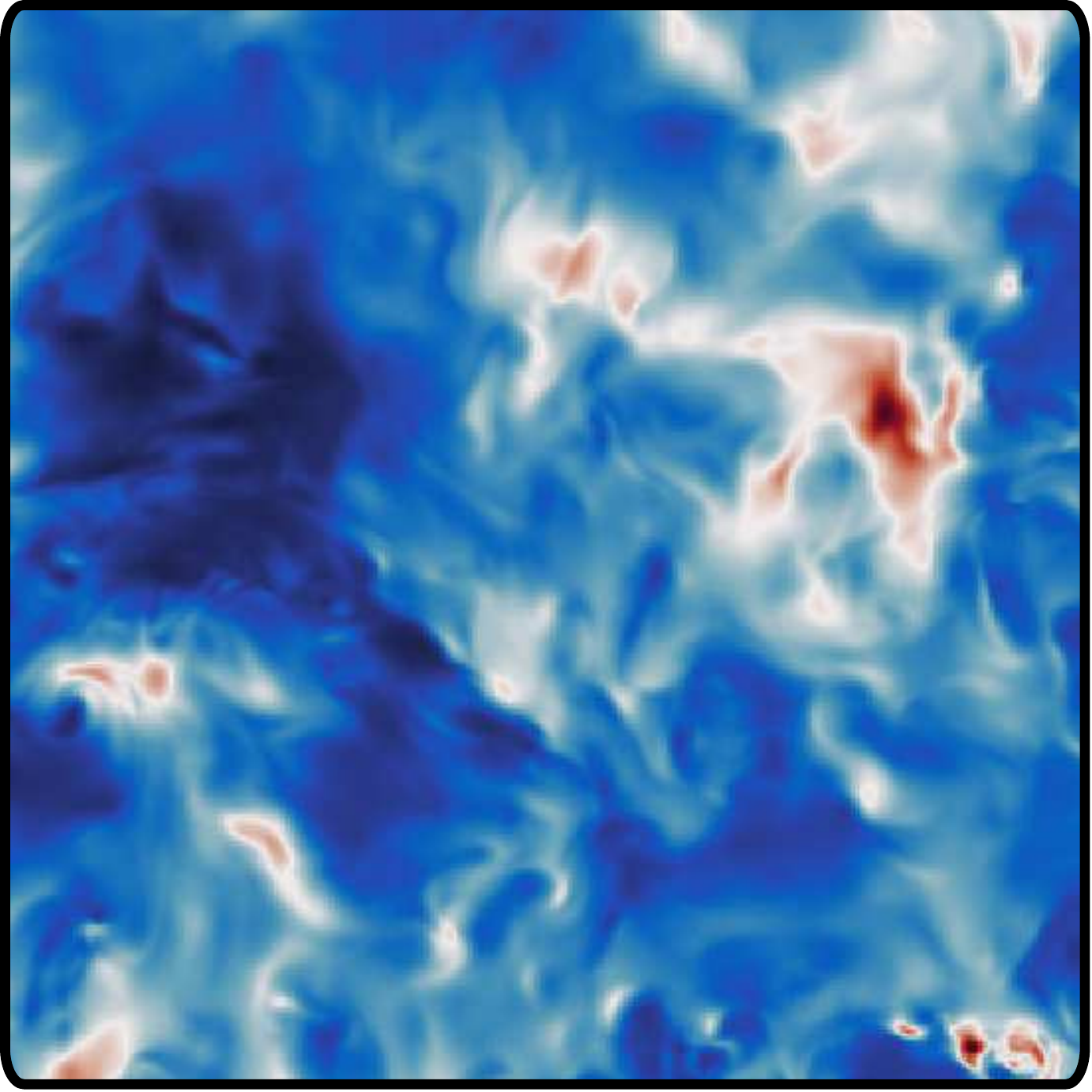}
		\caption{Snapshot}
	\end{subfigure}
	~
	\begin{subfigure}[t]{0.31\textwidth}
		\centering
		\DeclareGraphicsExtensions{.pdf}
		\includegraphics[width=1\textwidth]{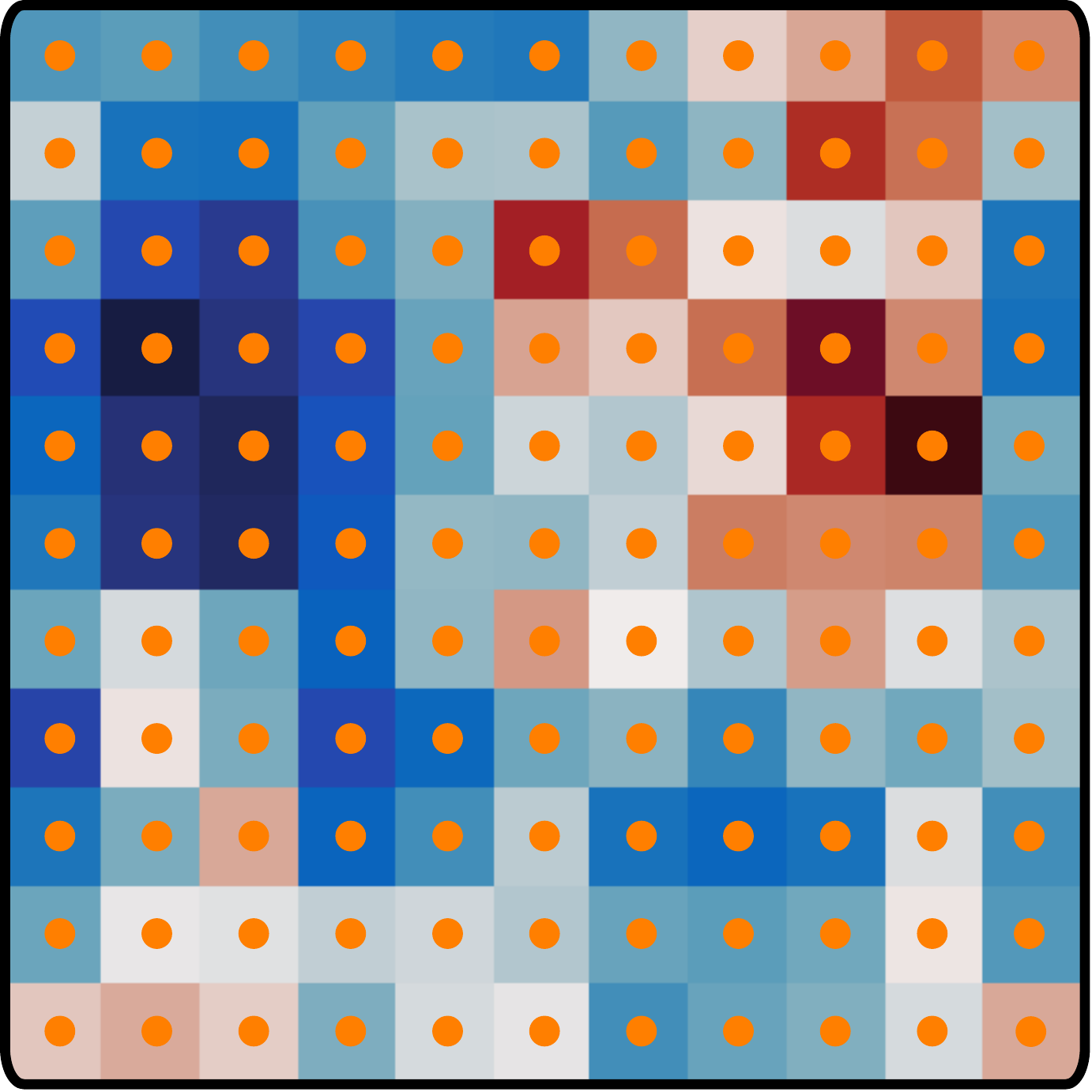}
		\caption{Low resolution}
		\label{fig:fig:isotropic_low_res}
	\end{subfigure}	
	~	
	\begin{subfigure}[t]{0.31\textwidth}
		\centering
		\DeclareGraphicsExtensions{.pdf}
		\includegraphics[width=1\textwidth]{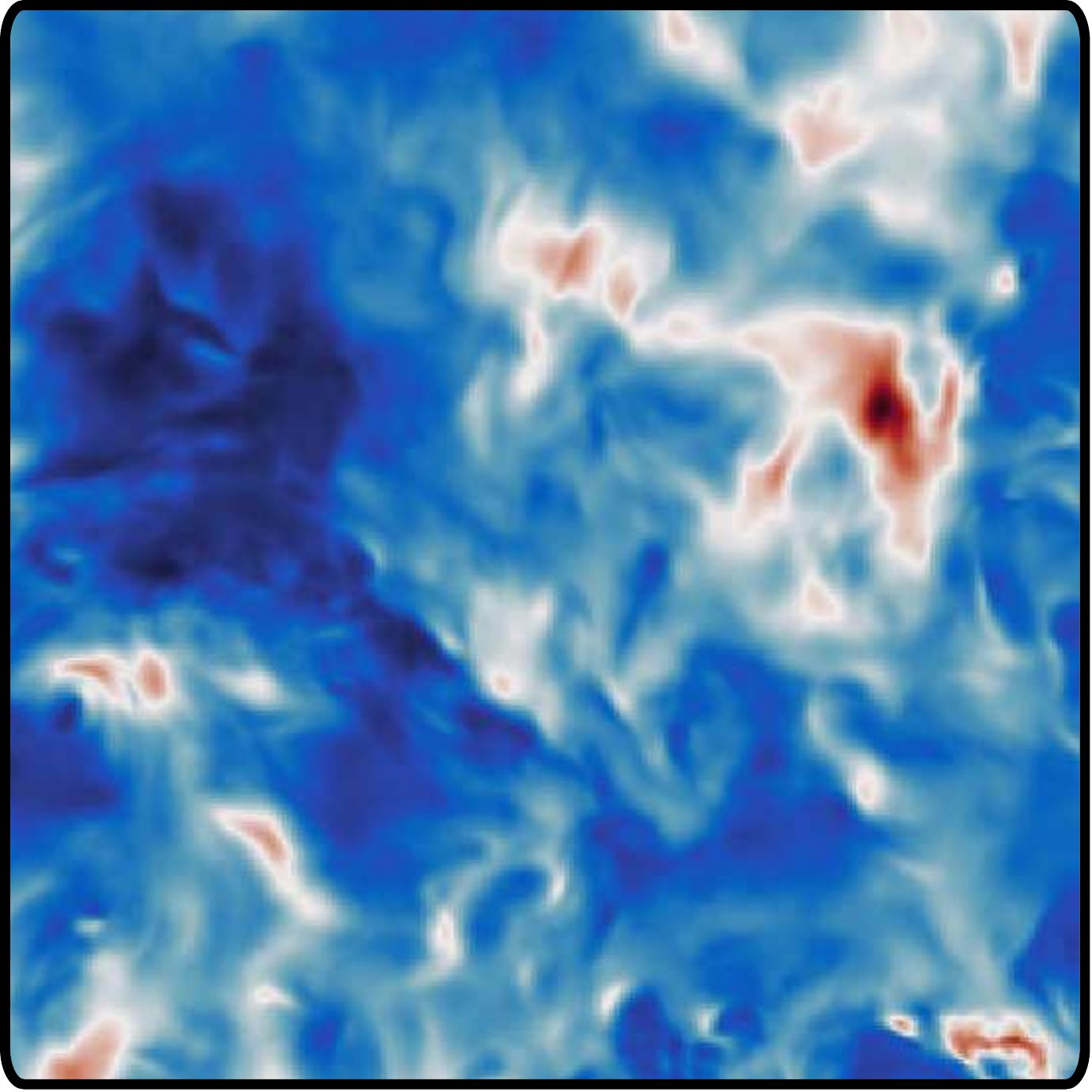}
		\caption{Shallow Decoder}
	\end{subfigure}

	\caption{Visual results for the turbulent isotropic flow using 121 subgrid-cell measurements. The interpolation error of the \textsc{shallow decoder} error is about $9.3\%$.}
	\label{fig:isotropic_vis}
\end{figure}

If the sensor measurements $\bobs$ are acquired on a coarse but regular grid, then the reconstruction task may be considered as a \emph{super-resolution} problem~\cite{Yang2010ieeetip,Freeman2002ieeecga,Callaham2018arxiv}.  
There are a number of direct applications of super-resolution in fluid mechanics centered around sub-gridscale modeling.  
Because many fluid flows are inherently multiscale, it may be prohibitively expensive to collect data that captures all spatial scales, especially for iterative optimization and real-time control~\cite{Brunton2015amr}.  
Inferring small-scale flow structures below the spatial resolution available is an important task in large eddy simulation (LES), climate modeling, and particle image velocimetry (PIV), to name a few applications.  
Deep learning has recently been employed for super-resolution in fluid mechanics applications with promising results~\cite{fukami2018super}.  
{ Note that our setting differs from the super-resolution problem. Here, we obtain first a low-resolution image by applying a mean filter to the high-dimensional snapshot. Then, we use a single sensor measurement per grid cell to form the inputs (illustrated in Figure~\ref{fig:fig:isotropic_low_res}). In contrast, super-resolution uses the low-resolution image as input.}

\begin{table}[!b]
	\centering
	\scalebox{0.95}{
		\begin{tabular}{lcccccccc} \toprule
			& {Grids} &  \multicolumn{2}{c}{Training Set}  &  \multicolumn{2}{c}{Test Set} \\ 
			\cmidrule{3-4}  \cmidrule{4-6} 
			& 	& {NME} & {NFE}  & {NME} & {NFE}
			\\
			\midrule
			{\textsc{shallow decoder}}  	& 36 & {0.029 (0.00)} & {0.041 (0.00)} & {0.071 (0.00)} & {0.101 (0.01)} \\			
			{\textsc{shallow decoder}}  	& 64 & {0.027 (0.00)} & {0.039 (0.00)} & {0.067 (0.00)} & {0.096 (0.00)} \\ 
			{\textsc{shallow decoder}}  	& 121 & {0.026 (0.00)} & {0.038 (0.00)} & {0.066 (0.00)} & {0.093 (0.00)} \\
			\bottomrule
			
	\end{tabular}}\vspace{+0.2cm}
	\caption{Flow reconstruction performance for estimating the isotropic flow. Results are averaged over $30$ runs with different sensor distributions, with standard deviations in parentheses.}
	\label{tab:iso_summary}	
\end{table}

\begin{figure}[!t]
	\centering
	\begin{subfigure}[t]{0.31\textwidth}
		\centering
		\DeclareGraphicsExtensions{.pdf}
		\includegraphics[width=1\textwidth]{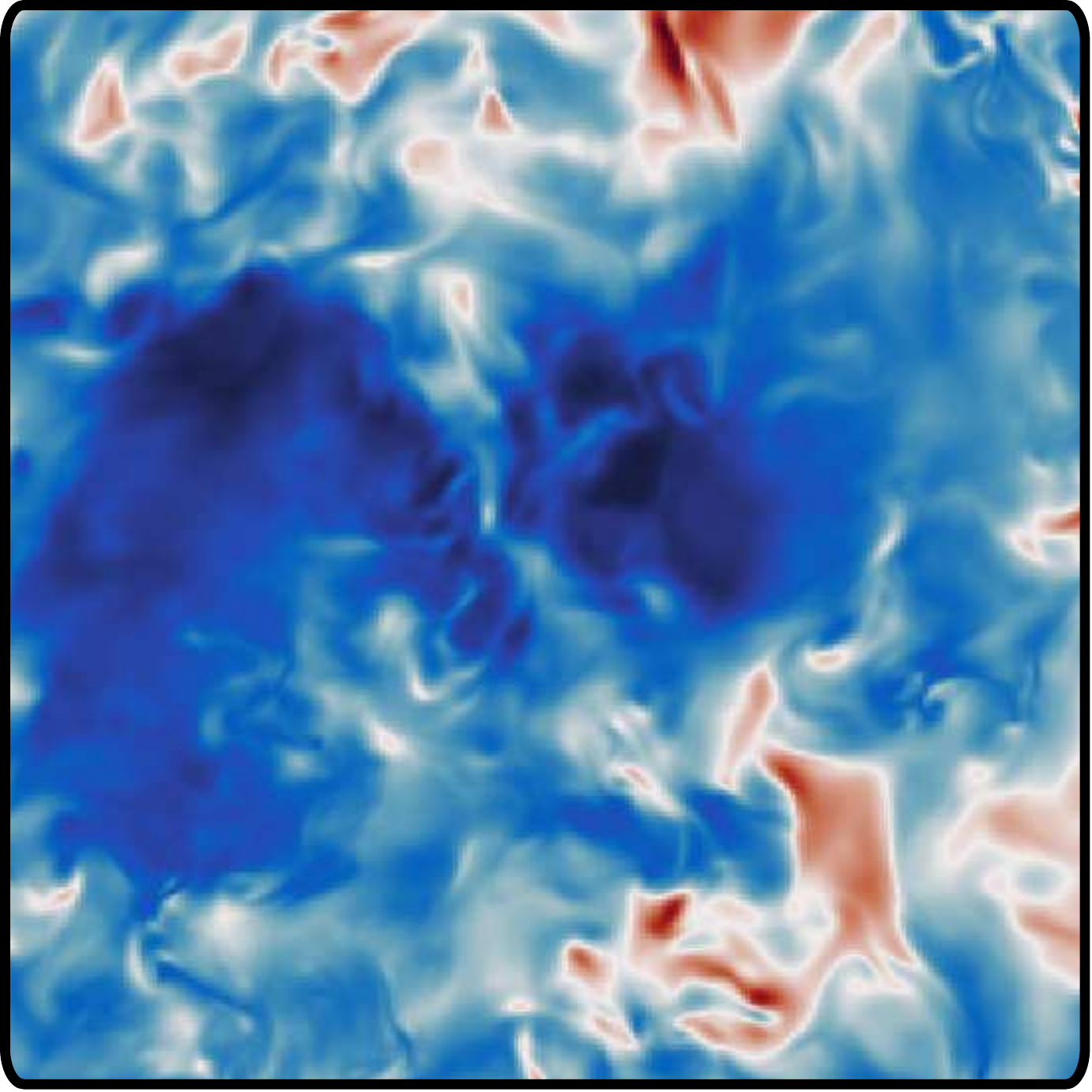}
		\caption{Test snapshot $t=1$}\label{fig:iso_extrapolation_close}
		\includegraphics[width=1\textwidth]{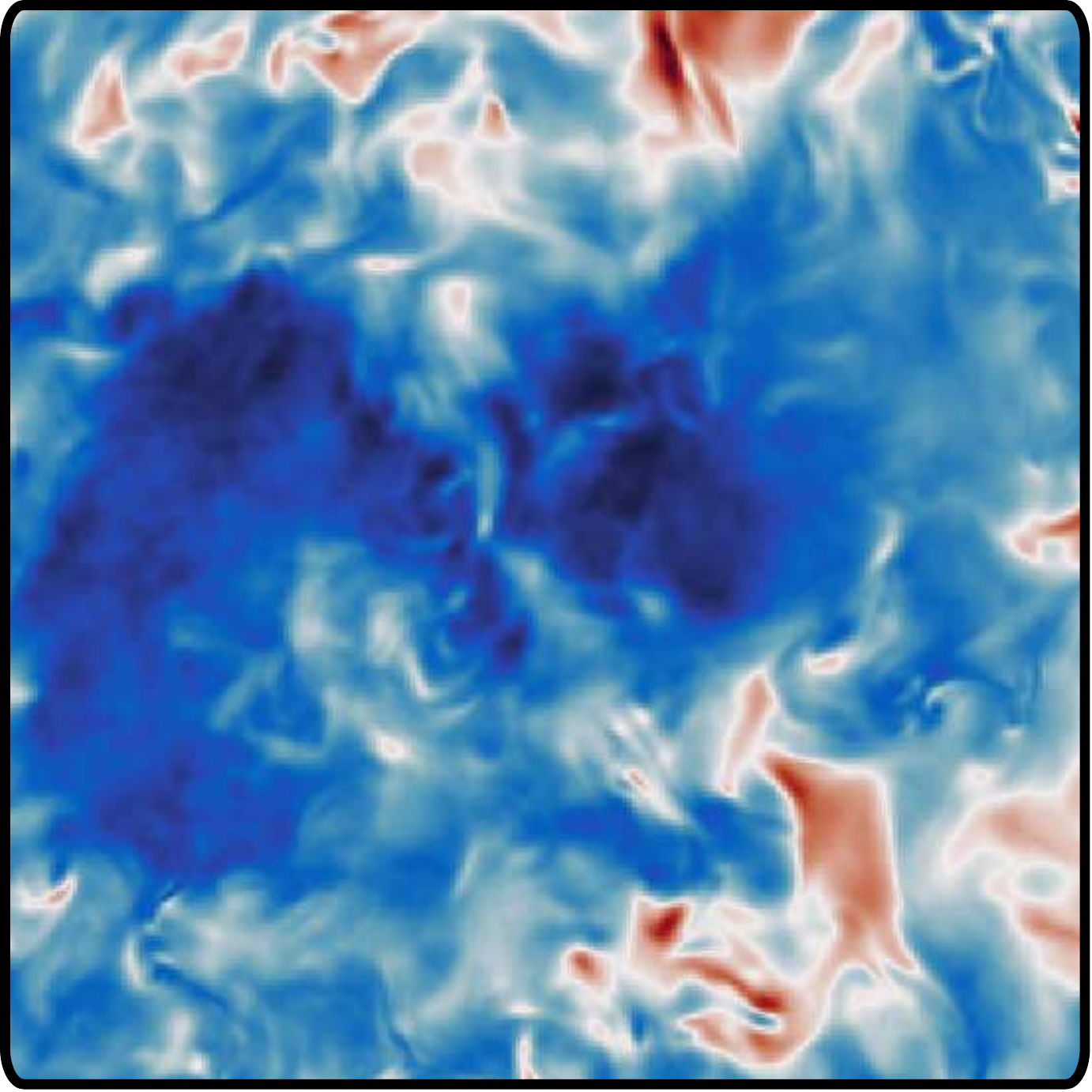}
		\caption{Reconstruction}		
	\end{subfigure}
	~
	\begin{subfigure}[t]{0.31\textwidth}
		\centering
		\DeclareGraphicsExtensions{.pdf}
		\includegraphics[width=1\textwidth]{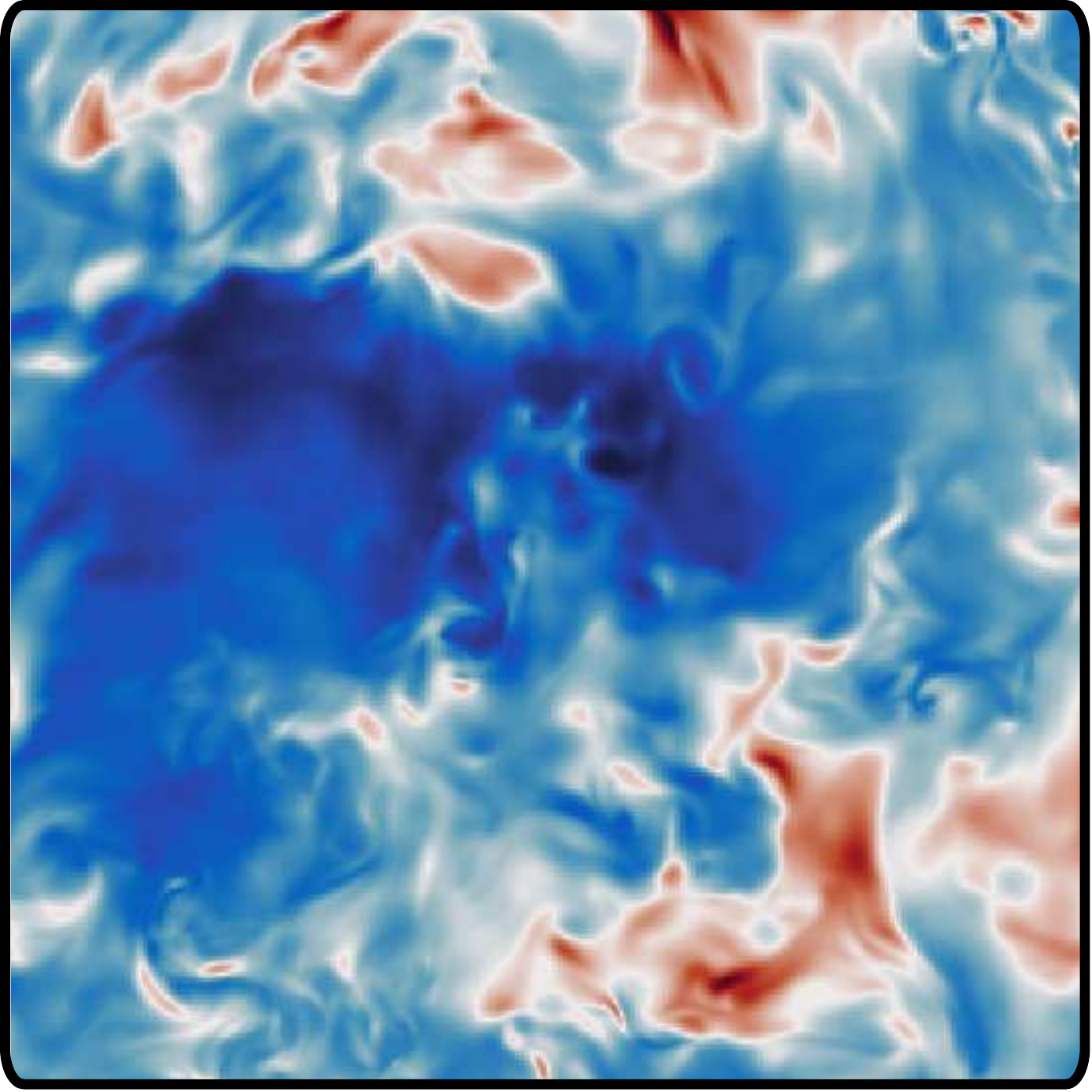}
		\caption{Test snapshot $t=20$}		\label{fig:iso_extrapolation_mid}
		\includegraphics[width=1\textwidth]{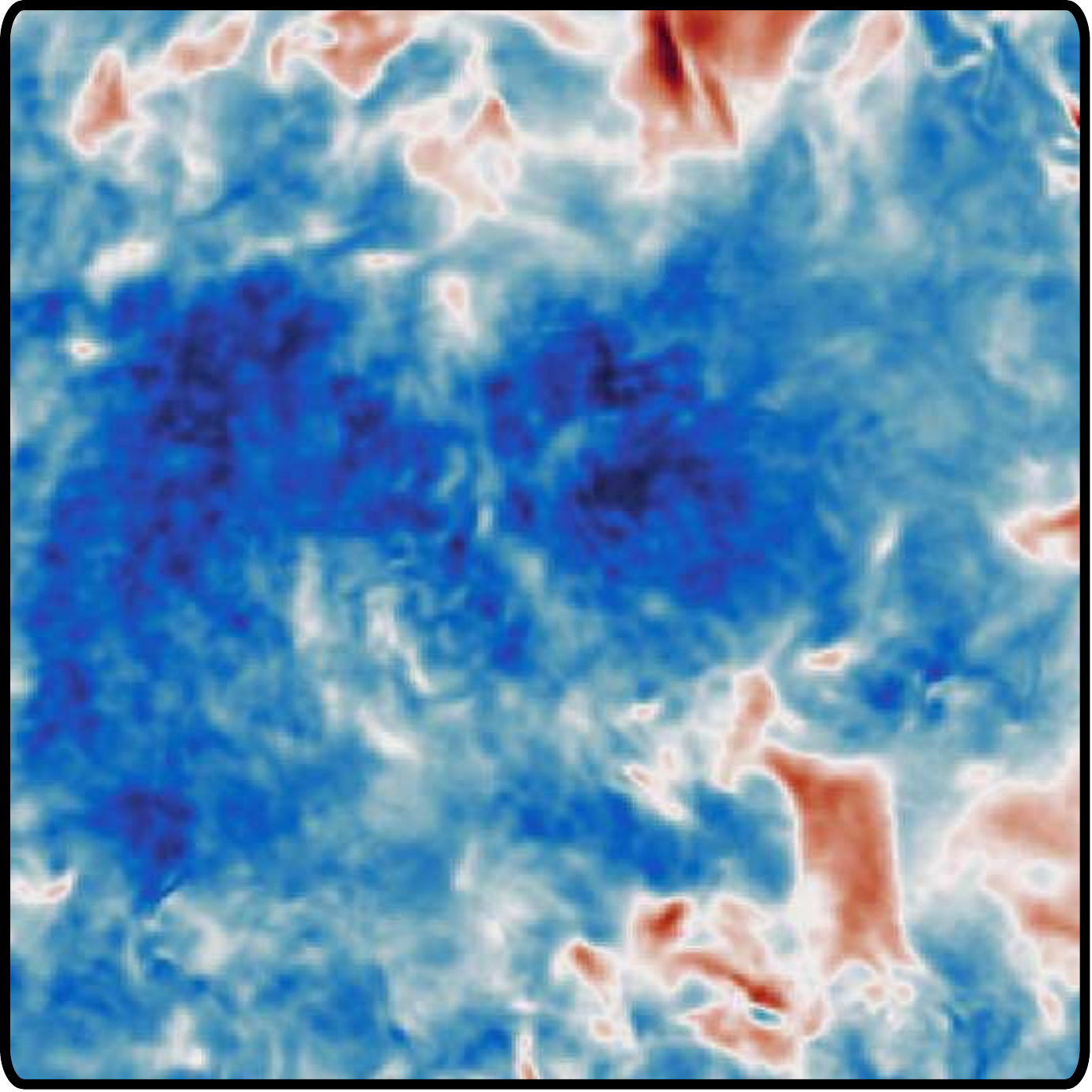}
		\caption{Reconstruction}
	\end{subfigure}	
	~	
	\begin{subfigure}[t]{0.31\textwidth}
		\centering
		\DeclareGraphicsExtensions{.pdf}
		\includegraphics[width=1\textwidth]{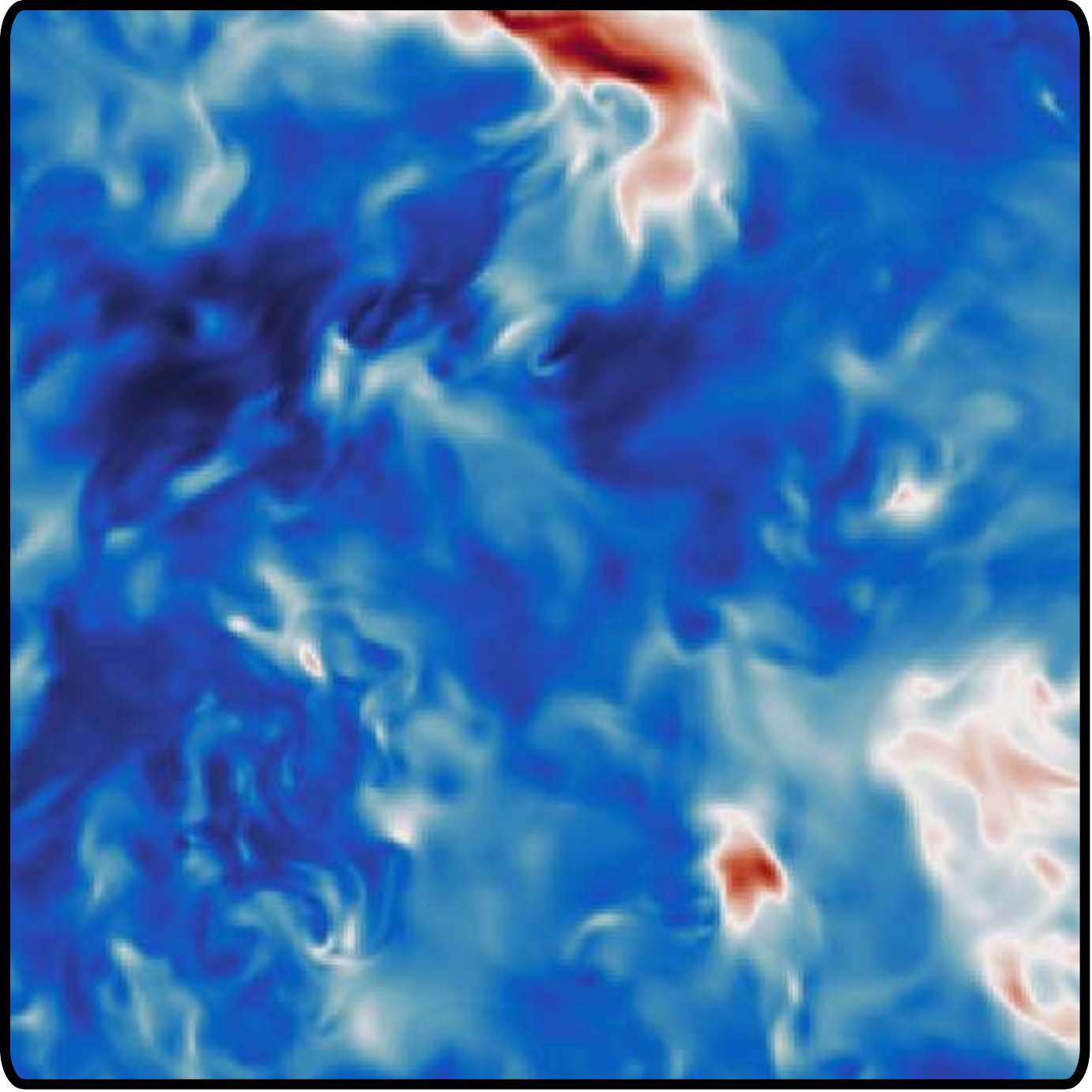}
		\caption{Test snapshot $t=50$}		\label{fig:iso_extrapolation_far}
		\includegraphics[width=1\textwidth]{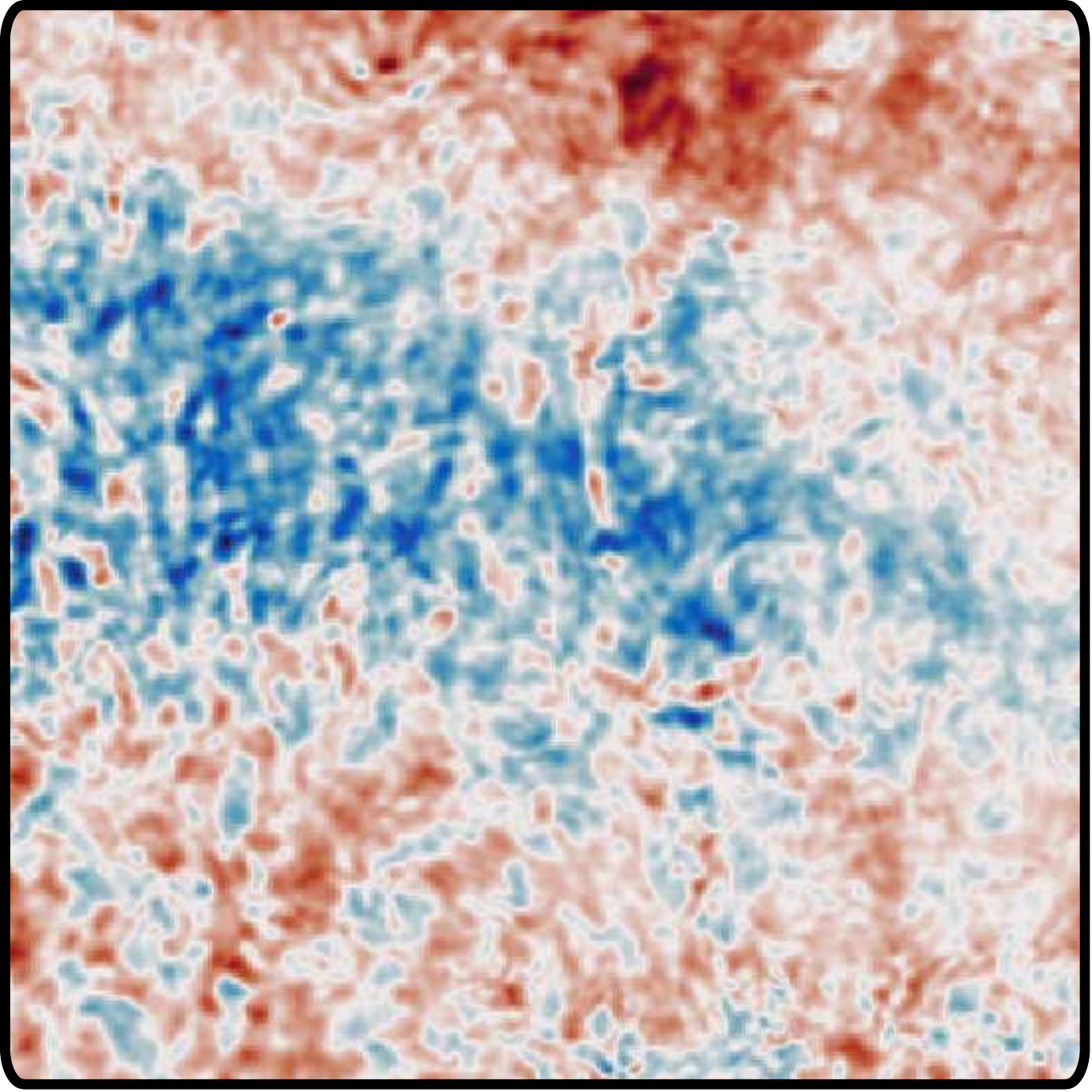}
		\caption{Reconstruction}
	\end{subfigure}	
	\caption{Visual results illustrating the limitation of the \textsc{shallow decoder} for extrapolation tasks. Flow fields sampled from or close to the statistical distribution describing the training examples can be reconstructed with high accuracy, as shown in (a) and (b). Extrapolation fails for fields which belong to a different statistical distribution, as shown in (e) and (f).}
	\label{fig:iso_extrapolation}
\end{figure}

First, we consider the within sample prediction task. In this case, we yield excellent results for the estimated high-dimensional flow fields, despite the challenging problem. Table~\ref{tab:iso_summary} quantifies the performance for varying numbers of sub-gridscale measurements. 
In addition, Figure~\ref{fig:isotropic_vis} provides some visual evidence for the good performance for this problem.

Next, we illustrate the limitation of the \textsc{shallow decoder}. Indeed, it is important to stress that the SD cannot be used for ``out of sample prediction tasks'' if the fluid flow is highly non-stationary.
To illustrate this issue, Figure~\ref{fig:iso_extrapolation} shows three flow fields at different temporal locations. First, Figure~\ref{fig:iso_extrapolation_close} shows a test example, which is close in time to the training set. In this case, the SD is able to reconstruct the flow field with high accuracy. The reconstruction quality drops for snapshots which are further away in time, as shown in Figure~\ref{fig:iso_extrapolation_mid}. 
Finally, Figure~\ref{fig:iso_extrapolation_far} shows that reconstruction fails if the test example is far away from the training set in time, \ie{}, the flow field is not drawn from the same statistical distribution as the training examples are.

\section{Discussion}    \label{sec:Sec_Discussion}

{
The emergence of sensor networks for global monitoring (\eg{}, ocean and atmospheric monitoring) requires new mathematical techniques that are capable of maximally exploiting sensors for state estimation and forecasting.  Emerging algorithms from the machine learning community can be integrated with many traditional scientific computing approaches to enhance sensor network capabilities.  
For many global monitoring applications, the placement of sensors can be prohibitively expensive, thus requiring learning techniques such as the one proposed here, which can exploit a reduction in the number of sensors while maintaining required performance characteristics.  

To partially address this challenge, we proposed a \textsc{shallow decoder} with two hidden layers for the problem of flow reconstruction.
The mathematical formulation presented is significantly different from what is commonly used in flow reconstruction problems, \eg{}, gappy interpolation with dominant POD modes.
Indeed, our experiments demonstrate the improved the enhanced robustness and accuracy of fluid flow field reconstruction by using our \textsc{shallow decoder}.

Future work aims to leverage the underlying laws of physics in flow problems to further improve the efficiency. In the context of flow reconstruction or, more generally, observation of a high-dimensional physical system, insights from the physics at play can be exploited~\cite{raissi2018hidden}. In particular, the dynamics of many systems do indeed remain low-dimensional and the trajectory of their state vector lies close to a manifold whose dimension is significantly lower than the ambient dimension.
Moreover, the features exploited from the shallow decoder network can also be integrated in reduced order models (ROMs) for forecasting predictions~\cite{benner2015survey}. 
In many high-dimensional systems where ROMs are used, the ability to generate low-fidelity models that can be rapidly simulated has revolutionized our ability to model such complex systems, especially in application of complex flow fields. The ability to rapidly generate low-rank feature spaces alternative to POD generates new possibilities for ROMs using limited sampling and limited data.  This aspect of the \textsc{shallow decoder} will be explored further in future work.
}

\section*{Acknowledgments}
LM gratefully acknowledges the support of the French Agence Nationale pour la Recherche (ANR) and Direction G\'en\'erale de l'Armement (DGA) via the \textit{FlowCon} project (ANR-17-ASTR-0022).  
SLB acknowledges support from the Army Research Office (ARO W911NF-17-1-0422). 
JNK acknowledges support form the Air Force Office of Scientific Research (FA9550-19-1-0011).  LM and JNK also acknowledge support from the  Air Force Office of Scientific Research 
(FA9550-17-1-0329).
MWM would like to acknowledge ARO, DARPA, NSF, and ONR for providing partial support for this work.
We  would  also  like  to  thank  Kevin Carlberg for valuable  discussions about flow reconstruction techniques.

\appendix

\section{Hyper-parameter search for the POD based methods}\label{sec:tuning}

{In the following, we provide results of our hyper-parameter search for determining the optimal tuning parameters for flow reconstruction.
We proceed by evaluating the reconstruction error of the \textsc{pod} and \textsc{pod plus} method for a plausible range of values. Here, we consider hard-threshold regularization for the \textsc{pod} method and ridge regularization for the \textsc{pod plus} method.
We run $30$ trails of the experiment, where we use a unique sensor location configuration at each trial.

Figure~\ref{fig:pod_flow} shows the results for the fluid flow past the cylinder. First, we show the results for the \textsc{pod plus} method in (a). Regularizing the solution improves the reconstruction accuracy and the effect of regularization on the reconstruction error is pronounced for an increasing number of sensors. (Note, that at the same time, while the reconstruction error is decreasing with an increasing numbers of sensors, finding the optimal tuning parameter becomes more difficult.) Next, we show the results for \textsc{pod} with hard-threshold regularization in (b). It can be seen, that the performance is on par with ridge regularization (plotted as black dashed line), where hard-threshold regularization shows to have a lower variance compared to ridge regularization. In contrast, the shallow decoder outperforms both the \textsc{pod} and the \textsc{pod plus} method, represented by a dashed read line. However, the performance gap between the POD-based methods and the shallow decoder is closing for an increased number of sensors. This is not surprising, since the flow past the cylinder represents a relatively simple problem where the \textsc{pod} method is known to provide good reconstruction results, given a sufficient large number of sensors.

Figure~\ref{fig:pod_flow_10sensors} and~\ref{fig:pod_flow_regul} show the results for the noisy flow past the cylinder and for the sea surface temperature data. Again, it can be seen that both ridge regularization and hard-threshold regularization performs on par, while the shallow decoder outperforms the POD-based methods.

\begin{figure}[!t]
	
	\centering
	\begin{subfigure}[t]{0.75\textwidth}
		\centering
		\DeclareGraphicsExtensions{.pdf}
		
		\begin{overpic}[width=0.31\textwidth]{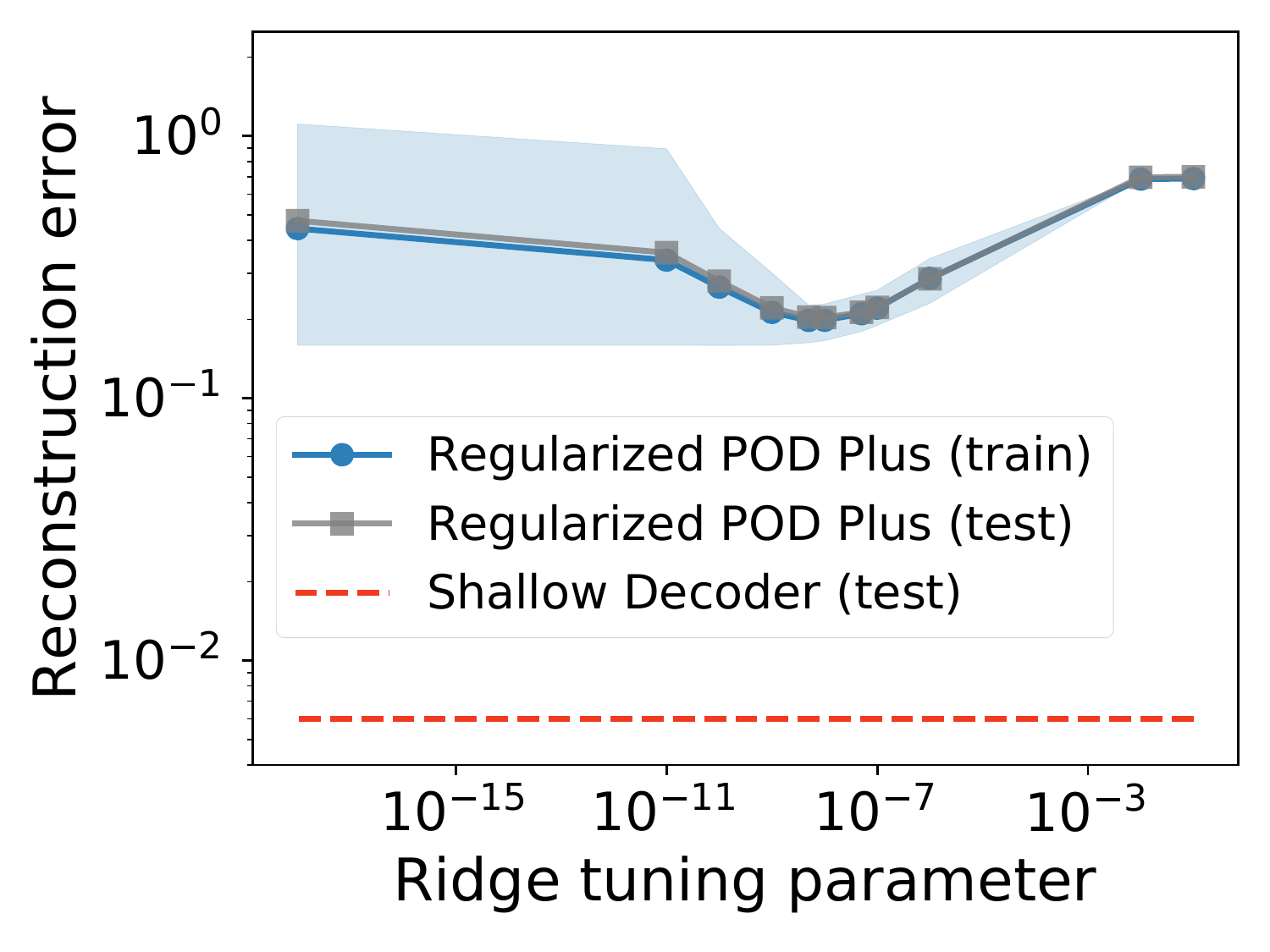} 
			\put(44, 76){\scriptsize 5 sensors}	
		\end{overpic}
		\begin{overpic}[width=0.31\textwidth]{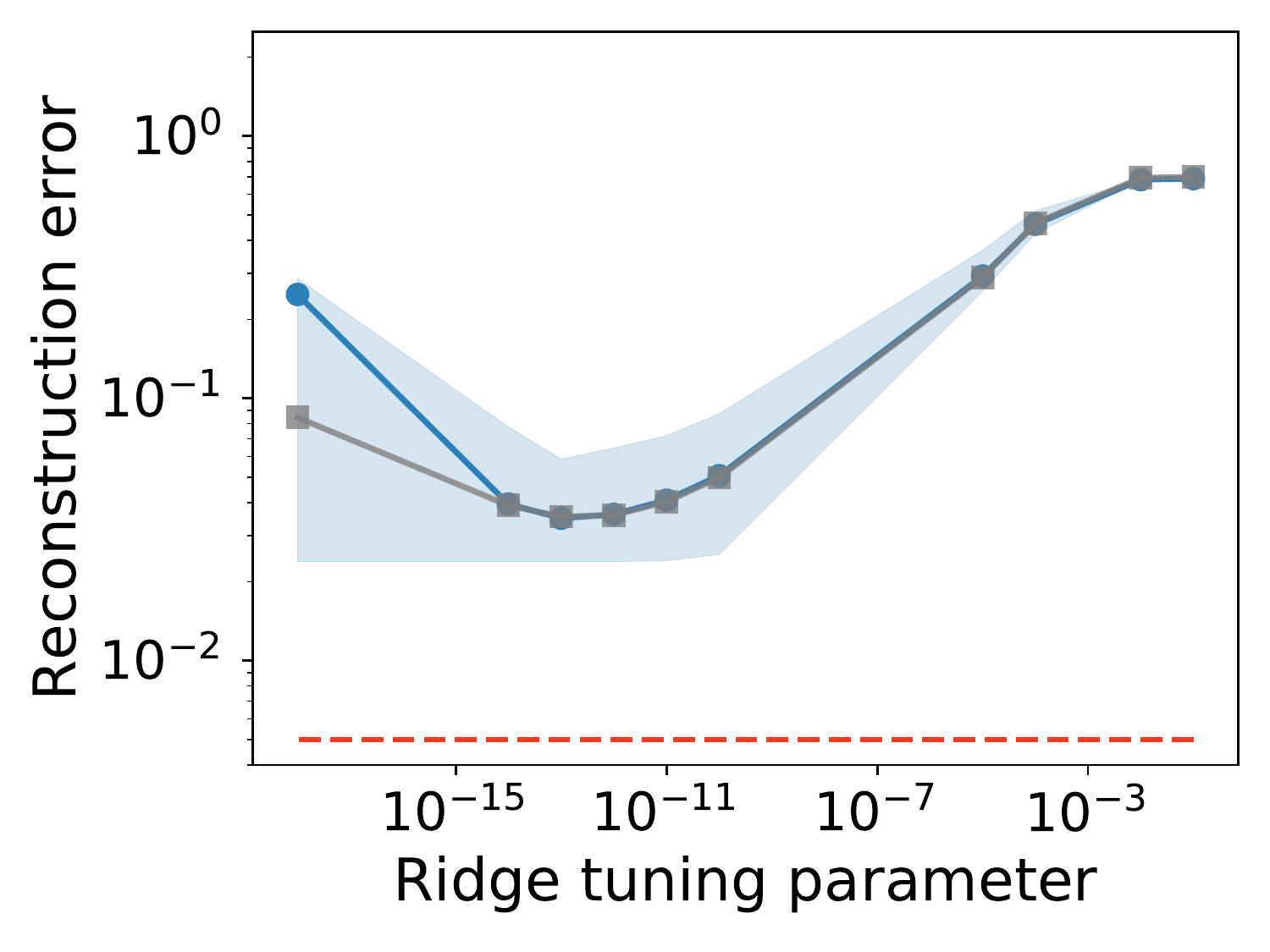} 
			\put(44, 76){\scriptsize 10 sensors}	
		\end{overpic}
		\begin{overpic}[width=0.31\textwidth]{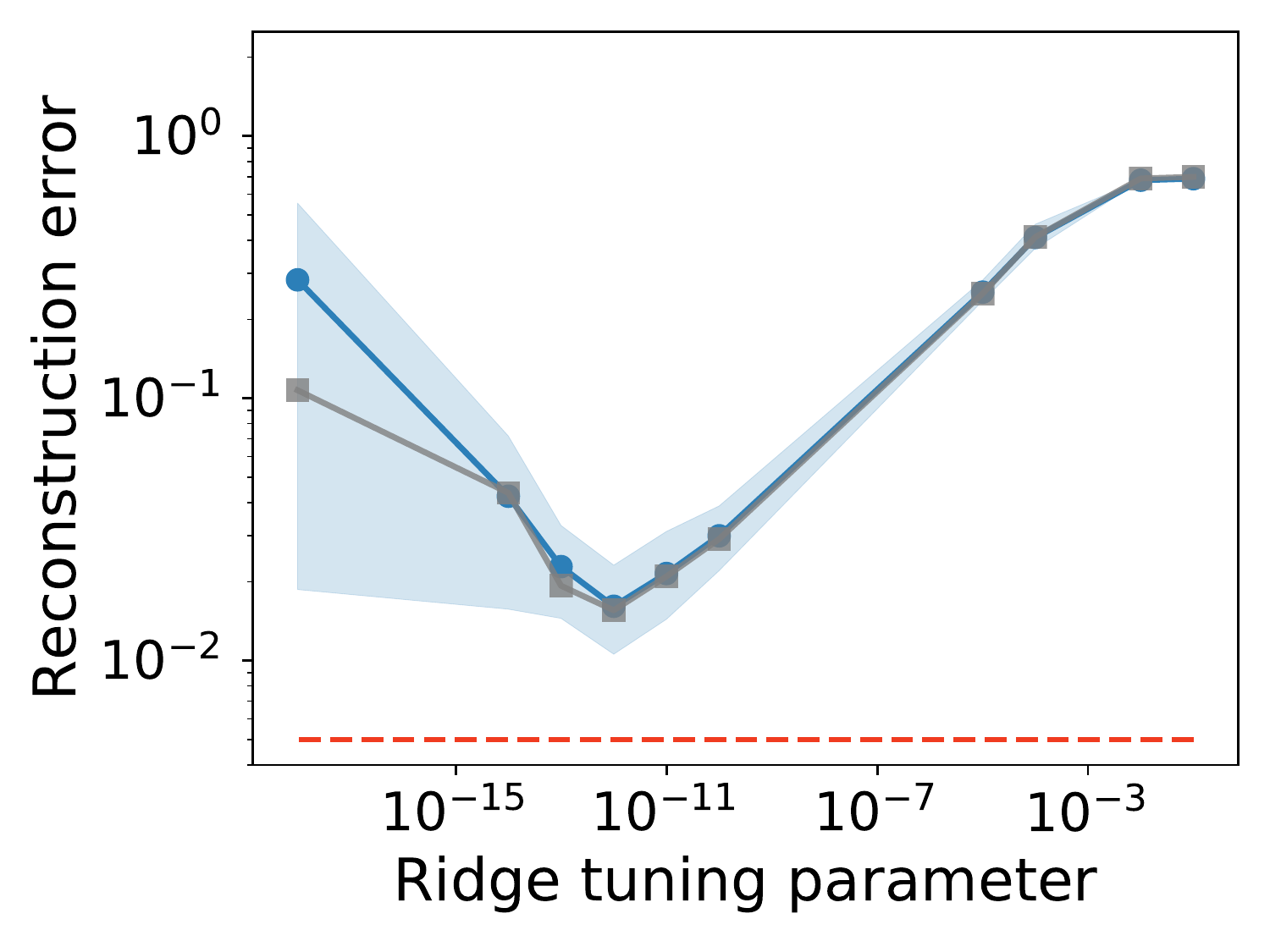} 
			\put(44, 76){\scriptsize 15 sensors}	
		\end{overpic}		
		\caption{POD Plus with ridge regularization.}
	\end{subfigure}\vspace{+0.5cm}
	
	\begin{subfigure}[t]{0.75\textwidth}
		\centering
		\DeclareGraphicsExtensions{.pdf}
		
		\begin{overpic}[width=0.31\textwidth]{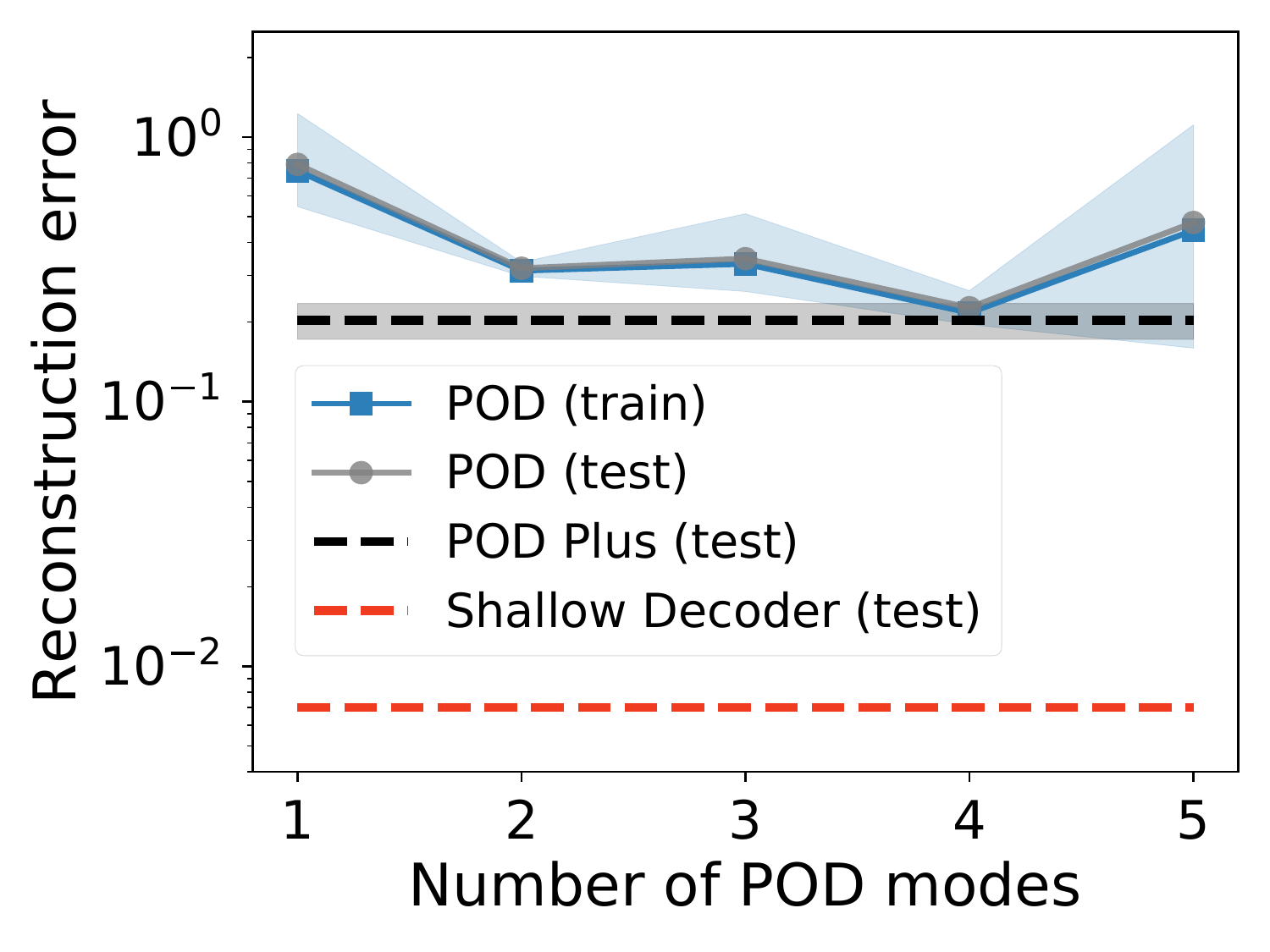} 
			\put(44, 76){\scriptsize 5 sensors}	
		\end{overpic}
		\begin{overpic}[width=0.31\textwidth]{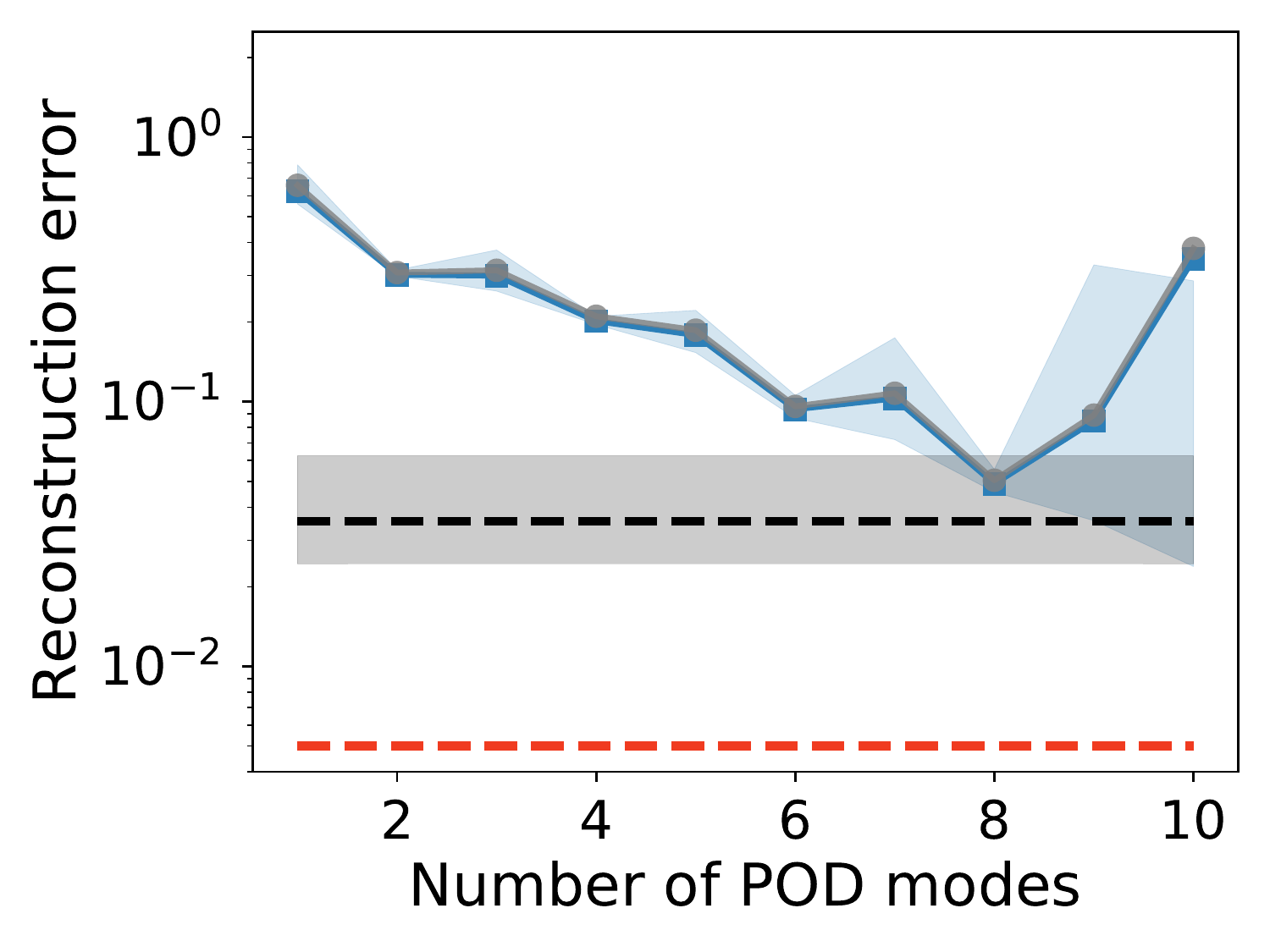} 
			\put(44, 76){\scriptsize 10 sensors}	
		\end{overpic}
		\begin{overpic}[width=0.31\textwidth]{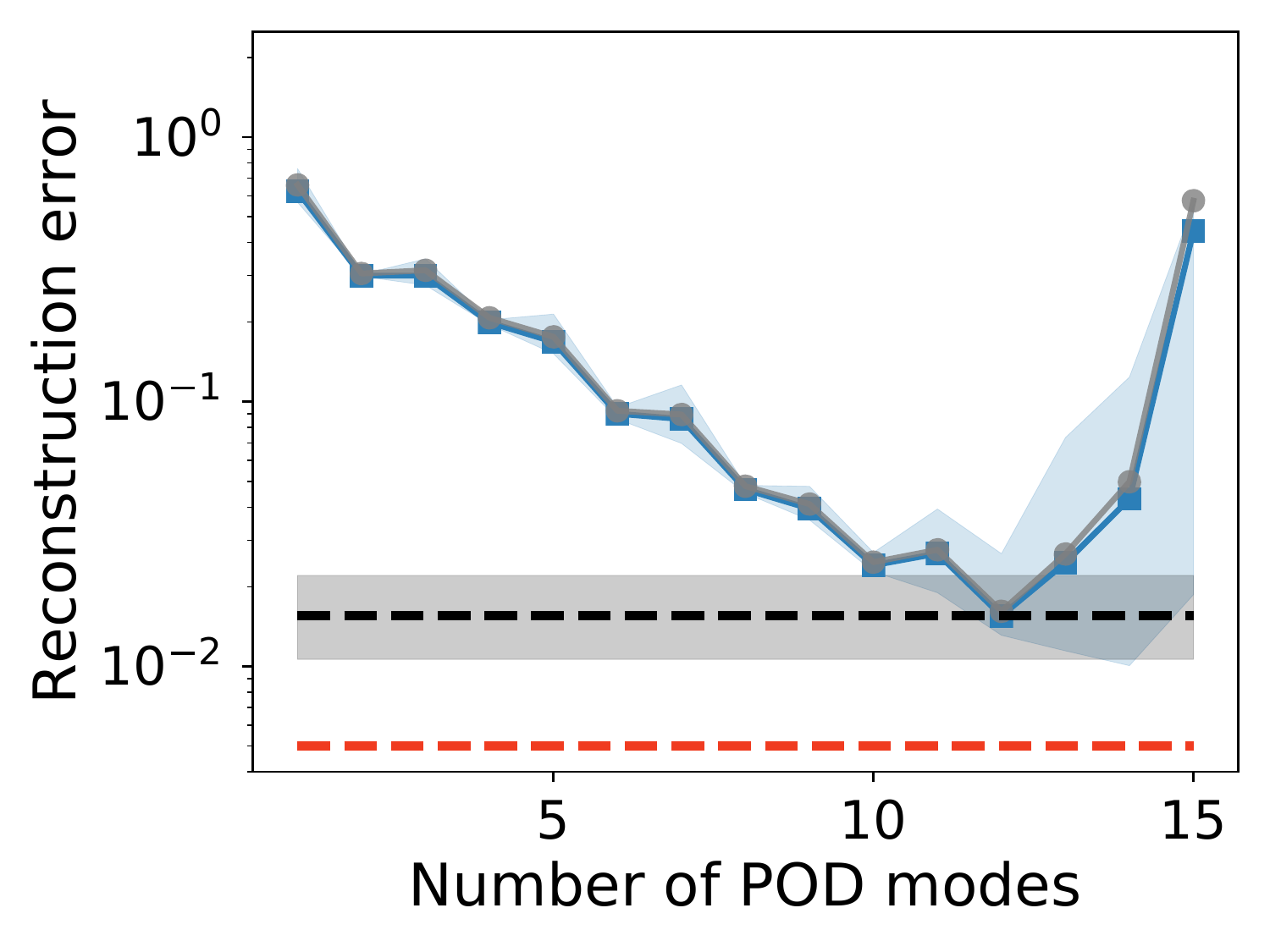} 
			\put(44, 76){\scriptsize 15 sensors}	
		\end{overpic}		
		\caption{POD with hard-threshold regularization.}
	\end{subfigure}
	
	\caption{Results of the hyper-parameter search for the flow past the cylinder. The results for the POD-PLUS method, using ridge regularization, are shown in (a), and the results for the POD method, using hard-threshold, are shown in (b). The shallow decoder outperforms the POD-based methods in all situations, while the performance gap closes for an increased number of sensors.}
	\label{fig:pod_flow}
\end{figure}

\begin{figure}[!t]
	\centering
	\begin{subfigure}[t]{0.48\textwidth}
		\centering
		\DeclareGraphicsExtensions{.pdf}
		
		\begin{overpic}[width=0.48\textwidth]{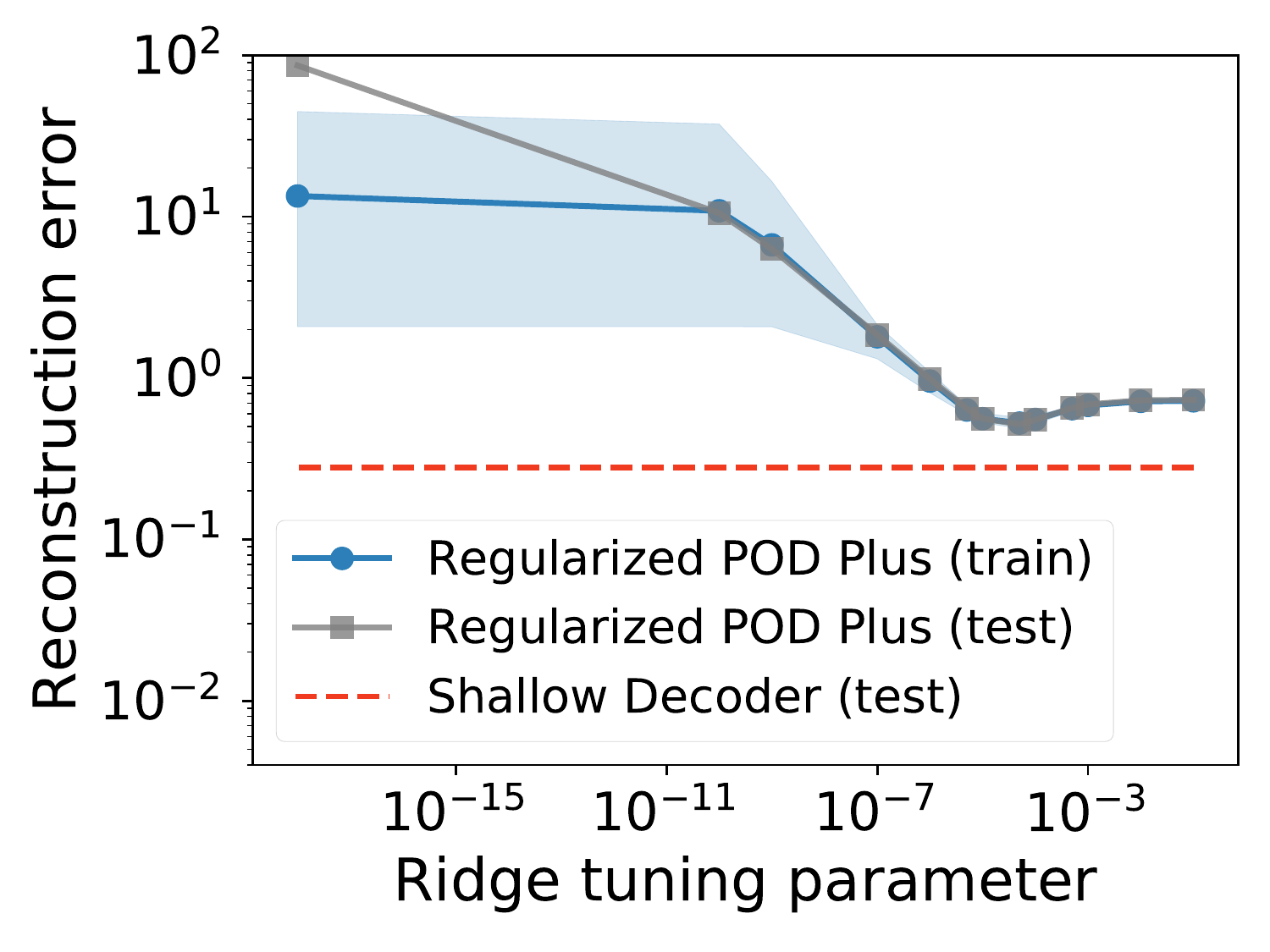} 
			\put(44, 76){\scriptsize SNR 10}	
		\end{overpic}
		\begin{overpic}[width=0.48\textwidth]{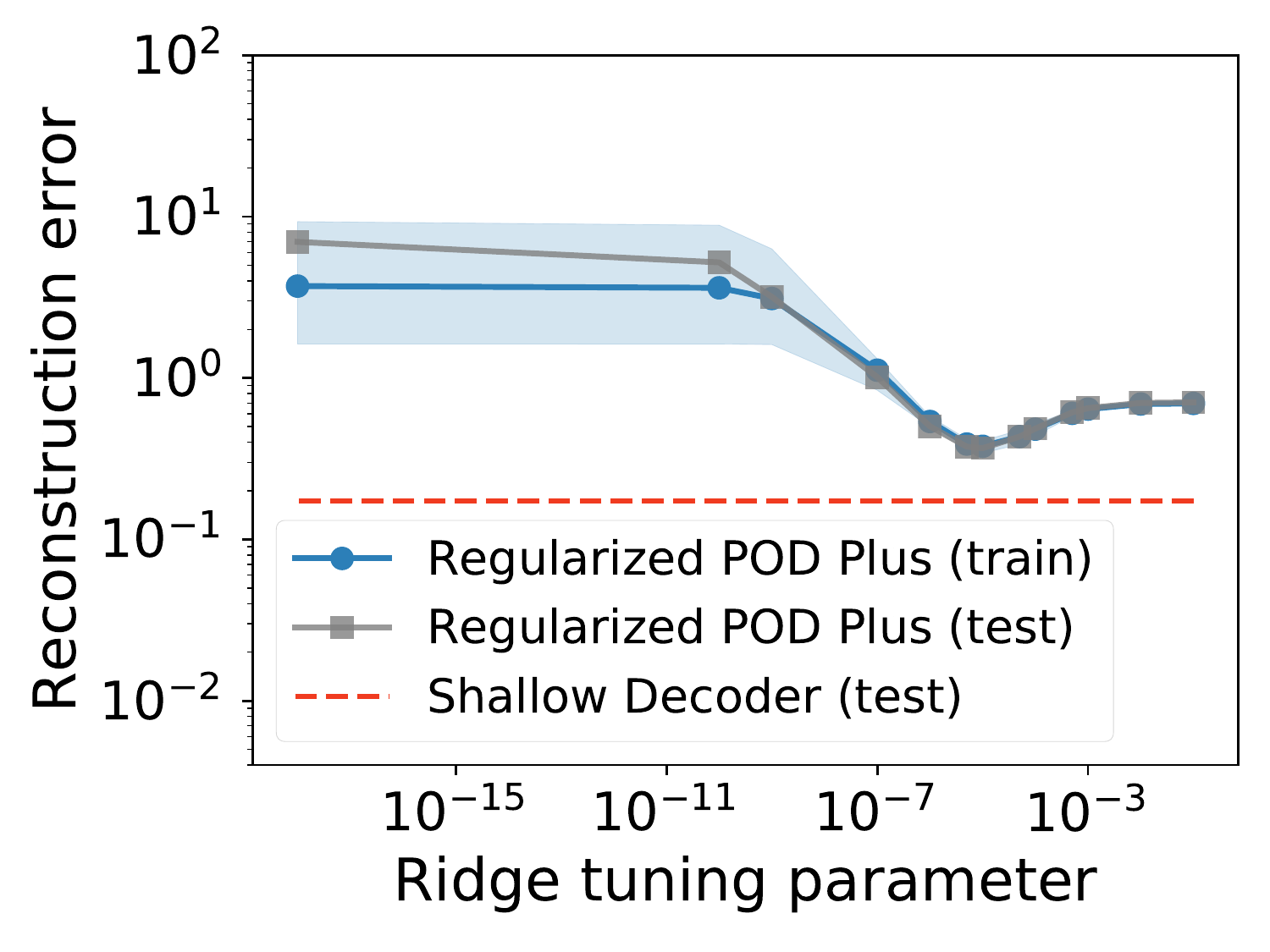} 
			\put(44, 76){\scriptsize SNR 50}	
		\end{overpic}
		\caption{POD Plus with ridge regularization.}
	\end{subfigure}
	~
	\begin{subfigure}[t]{0.48\textwidth}
	\centering
	\DeclareGraphicsExtensions{.pdf}

	\begin{overpic}[width=0.48\textwidth]{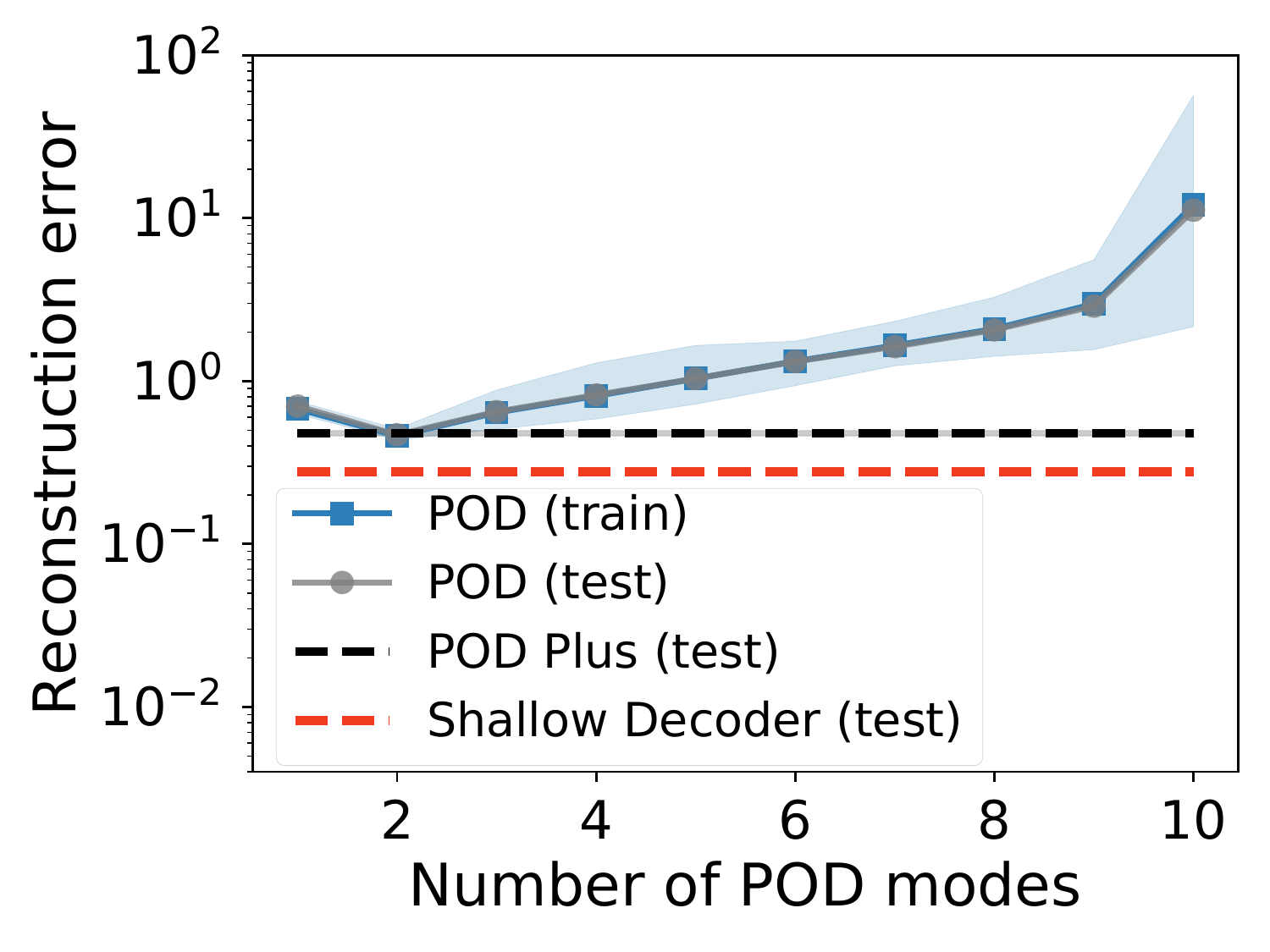} 
		\put(44, 76){\scriptsize SNR 10}	
	\end{overpic}
	\begin{overpic}[width=0.48\textwidth]{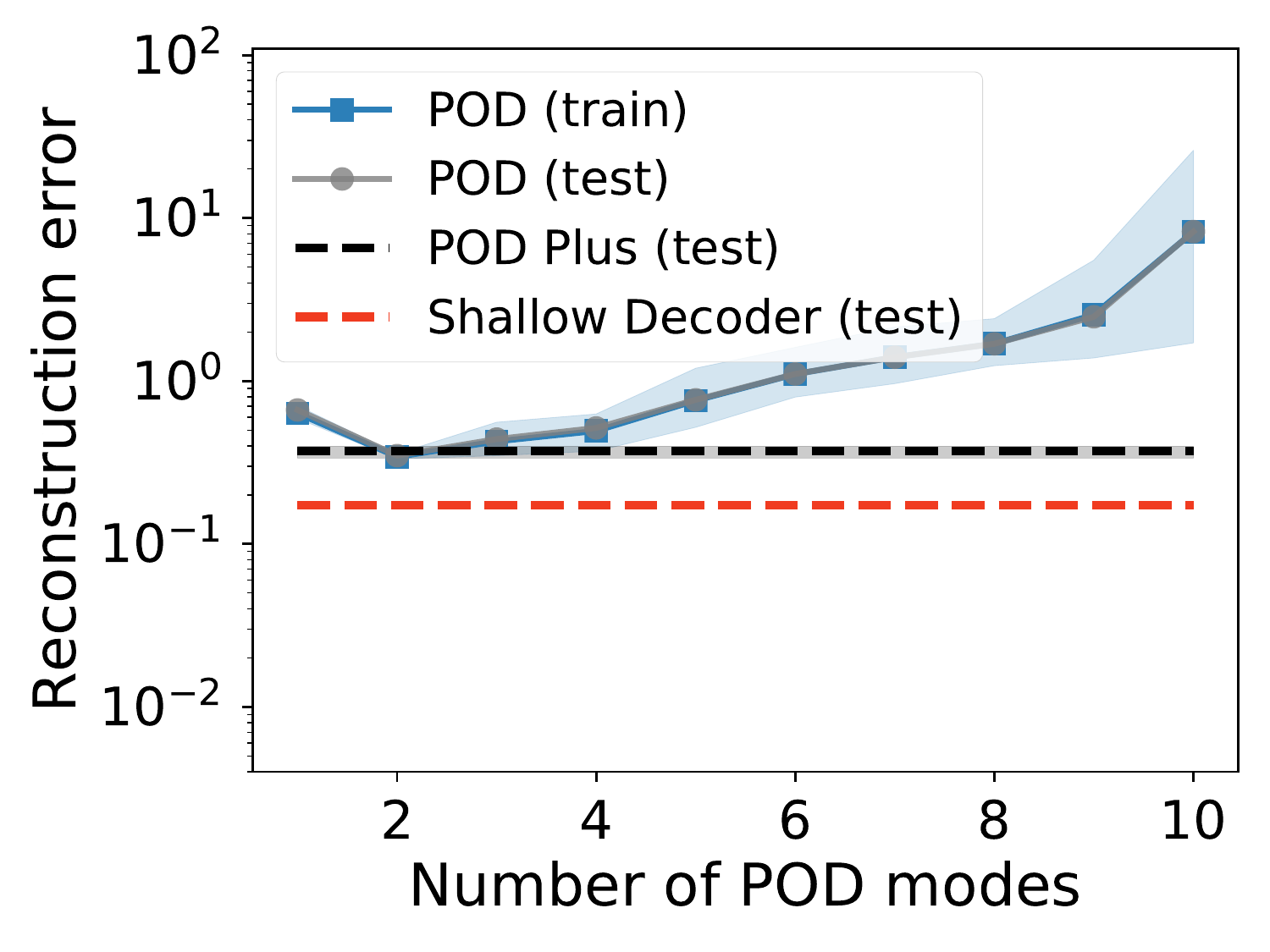} 
		\put(44, 76){\scriptsize SNR 50}	
	\end{overpic}
	\caption{POD with hard-threshold regularization.}
	\end{subfigure}
	
	\caption{Results of the hyper-parameter search for the noisy flow past the cylinder. Here we consider the signal-to-noise (SNR) ratios 10 and 50. Here we consider a setting with $10$ sensors.}
	\label{fig:pod_flow_10sensors}

\end{figure}

\begin{figure}[!t]
	
	\centering
\begin{subfigure}[t]{0.48\textwidth}
	\centering
	\DeclareGraphicsExtensions{.pdf}
	
	\begin{overpic}[width=0.48\textwidth]{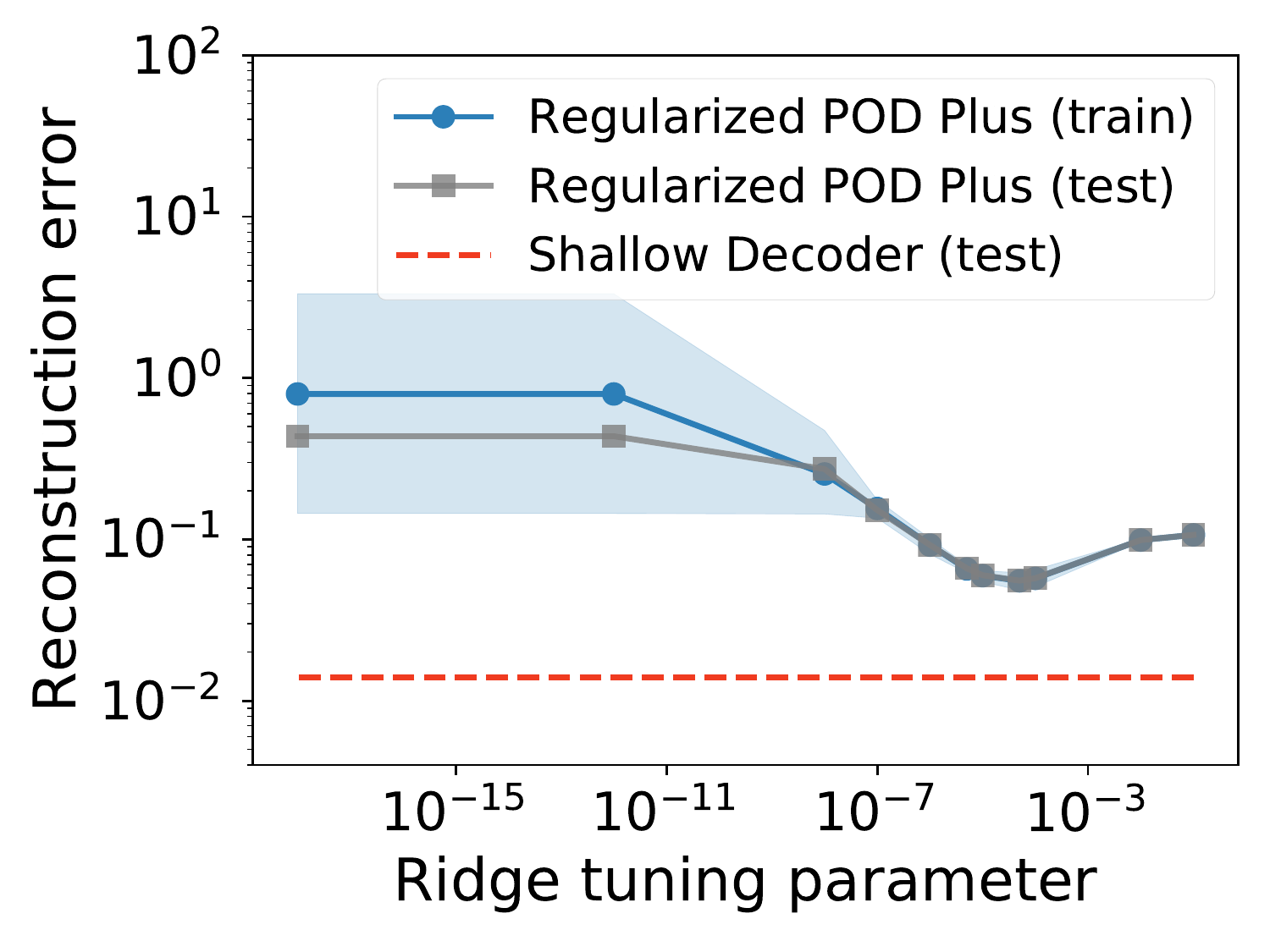} 
		\put(42, 76){\scriptsize 32 sensors}	
	\end{overpic}
	\begin{overpic}[width=0.48\textwidth]{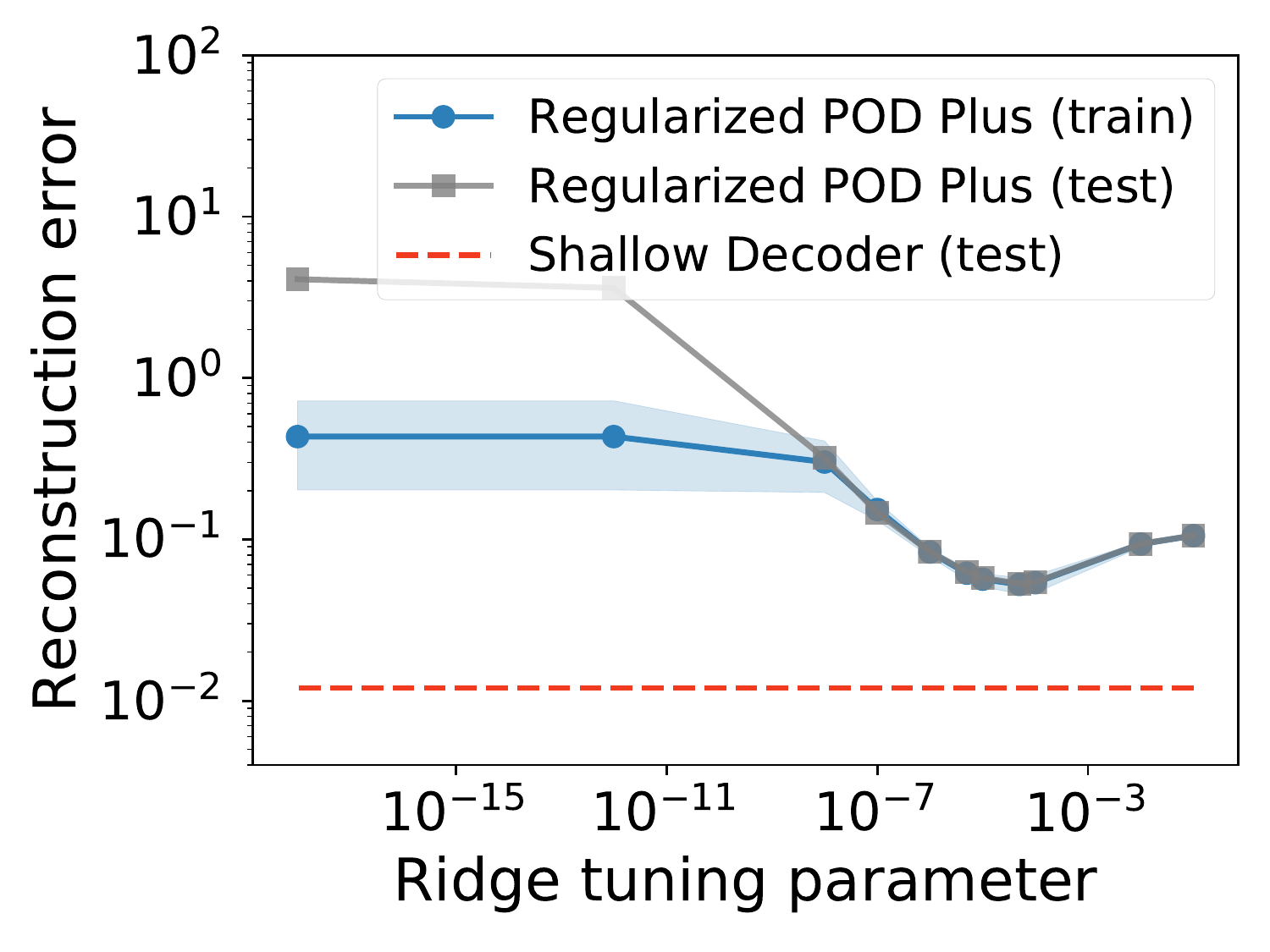} 
		\put(42, 76){\scriptsize 64 sensors}	
	\end{overpic}
	\caption{POD Plus with ridge regularization.}
	\end{subfigure}	
	~
	\begin{subfigure}[t]{0.48\textwidth}
	\centering
	\DeclareGraphicsExtensions{.pdf}
	
	\begin{overpic}[width=0.48\textwidth]{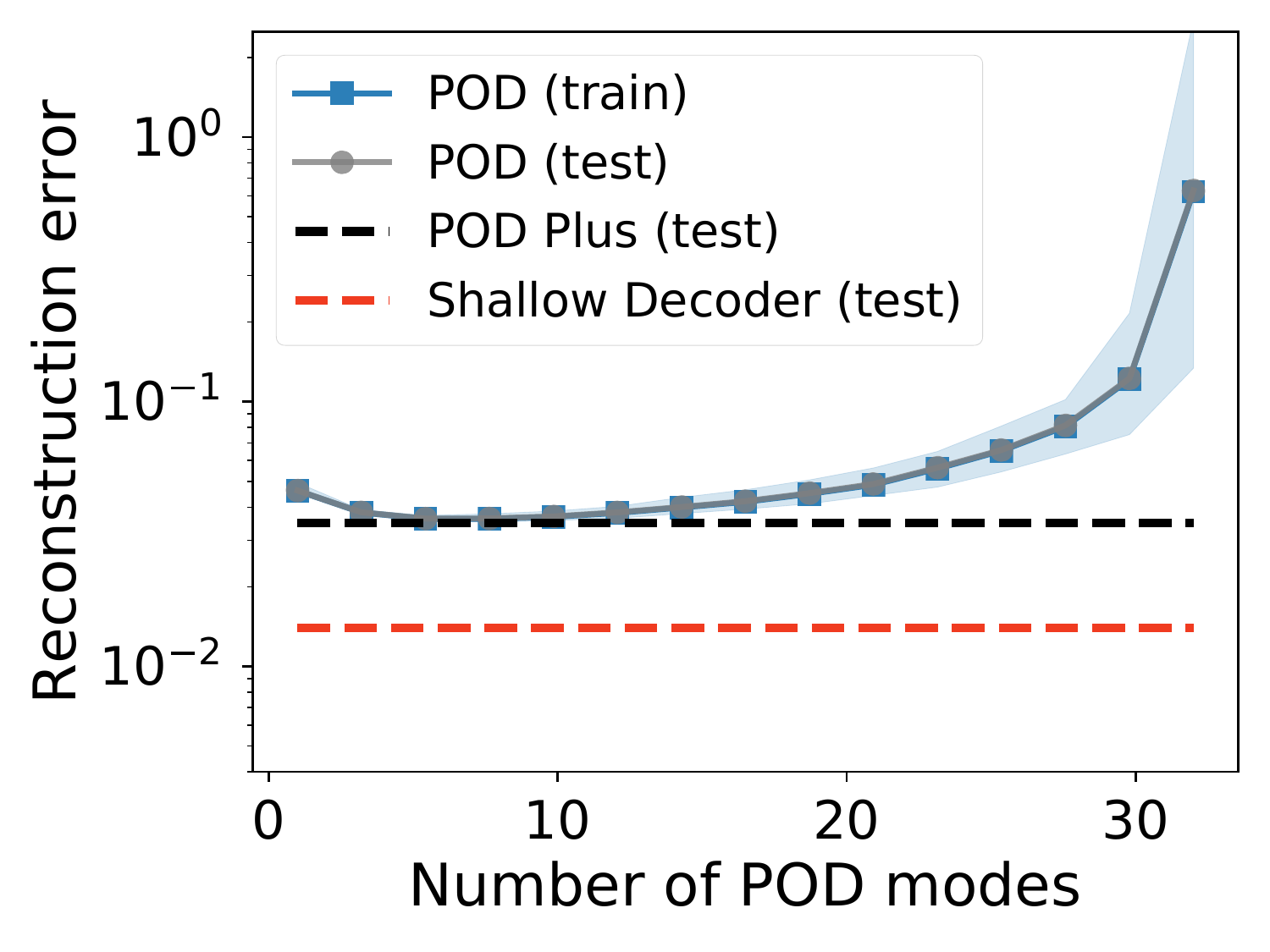} 
		\put(42, 76){\scriptsize 32 sensors}	
	\end{overpic}
	\begin{overpic}[width=0.48\textwidth]{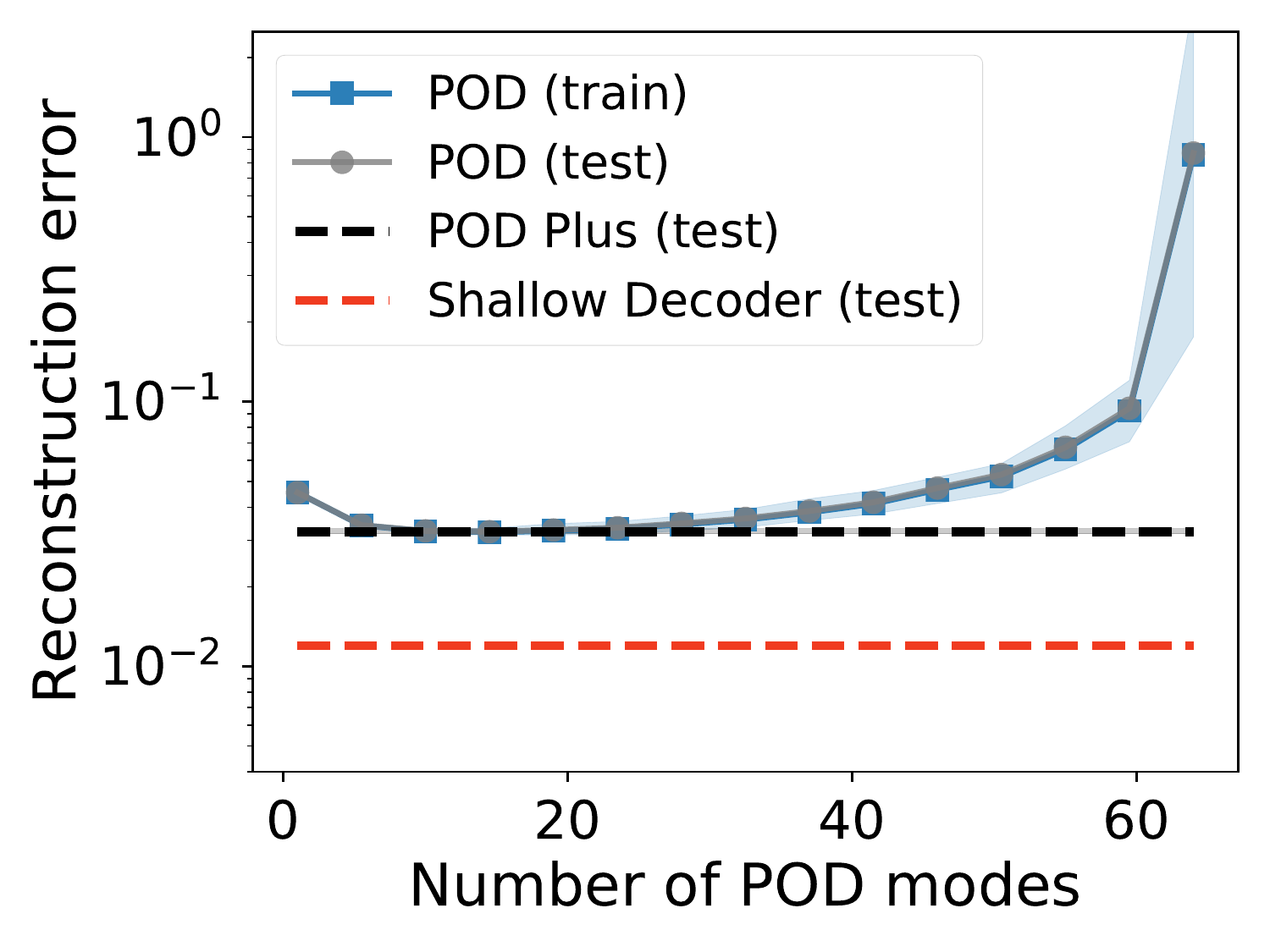} 
		\put(42, 76){\scriptsize 64 sensors}	
	\end{overpic}
	\caption{POD with hard-threshold regularization.}
	\end{subfigure}

	\caption{Results of the hyper-parameter search for the SST data. Here we consider a setting with $32$ and $64$ sensors. The shallow decoder outperforms the POD-based methods in all situations.}
	\label{fig:pod_flow_regul}

\end{figure}

}

\section{Singular spectrum analysis of reconstructed data} \label{sec:spectrum}

\begin{figure}[!b]
	
	\centering
	\begin{subfigure}[t]{0.9\textwidth}
		\centering
		\DeclareGraphicsExtensions{.pdf}
		
		\begin{overpic}[width=0.48\textwidth]{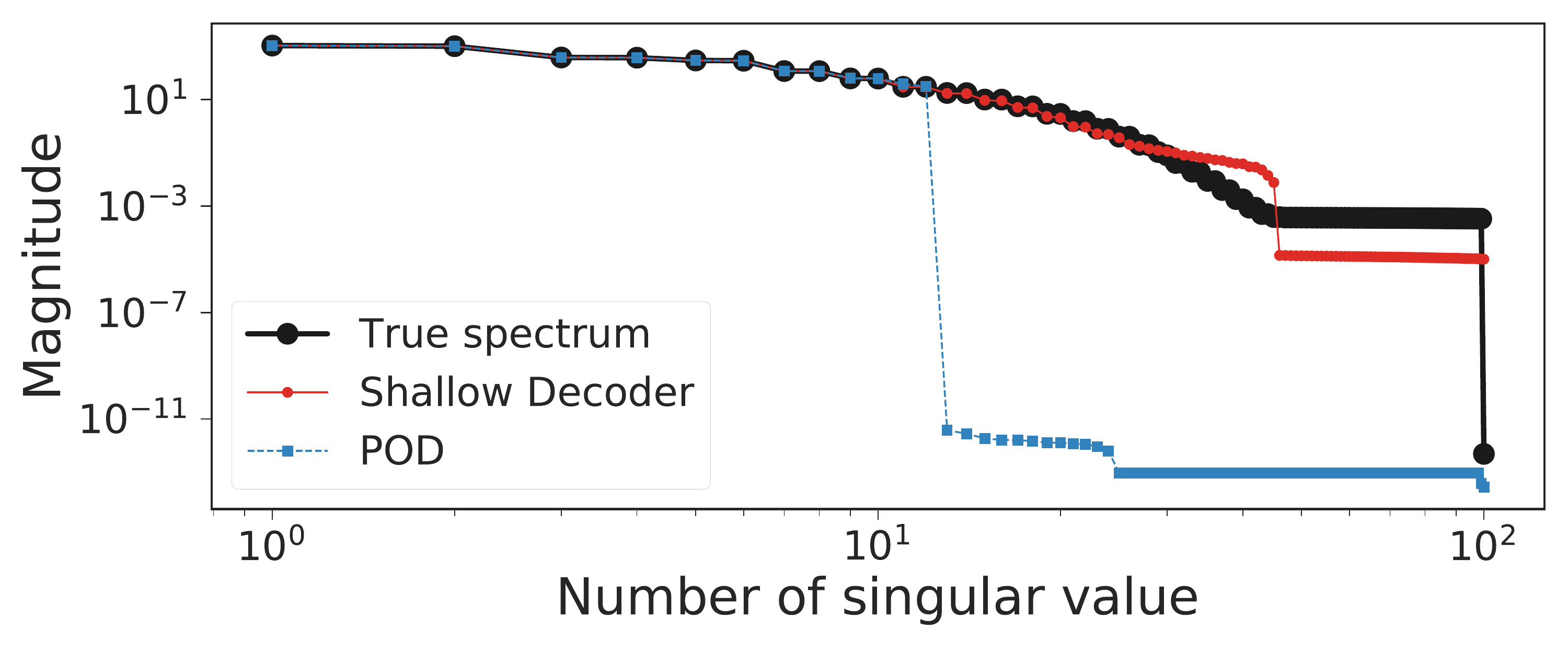} 
		\end{overpic}
		\begin{overpic}[width=0.48\textwidth]{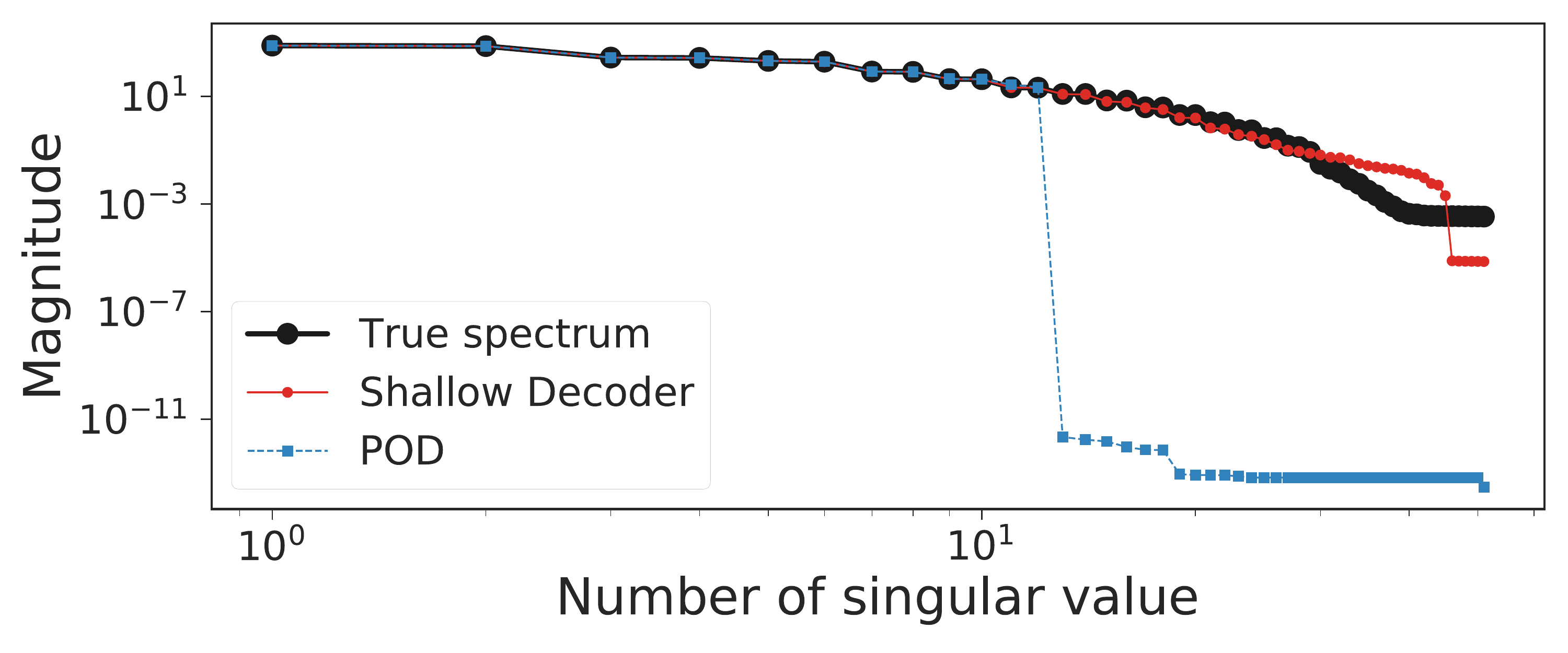} 
		\end{overpic}
		\caption{Fluid flow behind the cylinder. We use $15$ sensor location for reconstruction.}
	\end{subfigure}	
	~
	\begin{subfigure}[t]{0.9\textwidth}
		\centering
		\DeclareGraphicsExtensions{.pdf}
		
		\begin{overpic}[width=0.48\textwidth]{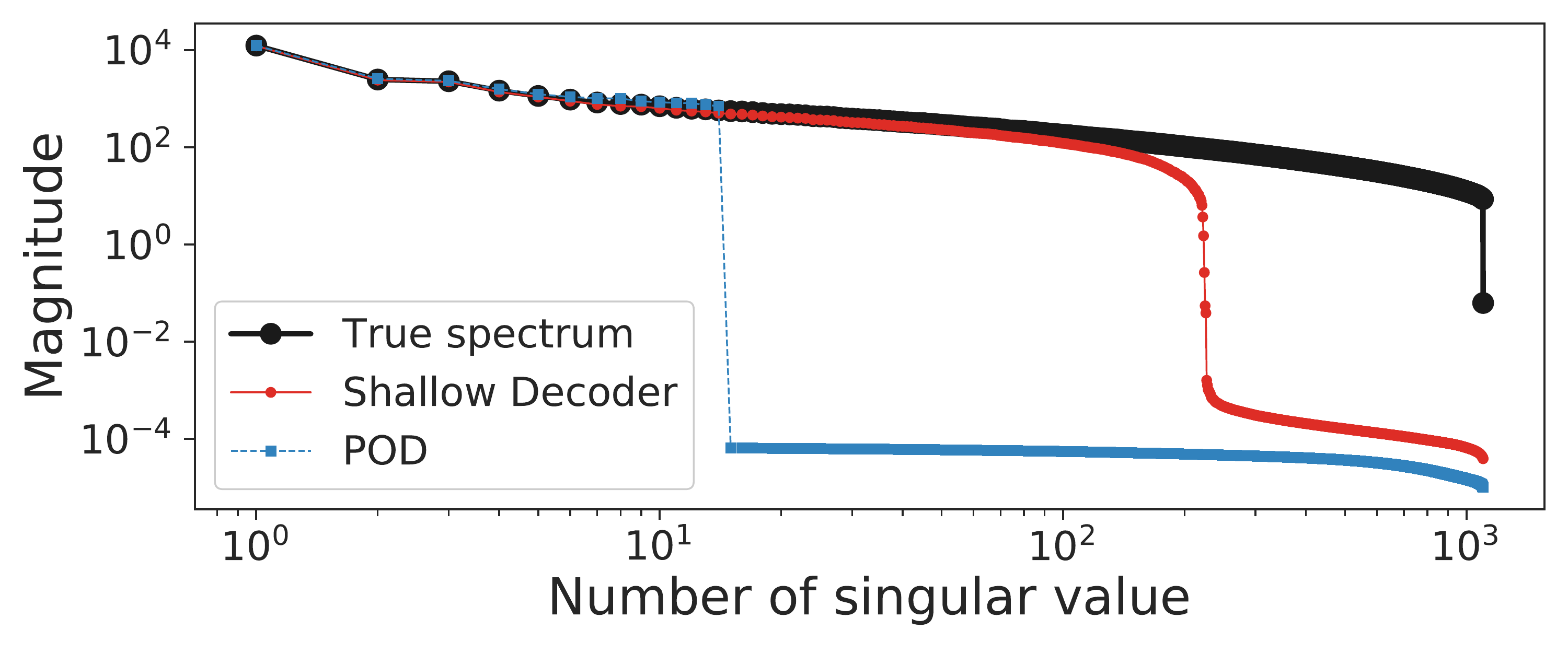} 
		\end{overpic}
		\begin{overpic}[width=0.48\textwidth]{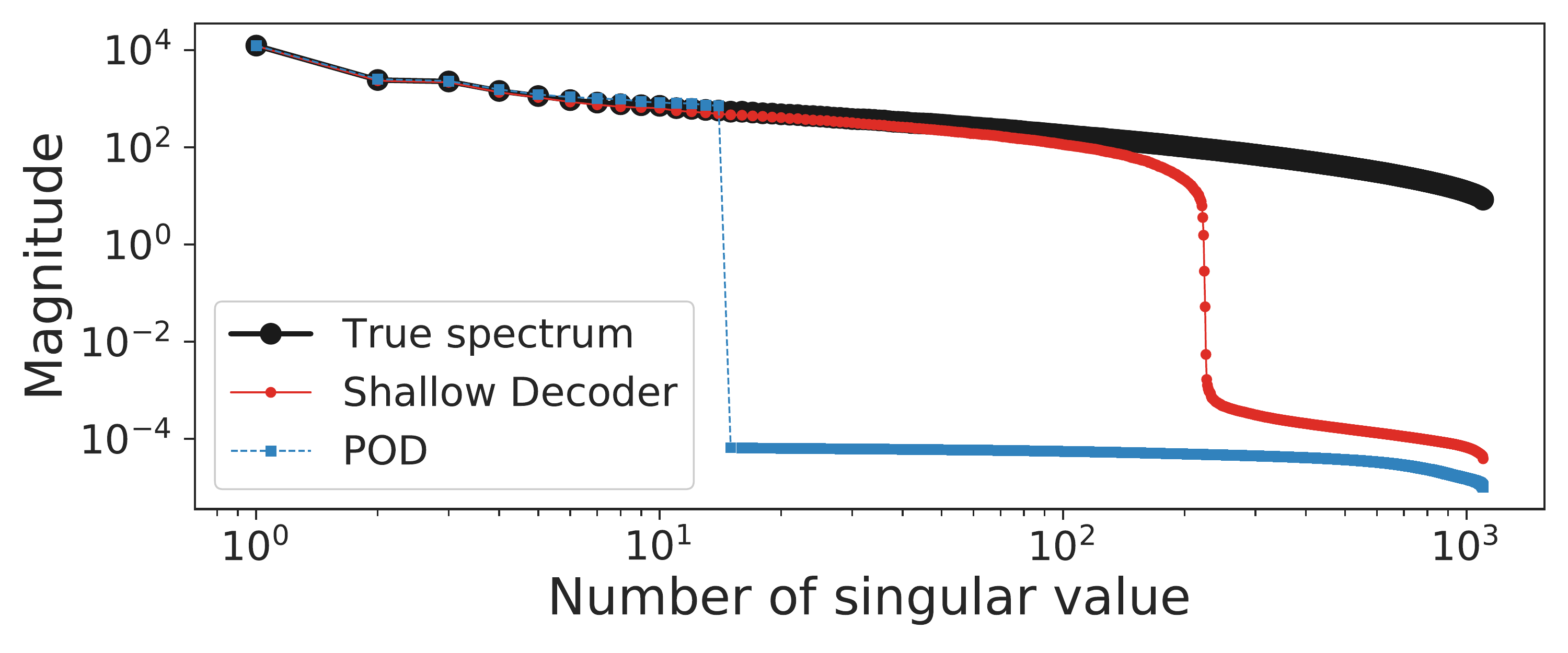} 
		\end{overpic}
		\caption{Sea surface temperature data. We use $64$ sensor location for reconstruction.}
	\end{subfigure}	
	~
	\begin{subfigure}[t]{0.9\textwidth}
		\centering
		\DeclareGraphicsExtensions{.pdf}
		
		\begin{overpic}[width=0.48\textwidth]{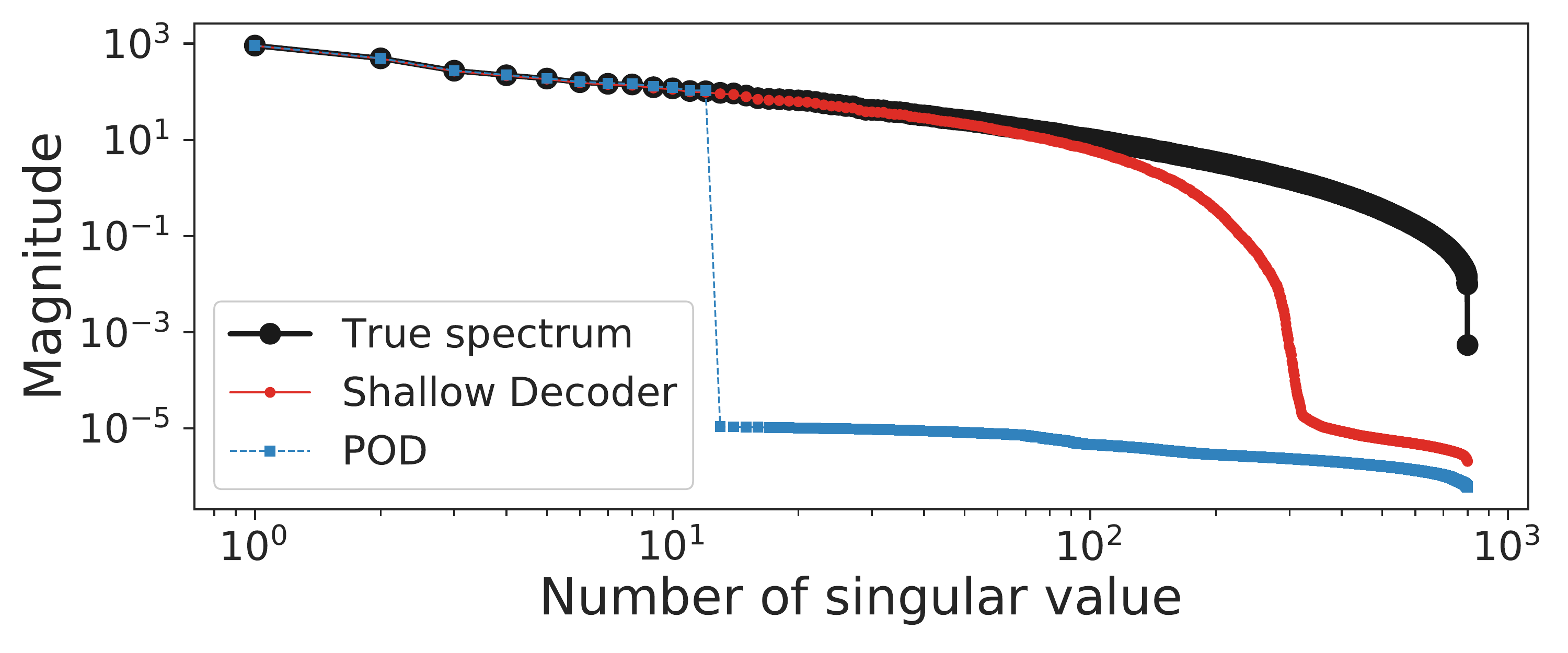} 
		\end{overpic}
		\begin{overpic}[width=0.48\textwidth]{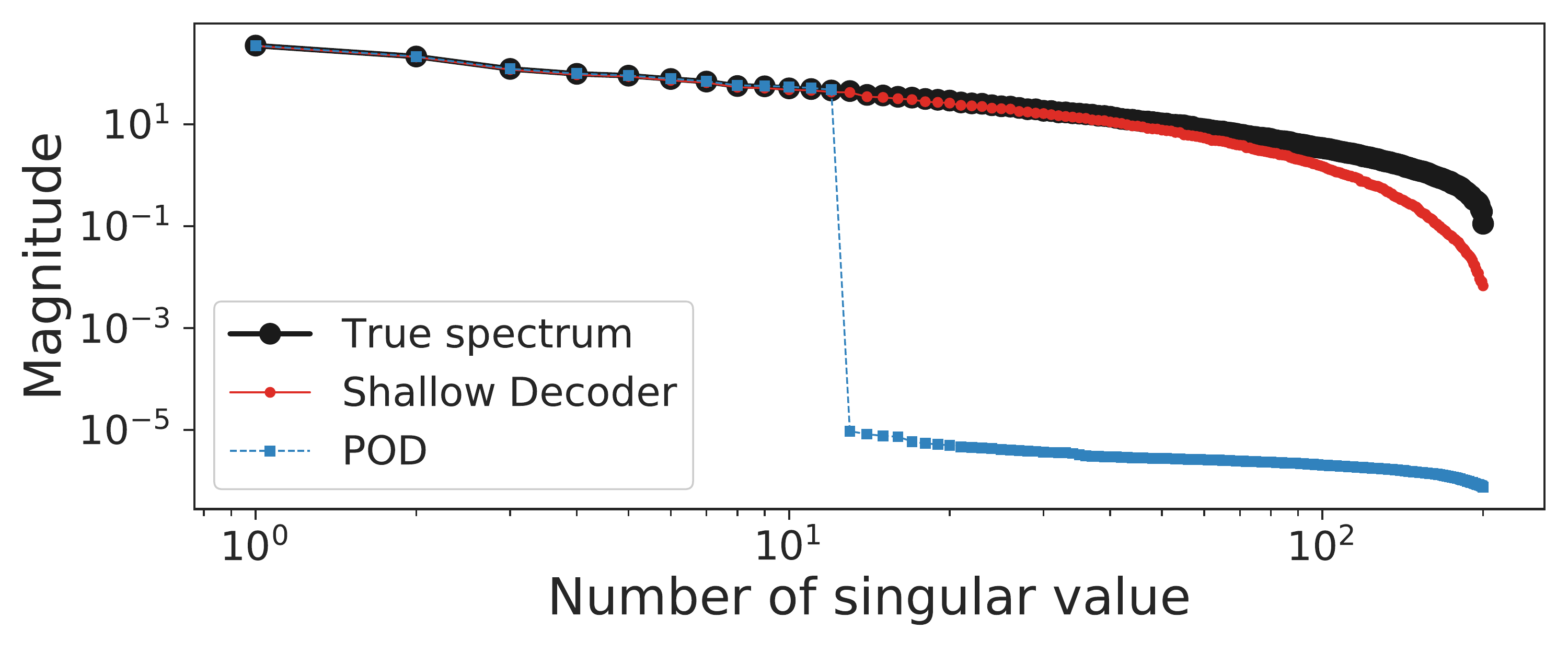} 
		\end{overpic}
		\caption{Turbulent flow. We use $128$ sub-gridscale measurements for reconstruction.}
	\end{subfigure}		
	
	\caption{Singular spectrum analysis of the reconstructed data and ground truth. The left column shows the training data and the right column the test data.  The POD-based method uses the optimal $k^*$ truncation to reconstruct the data. Here we consider in-sample test data. }
	\label{fig:spectrum_analysis}
	
\end{figure}

Here we provide additional results that show the singular value spectrum of the reconstructed training and test data. As reference we also show the spectrum of the ground truth data. Figure~\ref{fig:spectrum_analysis} shows the results for (a) the fluid flow behind the cylinder, (b) the sea surface data, and (c) the turbulent flow. The singular value spectrum of the reconstructed data helps us to compare the performance between the POD-based method and our shallow decoder.

For all problems that we consider, it can be seen that the shallow decoder captures more fine-scale information as compared to the truncated POD-based method. Note, that we consider the case where the training and test data are sampled from the same distribution.

\begin{table}[!b]
	\centering\scalebox{0.9}{
		\begin{tabular}{cccccccc} \toprule
			& Layer     & Weight size   & Input Shape         & Output Shape  & Activation & Batch Norm. & Dropout\\
			\midrule
			& FC   & sensors $\times$ 35  &  sensors    & 35   &  ReLU &   True &   -\\
			& FC   & 35 $\times$ 40     &  25      	& 40       &  ReLU & True &   -\\
			& FC   & 40 $\times$ 76,416   &  40        	& 76,416   &  Linear  &   - &   -\\ \bottomrule 
	\end{tabular}}\vspace{+0.2cm}
	\caption{Architecture of the SD for the flow behind the cylinder. The batch size is set to $32$. Here, we set the dropout rate to $0.1$ for the noisy situation. {We use a small amount of weight decay $\lambda=1\text{e-}{7}$.} }
	\label{tab:arch_model1}
\end{table}

\begin{table}[!b]\vspace{-0.1cm}
	\centering\scalebox{0.9}{
		\begin{tabular}{cccccccc} \toprule
			& Layer             & Weight size                          & Input Shape         & Output Shape  & Activation & Batch Norm. & Dropout\\
			\midrule
			& FC   & sensors $\times$ 350  &  sensors   & 350   &  ReLU &   True &   $0.1$\\
			& FC   & 350 $\times$ 400     &  350      	& 400       &  ReLU & True &   -\\
			& FC   & 400 $\times$ 44,219   &  400       & 44,219   &  Linear  &   - &   - \\ \bottomrule 
	\end{tabular}}\vspace{+0.2cm}
	\caption{Architecture of the SD for the SST dataset. Here, the batch size is set to $200$.}
	\label{tab:arch_model2}
\end{table}

\begin{table}[!b]\vspace{-0.1cm}
	\centering\scalebox{0.9}{	
		\begin{tabular}{cccccccc} \toprule
			& Layer             & Weight size                          & Input Shape         & Output Shape  & Activation & Batch Norm. & Dropout\\
			\midrule
			& FC   & sensors $\times$ 350  &  sensors   & 350   &  ReLU &   True & $0.1$ \\
			& FC   & 350 $\times$ 400     &  350      	& 400       &  ReLU & True &   - \\
			& FC   & 400 $\times$ 122,500   &  400       & 122,500   &  Linear  &   -  &   -\\ \bottomrule 
	\end{tabular}}\vspace{+0.2cm}
	\caption{Architecture of the SD for isotropic flow. Here, the batch size is set to $200$.}
	\label{tab:arch_model3}
\end{table}

\section{Setup for our empirical evaluation} \label{sec:setup}

Here, we provide details about the concrete network architectures of the shallow decoder, which are used for the different examples. The networks are implemented in Python using PyTorch; and research code for flow behind the cylinder is available via \url{https://github.com/erichson/ShallowDecoder}.
Tables~\ref{tab:arch_model1}--~\ref{tab:arch_model3} show the details. For each example we use a similar architecture design. The difference is that we use a slightly wider design (more neurons per layer) for the SST dataset and the isotropic flow. That is because we are using a larger number of sensors for these two problems, and thus we need to increase the capacity of the network. In each situation, the learning rate is set to $1\text{e-}{2}$ with a scheduled decay rate of $0.3$. Further, 
we use a small amount of weight decay $\lambda=1\text{e-}{7}$ to regularize the network.

\clearpage
{\normalsize
\bibliographystyle{plainnat} 
\bibliography{main}   
}

\end{document}